\documentstyle[psfig]{mn}

\def\etal{{\it et al.\ }}
\def\eg{{\it e.g.\ }}

\def\ie{{\it i.e.\ }}

\def\spose#1{\hbox to 0pt{#1\hss}}
\def\approxlt{\mathrel{\spose{\lower 3pt\hbox{$\sim$}}
	\raise 2.0pt\hbox{$<$}}}
\def\approxgt{\mathrel{\spose{\lower 3pt\hbox{$\sim$}}
	\raise 2.0pt\hbox{$>$}}}
\def\approxpropto{\mathrel{\spose{\lower 3pt\hbox{$\sim$}}
	\raise 2.0pt\hbox{$\propto$}}}
\mathchardef\twiddle="2218

\def\multleft#1{\hbox to size{\vbox {\halign {\lft{##}\cr #1}}\hfill}\par}
\def\multright#1{\hbox to size{\vbox {\halign {\rt{##}\cr #1}}\hfill}\par}

\def\today{\ifcase\month\or January\or February\or March\or April\or May\or
      June\or July\or August\or September\or October\or November\or December\fi
      \space\number\day, \number\year}
\def\<{\thinspace}

\def\apc{\rm atom cm$^{-2}$}

\def\cm{{\rm\thinspace cm}}
\def\erg{{\rm\thinspace erg}}

\def\keV{{\rm\thinspace keV}}

\def\km{{\rm\thinspace km}}

\def\Mpc{{\rm\thinspace Mpc}}
\def\Msun{\hbox{$\rm\thinspace M_{\odot}$}}

\def\s{{\rm\thinspace s}}
\def\yr{{\rm\thinspace yr}}

\def\ergpcmsqps{\hbox{$\erg\cm^{-2}\s^{-1}\,$}}

\def\ergps{\hbox{$\erg\s^{-1}\,$}}

\def\kmps{\hbox{$\km\s^{-1}\,$}}

\def\Msunpyr{\hbox{$\Msun\yr^{-1}\,$}}

\def\psqcm{\hbox{$\cm^{-2}\,$}}

\def\kmpspMpc{\hbox{$\kmps\Mpc^{-1}$}}

\def\apc{\rm atom cm$^{-2}$}
\title[Intrinsic absorption and cooling gas 
in cooling flows]
{The spatial distributions of cooling gas and intrinsic X-ray absorbing
material in cooling flows}
\author[S.W. Allen and A.C. Fabian ]
{\parbox[]{6.in} {S.W. Allen and A.C. Fabian \\
\footnotesize
Institute of Astronomy, Madingley Road, Cambridge CB3 OHA\\
}}

\begin{document}


\maketitle

\begin{abstract}
We present the results from a study of the spatial distributions of
cooling gas and intrinsic X-ray absorbing material in a sample of 
nearby, X-ray bright cooling flow
clusters observed with the Position Sensitive Proportional Counter (PSPC)
on ROSAT. Our method of analysis employs X-ray colour profiles,
formed from ratios of the surface brightness profiles of the clusters in
selected energy bands, and an adapted version of the deprojection code of
Fabian \etal (1981). We show that all of the cooling flow
clusters in our sample exhibit significant central concentrations of
cooling gas. At larger radii the clusters
appear approximately isothermal. In detail, the spatial distributions and 
emissivity of the cooling
material are shown to be in excellent agreement with the predictions 
from the deprojection code, and can be used
to constrain the ages of the cooling flows. 
The X-ray colour profiles also indicate substantial levels of 
intrinsic X-ray absorption in the clusters. 
The intrinsic absorption increases with decreasing
radius, and is confined to the regions occupied by the cooling flows. 
We explore a range of likely spatial distributions for the absorbing gas 
and discuss the complexities involved in the measurement 
of column densities from X-ray data. We show 
that the application of simple spectral models, in which the intrinsic
absorber is treated 
as a uniform foreground screen, will naturally lead to 
significant underestimates of the true amounts of absorption. 
Comparison of our results with previously reported 
observations made with Einstein Observatory 
Solid State Spectrometer (White \etal 1991) shows reasonable agreement, 
but requires that the absorbing material only
partially covers the X-ray emitting regions. 
The masses of absorbing gas in the central (30 arcsec radius)
regions of the clusters, calculated under the assumption of solar
metallicity in the absorbing material, can be accumulated
by the cooling flows on timescales of a few $10^8$
years, which are much less than the ages of the flows. This implies that most of 
the material deposited by the cooling flows in these regions cannot remain 
in X-ray absorbing gas. The results presented in this paper 
provide strong support for  the standard
model of inhomogeneous cooling flows in clusters of galaxies. 

\end{abstract}


\section{Introduction}

\begin{table*}
\vskip 0.2truein
\begin{center}
\caption{Observation summary}
\vskip 0.2truein
\begin{tabular}{ c c l c l r c c r }
\multicolumn{1}{c}{} &
\multicolumn{1}{c}{} &
\multicolumn{1}{c}{} &
\multicolumn{1}{c}{} &
\multicolumn{2}{c}{OBSERVATIONS} &
\multicolumn{1}{c}{} &
\multicolumn{2}{c}{X-RAY CENTROID (J2000)} \\                            
 cluster     & ~ &    z     & ~ &       Date     &    Exposure (s)    & ~
&     R.A.                              &    Dec.~~~~                \\  
\hline                                                                                                                               
&&&&&&&& \\                                                                                                                         
Abell 85     & ~ &  0.0518  & ~ &   1992 Jul 01  &     10086           & ~ &   $00^{\rm h}41^{\rm m}50.5^{\rm s}$ &  $-09^{\circ}17'58''$  \\  
Abell 3112   & ~ &  0.0746  & ~ &   1992 Dec 17  &     7600            & ~ &   $03^{\rm h}17^{\rm m}58.4^{\rm s}$ & $-44^{\circ}14'27''$   \\  
Abell 426    & ~ &  0.0183  & ~ &   1992 Feb 02  &     4787            & ~ &   $03^{\rm h}19^{\rm m}48.5^{\rm s}$ &  $41^{\circ}30'27''$   \\  
Abell 478    & ~ &  0.088   & ~ &   1991 Aug 31  &     22139           & ~ &   $04^{\rm h}13^{\rm m}24.9^{\rm s}$ &  $10^{\circ}28'04''$   \\  
Abell 496    & ~ &  0.0320  & ~ &   1991 Mar 06  &     8852            & ~ &   $04^{\rm h}33^{\rm m}37.9^{\rm s}$ &  $-13^{\circ}15'43''$  \\  
Abell 644    & ~ &  0.0704  & ~ &   1993 Apr 27  &     10285           & ~ &   $08^{\rm h}17^{\rm m}25.3^{\rm s}$ & $-06^{\circ}49'59''$   \\  
Hydra A      & ~ &  0.0522  & ~ &   1992 Nov 08  &     18070           & ~ &   $09^{\rm h}18^{\rm m}06.2^{\rm s}$ &  $-12^{\circ}05'41''$  \\  
Coma         & ~ &  0.0232  & ~ &   1991 Jun 17  &     22108           & ~ &   $12^{\rm h}59^{\rm m}46.1^{\rm s}$ &  $27^{\circ}56'21''$   \\  
Virgo        & ~ &  0.0043  & ~ &   1992 Dec 17  &     9961            & ~ &   $12^{\rm h}30^{\rm m}49.8^{\rm s}$ &  $12^{\circ}23'32''$   \\  
Centaurus    & ~ &  0.0104  & ~ &   1994 Jul 05  &     3192            & ~ &   $12^{\rm h}48^{\rm m}48.7^{\rm s}$ &  $-41^{\circ}18'44''$   \\  
Abell 1795   & ~ &  0.0634  & ~ &   1992 Jan 04  &     36515           & ~ &   $13^{\rm h}48^{\rm m}53.1^{\rm s}$ &  $26^{\circ}35'34''$   \\  
Abell 2029   & ~ &  0.0767  & ~ &   1992 Aug 10  &     12353           & ~ &   $15^{\rm h}10^{\rm m}55.7^{\rm s}$ &  $05^{\circ}44'39''$    \\  
MKW3s        & ~ &  0.043   & ~ &   1992 Aug 15  &     9802            & ~ &   $15^{\rm h}21^{\rm m}51.7^{\rm s}$ &  $07^{\circ}42'09''$   \\  
Abell 2199   & ~ &  0.0309  & ~ &   1993 Jul 25  &     40290           & ~ &   $16^{\rm h}28^{\rm m}38.4^{\rm s}$ &  $39^{\circ}33'00''$   \\  
Cyg A        & ~ &  0.057   & ~ &   1993 Oct 10  &     9447            & ~ &   $19^{\rm h}59^{\rm m}28.4^{\rm s}$ & $40^{\circ}43'59''$   \\  
Sersic 159   & ~ &  0.0556  & ~ &   1993 May 28  &     13137           & ~ &   $23^{\rm h}13^{\rm m}58.3^{\rm s}$ &  $-42^{\circ}43'35''$  \\  
Abell 2597   & ~ &  0.0824  & ~ &   1991 Nov 27  &     7094            & ~ &   $23^{\rm h}25^{\rm m}20.0^{\rm s}$ &  $-12^{\circ}07'31''$  \\  
Abell 4059    & ~ &  0.0478  & ~ &   1991 Nov 21  &     5514            & ~ &   $23^{\rm h}57^{\rm m}02.0^{\rm s}$ & $-34^{\circ}36'23''$   \\  


&&&&&&&& \\                                                                                                                         
\end{tabular}
\end{center}
\parbox {7in}
{Notes: Summary of the PSPC observations. From left to right we list the 
names, 
redshifts, observation dates, effective exposure times 
(after corrections for satellite dead
time) and the coordinates of the peaks of the
X-ray emission (in the $0.4-2.0$ keV band) from the clusters.}
\end{table*}

Clusters of galaxies are luminous X-ray sources. The bulk of the X-ray
emission arises from bremsstrahlung  and line radiation processes in the 
IntraCluster Medium (ICM). 
In the most luminous systems, the ($2-10$ keV) luminosity 
exceeds $10^{45}$ \ergps and the emission extends to radii $> 2$ Mpc. 
The total mass of the ICM exceeds that of the galaxies 
by factors of $1-5$. In the outer regions of clusters ($r \approxgt 1$ Mpc) 
the ICM is highly diffuse ($n_e \approxlt 10^{-4}$ cm$^{-3}$)  and 
the cooling time far exceeds the Hubble time ($t_{\rm cool} 
\approxgt 10^{11}$ yr).
In the central regions of most
clusters, however, the density of the ICM rises sharply and 
the cooling time inferred is significantly less than the Hubble time 
($t_{\rm cool} \approxlt 10^9$ yr;
\eg Edge, Stewart \& Fabian 1992).  In the absence of forces other than thermal
pressure  and gravity, the cooling of the ICM will lead to a slow net inflow
of material towards the cluster centre; a process known as a  {\it cooling
flow}. (See Fabian 1994 for a review of the theory and observations of
cooling flows.) 

Within the idealized model of a {\it homogeneous} cooling flow (in
which all of the gas at any particular radius has the same temperature and
density), all of the cooling gas flows to the centre of the system where
it is `deposited', having radiated away its thermal energy. However, 
observations
of clusters of galaxies show that cooling flows are not
homogeneous.  The X-ray surface brightness profiles of clusters, although
sharply peaked, are not as peaked as they would be for homogeneous flows. 
Rather than all of the cool 
gas flowing to the centres of the clusters, the data show that
cooled material is deposited throughout the central few tens to hundreds of
kpc. Typically the deposition occurs with ${\dot M}(r)
\propto r$, where ${\dot M}(r)$ is the integrated mass deposition rate
within radius $r$.  Spatially-resolved X-ray spectroscopy of clusters
also demonstrates the presence of a range of
temperature and density phases in the central regions of cooling flow
clusters (\eg Fabian \etal 1994; Fukazawa \etal 1994; 
Allen, Fabian \& Kneib 1996; Allen \etal
1996b; Fabian \etal 1996).  The X-ray spectra and imaging data firmly 
require that
cooling flows are {\it inhomogeneous}.

For some years, the primary uncertainty with the standard inhomogeneous 
model of cooling flows (Fabian 1994) was the fate of the cooled matter 
(see \eg Fabian, Nulsen \& Canizares 1991). No identification 
of the large masses of cooled gas, which should presumably be accumulated by the flows 
over their lifetimes, was  made. However, a study 
of 21 clusters observed with the Solid State
Spectrometer (SSS) on the Einstein Observatory by White
\etal (1991) showed that 
large masses of 
X-ray absorbing material are common in cooling flows. 
These authors also showed that significant absorption 
is not present in non cooling-flow clusters.  
Spatially-resolved X-ray spectroscopy of the
massive cooling flow cluster Abell 478 with ROSAT (Allen \etal 1993)
showed that the spatial distribution of the intrinsic absorbing material
in this cluster is well-matched to that of the cooling flow.  Allen \etal
(1993) also demonstrated that the mass of absorbing material is in good agreement
with the mass likely to have been accumulated by the cooling flow over
its lifetime. The data thus suggest the X-ray absorbing
material as a plausible sink for much of the cooled gas deposited by cooling flows.
Recent spectral studies of cooling flows with ASCA (Fabian \etal 1994;
Fabian \etal 1996) have further 
confirmed the presence of cooling gas and intrinsic absorbing
material in cooling flows. However, to date, the constraints on the 
spatial distributions of these
components remain poor. 

In this paper, we present detailed results on the 
distributions of cool gas and
intrinsic X-ray absorbing material in a sample of 17 cooling flow clusters
observed with the Position Sensitive Proportional Counter (PSPC) on
ROSAT. Our target clusters were selected on the basis of the following
criteria; that they be X-ray bright ($F_{\rm X,2-10} > 1.7 \times 10^{-11} 
\ergpcmsqps$;  Edge \etal 1990),  relatively
nearby ($z < 0.1$; allowing them to be resolved on scales of
tens of kpc with the PSPC), and contain large cooling flows (${\dot M} >
100$\Msunpyr). We also
include two lower-luminosity (smaller ${\dot M}$) systems, 
the Centaurus cluster and the Virgo cluster, since they are 
previously well-studied and since good X-ray data exist for them. 
Finally, we included one non-cooling flow cluster, the Coma cluster,  
in our sample as a control.

The large intrinsic column densities ($\Delta N_{\rm H} \sim 10^{21}$
\apc) in cooling flows inferred from the Einstein SSS (White
\etal 1991) and ASCA data (Fabian \etal 1994) must have a
significant effect on the flux detected with the PSPC at energies below
about 0.8~keV. Indeed, few X-rays can escape the absorbed region at
energies below the carbon edge in the detector window
at 0.28~keV. However, this does not mean that no X-rays escape the
cluster in the carbon band, since there is always a large contribution
from the outer, less intrinsically absorbed,  parts of the cluster 
along any line of sight.
Given the spectral resolution of the PSPC [$\Delta E/E = 0.43(E/0.93
{\rm keV})^{-0.5}$] and the energy-dependent effects of absorption, 
one might expect that 
the lowest energy range observable with the PSPC (0.1--0.4~keV) 
would be optimum for absorption studies. Unfortunately, however, 
the spatial resolution of the PSPC is substantially 
worse in this band than at higher
energies. A significant fraction of the counts from a point source are
scattered to radii $> 1$ arcmin in the 0.1--0.2~keV band, and in 
the 0.2--0.4~keV band  the
Point-Spread-Function (PSF) is still almost a factor of $\sim 2$ larger
than at higher (0.4--2.0 keV) energies (Hasinger \etal 1992). In order to make
meaningful spectra, the spatial regions selected must be at least as
large as the largest relevant PSF. Since bright cooling flows typically 
have radii of only one or two arcmin, it is then difficult to 
make spatially-resolved studies of cooling flows with the PSPC 
using data from the softest band. Certainly, any study of the distribution of
absorbing matter from such an exercise is very limited.

The problems with the PSF are compounded by the large number of 
parameters required to
model the spectrum of an absorbed cooling flow, and the small number of
emission-independent spectral bands of the PSPC. Even a simple model for
an absorbed cooling flow involves 5 free parameters; the normalization,
abundance and temperature of the ambient cluster gas, the cooling 
rate in the cluster and the column density of intrinsic 
absorption. (This assumes that cluster redshift and Galactic column
density are known beforehand.) Although the counts in each of the 256
PSPC spectral channels are statistically independent, the limited
spectral resolution of the detector means that there are only about 5
independent spectral bands from a modelling point of view. 
Consequently, most spectral
analyses with ROSAT data employ simple models of gas at a single
temperature and with a uniform screen of absorption. We term these
`single-phase' models. [Temperatures and column
densities determined with single-phase models 
should only be interpreted as some
emission/detector-weighted mean over a range of temperatures or column
densities at any projected radius.] Only with the much higher spectral 
resolution of instruments such as the Einstein Observatory SSS 
and the ASCA Solid-state Imaging Spectrometers (SIS) can
more detailed spectral fits be attempted.

In this paper we make use of X-ray colour profiles, rather than 
individual spectra, 
to study the spatial distributions of 
cool gas and absorbing material in clusters. 
By restricting our study to the 0.4--2.0~keV band, where the
FWHM of the PSF is approximately constant at 25~arcsec, we can make optimum
use of the spatial resolution of the PSPC. Emission lines,
principally due to Fe-L transitions characteristic of gas cooling through 
temperatures of $\sim 1$ keV, are strongest between energies of 
0.8 and 1.4~keV. Intrinsic column densities at the level of $\sim 10^{21}$
\apc~should have a marked effect on the spectrum below about
0.8~keV. The remaining 1.4--2.0 keV band is comparatively 
little-affected 
by emission lines and absorption. 
We have therefore divided the 0.4--2.0 keV band into bands B, covering the 
range 0.41--0.79~keV, C from 0.80--1.39~keV and D from 1.4--2.0~keV.
The ratio of B to D should be sensitive to absorption, and the ratio of 
C to D to
the presence of cool gas (where `cool' implies a temperature
significantly less than that of the bulk of the ICM). 
By creating ratios of the surface brightness profiles of the
clusters in these bands, we parameterize the amounts of cool
gas and absorbing material in the clusters and, to the limits of the
spatial resolution of the PSPC, constrain the spatial distributions of
these components. We then compare these results to the predictions from
standard cooling flow models.

The structure of the paper is as follows. In Section 2 we summarize the
observations and data reduction. In Section 3 we present the results 
from a standard 
deprojection analysis of the data. In Section 4 
we present the results on the X-ray colour profiles and relate them to
expectations from simulated cluster spectra. In Section 5 we compare the
observed colour profiles to  the predictions from the deprojection
analysis and determine the results on the distributions of cool gas
and X-ray absorbing material in the clusters. In Section 6 
we discuss the complexities
associated with the measurement of intrinsic column densities with PSPC
data. Section 7 summarizes our conclusions. 
Throughout this paper, $H_0$=50
\kmpspMpc, $\Omega = 1$ and $\Lambda = 0$ are assumed.

\section{Observations}

\begin{figure}
\hbox{
\hspace{-0.7cm}\psfig{figure=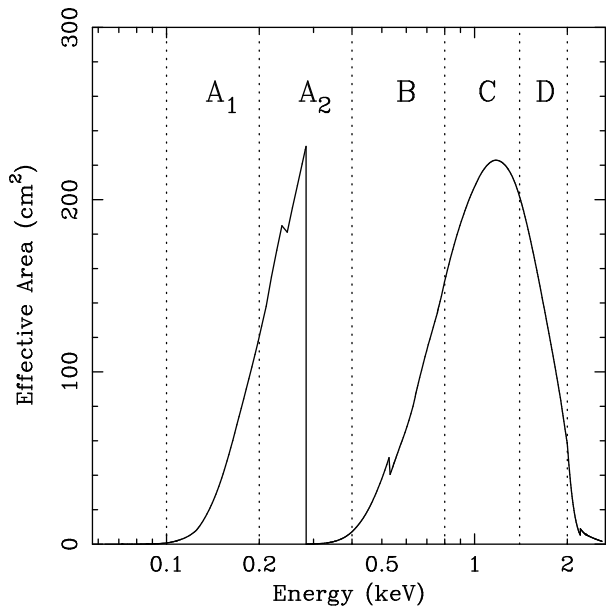,width=0.7\textwidth,angle=270}
}
\caption{Representation of the energy bands used in the X-ray colour
profile analysis, overlaid on the effective area curve for the PSPC.}
\end{figure}

The PSPC observations were carried out between 1991 March 6
and 1994 July 5. Exposure times range from 3192s for the Centaurus cluster 
to 40290s for Abell 2199.  
Details of the individual
cluster observations are summarized in Table 1. 
In cases where more than one observation of a source was made, only
the longest of the individual exposures has been used. This minimizes
systematic errors in the central surface brightness profiles 
that could be introduced by mosaicing. 
For the Coma cluster, for which four individual observations were made 
at different pointing positions, we have used the data set that
places the peak of the cluster emission closest to the centre of the PSPC 
field of
view. For the Centaurus cluster, we have used only the data from the
3.2ks observation in 1994 July, which was less-affected by scattered
solar X-ray emission than earlier observations. 

The data were reduced using the Starlink ASTERIX package. 
X-ray images with $15 \times 15$ arcsec$^2$ pixels  were extracted 
in each of the 
energy bands listed in Table 2. Regions affected by
emission from point sources were masked out and ignored.
Corrections for 
telescope vignetting were made and the data were background subtracted
using circular regions of radius 6
arcmin, offset by $\sim 40$ arcmin from the centres of the fields (except
in the case of the Virgo cluster, where a 3 arcmin radius region offset
by 45 arcmin was used). 
Accurate centroids for the X-ray emission were determined from the full 
0.4--2.0 keV (Band F) images. These are also listed in Table 1. 
X-ray surface brightness profiles were then extracted, about these
centroids, in each energy band. 

\begin{table}
\vskip 0.2truein
\begin{center}
\caption{Definition of the energy bands}
\vskip 0.2truein
\begin{tabular}{ c c c  }
  Band      & ~ &   Energy range (keV) \\  
\hline                                          
&& \\
   $A_1$    & ~ &   $0.10 - 0.40$ \\
   $A_2$    & ~ &   $0.20 - 0.40$ \\
   B        & ~ &   $0.41 - 0.79$ \\
   C        & ~ &   $0.80 - 1.39$ \\
   D        & ~ &   $1.40 - 2.00$ \\
   F        & ~ &   $0.41 - 2.00$ \\
&& \\
\end{tabular}
\end{center}
\parbox {3.2in}

\end{table}

\section{Deprojection analysis}

\begin{table*}
\vskip 0.2truein
\begin{center}
\caption{Cooling flow results}
\vskip 0.2truein
\begin{tabular}{ c c c c  c c c c c}
            & ~ &     $kT$             & $N_{\rm H}$     & ~ & $r_{\rm cool}$       &  $r_{\rm cool}$           &    $t_{\rm cool}$          & ${\dot M}$   \\
            & ~ &     (keV)            & ($10^{20}$ \apc) & ~ &     (kpc) &     (arcmin)               &    ($10^{9}$ yr)             &   (\Msunpyr)   \\  
\hline                                                                                                                                            
&&&&&&& \\                                                                                                                                                            
Abell 85    & ~ & $6.6^{+1.8}_{-1.4}$  & 2.96     & ~ & $137^{+8}_{-34}$     &  $1.66^{+0.10}_{-0.41}$   &   $2.85^{+0.15}_{-0.11}$   &  $187^{+16}_{-42}$ \\  
Abell 3112  & ~ & $4.1^{+2.3}_{-1.1}$  & 2.78     & ~ & $251^{+71}_{-46}$    &  $2.19^{+0.62}_{-0.40}$   &   $2.13^{+0.07}_{-0.08}$   &  $406^{+75}_{-57}$ \\  
Abell 426   & ~ & $6.3^{+0.2}_{-0.2}$  & 13.7     & ~ & $169^{+9}_{-6}$      &  $5.46^{+0.29}_{-0.19}$   &   $1.06^{+0.06}_{-0.05}$   &  $456^{+16}_{-19}$ \\  
Abell 478   & ~ & $7.0^{+0.7}_{-0.5}$  & 20.0     & ~ & $213^{+19}_{-48}$    &  $1.61^{+0.14}_{-0.36}$   &   $2.59^{+0.06}_{-0.06}$   &  $717^{+57}_{-107}$ \\  
Abell 496   & ~ & $4.8^{+0.9}_{-0.8}$  & 4.27     & ~ & $105^{+14}_{-13}$    &  $1.99^{+0.26}_{-0.25}$   &   $1.84^{+0.08}_{-0.09}$   &  $85^{+10}_{-9}$   \\  
Abell 644   & ~ & $8.1^{+2.6}_{-2.5}$  & 8.61     & ~ & $200^{+50}_{-61}$    &  $1.83^{+0.46}_{-0.56}$   &   $8.28^{+1.27}_{-0.80}$   &  $244^{+69}_{-93}$ \\  
Hydra A     & ~ & $3.6^{+0.8}_{-0.5}$  & 4.84     & ~ & $161^{+26}_{-15}$    &  $1.93^{+0.31}_{-0.18}$   &   $1.63^{+0.03}_{-0.04}$   &  $267^{+48}_{-30}$ \\  
Coma        & ~ & $8.0^{+0.3}_{-0.3}$  & 0.90     & ~ & $0^{+29}_{-0}$       &  $0^{+0.75}_{-0}$         &   $18.4^{+21.1}_{-5.9}$    &  $0^{+1}_{-0}$     \\  
Virgo       & ~ & $2.4^{+0.3}_{-0.3}$  & 1.71     & ~ & $98^{+28}_{-2}$      &  $13.16^{+3.76}_{-0.27}$  &   $0.21^{+0.01}_{-0.01}$   &  $37^{+28}_{-6}$ \\  
Centaurus   & ~ & $3.6^{+0.4}_{-0.4}$  & 8.87     & ~ & $87^{+7}_{-29}$      &  $4.88^{+0.39}_{-1.63}$   &   $0.70^{+0.06}_{-0.04}$   &  $27^{+5}_{-7}$ \\  
Abell 1795  & ~ & $5.1^{+0.4}_{-0.5}$  & 1.11     & ~ & $192^{+32}_{-18}$    &  $1.93^{+0.32}_{-0.18}$   &   $2.14^{+0.05}_{-0.04}$   &  $447^{+50}_{-30}$ \\  
Abell 2029  & ~ & $7.1^{+2.0}_{-1.4}$  & 3.05     & ~ & $200^{+6}_{-53}$     &  $1.70^{+0.05}_{-0.45}$   &   $2.76^{+0.10}_{-0.08}$   &  $590^{+27}_{-96}$ \\  
MKW3s       & ~ & $2.8^{+0.7}_{-0.4}$  & 2.86     & ~ & $154^{+73}_{-32}$    &  $2.21^{+1.05}_{-0.46}$   &   $2.17^{+0.15}_{-0.17}$   &  $161^{+67}_{-29}$ \\  
Abell 2199  & ~ & $4.7^{+0.4}_{-0.4}$  & 0.87     & ~ & $142^{+24}_{-27}$    &  $2.78^{+0.47}_{-0.53}$   &   $1.94^{+0.04}_{-0.04}$   &  $162^{+26}_{-30}$ \\  
Cygnus A    & ~ & $4.1^{+4.3}_{-1.3}$  & 32.8     & ~ & $157^{+46}_{-44}$    &  $1.74^{+0.51}_{-0.49}$   &   $1.39^{+0.05}_{-0.06}$   &  $320^{+71}_{-32}$ \\  
Sersic 159  & ~ & $2.9^{+1.1}_{-0.8}$  & 1.80     & ~ & $167^{+33}_{-11}$    &  $1.89^{+0.37}_{-0.12}$   &   $1.75^{+0.06}_{-0.06}$   &  $231^{+11}_{-10}$ \\  
Abell 2597  & ~ & $> 3.8$              & 2.46     & ~ & $162^{+57}_{-6}$     &  $1.29^{+0.46}_{-0.05}$   &   $2.44^{+0.11}_{-0.07}$   &  $299^{+32}_{-17}$ \\  
Abell 4059  & ~ & $3.5^{+0.6}_{-0.6}$  & 1.61     & ~ & $134^{+77}_{-38}$    &  $1.74^{+1.00}_{-0.49}$   &   $3.53^{+0.56}_{-0.48}$   &  $115^{+57}_{-37}$ \\  
&&&&&&& \\                                                         
     
\end{tabular}
\end{center}

\parbox {7in}
{Notes: A summary of the results from the deprojection analyses of the F band
images. Temperature ($kT$) data are from Edge et al (1990) except for Abell 426 
(Allen \etal 1992) and Abell 478 (Allen \etal 1993). $N_{\rm H}$ values are 
Galactic column densities from Stark \etal (1992) except for Abell 478 which is 
from Johnstone \etal (1992).
Cooling radii ($r_{\rm cool}$) are quoted in both kpc and arcmin. Cooling
times ($t_{\rm cool}$) for the  central (30 arcsec radius) bin are in $10^9$
\yr, and  mass deposition rates (${\dot M}$ ) in \Msunpyr. Errors on the
cooling times are the 10 and 90 percentile values from 100 Monte Carlo
simulations. The upper and lower confidence limits on the cooling radii are the
points where the 10 and 90 percentiles exceed and become less than the Hubble
time, respectively. Errors on the mass deposition rates are the 90 and 10
percentile values at the upper and lower limits for the cooling radius. } 
\end{table*}

The F band (0.4--2.0 keV) images of the clusters have been analysed  
using an updated version of the  deprojection code of Fabian \etal 
(1981; see also Thomas, Fabian \& Nulsen 1987).  Under assumptions of
spherical symmetry and hydrostatic equilibrium in the ICM, the deprojection
technique can be used to study the properties of the intracluster gas  (\eg
density, pressure, temperature, cooling rate) as a function of radius. The
deprojection method requires that either the total mass profile (which defines
the pressure profile)  or the gas temperature profile be specified.  Following 
ASCA observations of a number of the clusters included in  the sample 
(Fabian \etal 1996; Ohashi \etal 1996),  and the results from the combined 
X-ray and gravitational lensing study of the cooling-flow cluster PKS0745-191 by
Allen \etal (1996a), we assume that the mass-weighted  temperature profiles
in the clusters remain approximately constant, at the temperatures  listed in
Table 3. Note that the assumption of a constant mass-weighted
temperature profile is consistent with ROSAT results
which often show a drop in the {\it emission-weighted} temperatures
in the central regions of nearby cooling-flow clusters. Such measurements 
will naturally result from the presence of high emissivity cooling gas,
distributed throughout the (inhomogeneous) cooling flow with ${\dot M}
\approxpropto r$, even when the mass-weighted temperature profile 
of the inflowing material remains constant (\eg Allen \etal 1996a,b).

The primary  results on the cooling flows in the clusters; the cooling
times in the central 30 arcsec bin, the cooling radii 
(the radii at which the cooling  time first exceeds the
Hubble time) and the integrated mass deposition rates within the cooling
radii, are listed in Table 3.  In all cases the mass
deposition from the cooling flows is distributed throughout the cooling radii
of the clusters,  with ${\dot M} \approxpropto r$.

\section{The X-ray colour profiles}

\begin{figure*}
\hbox{
\hspace{1cm}\psfig{figure=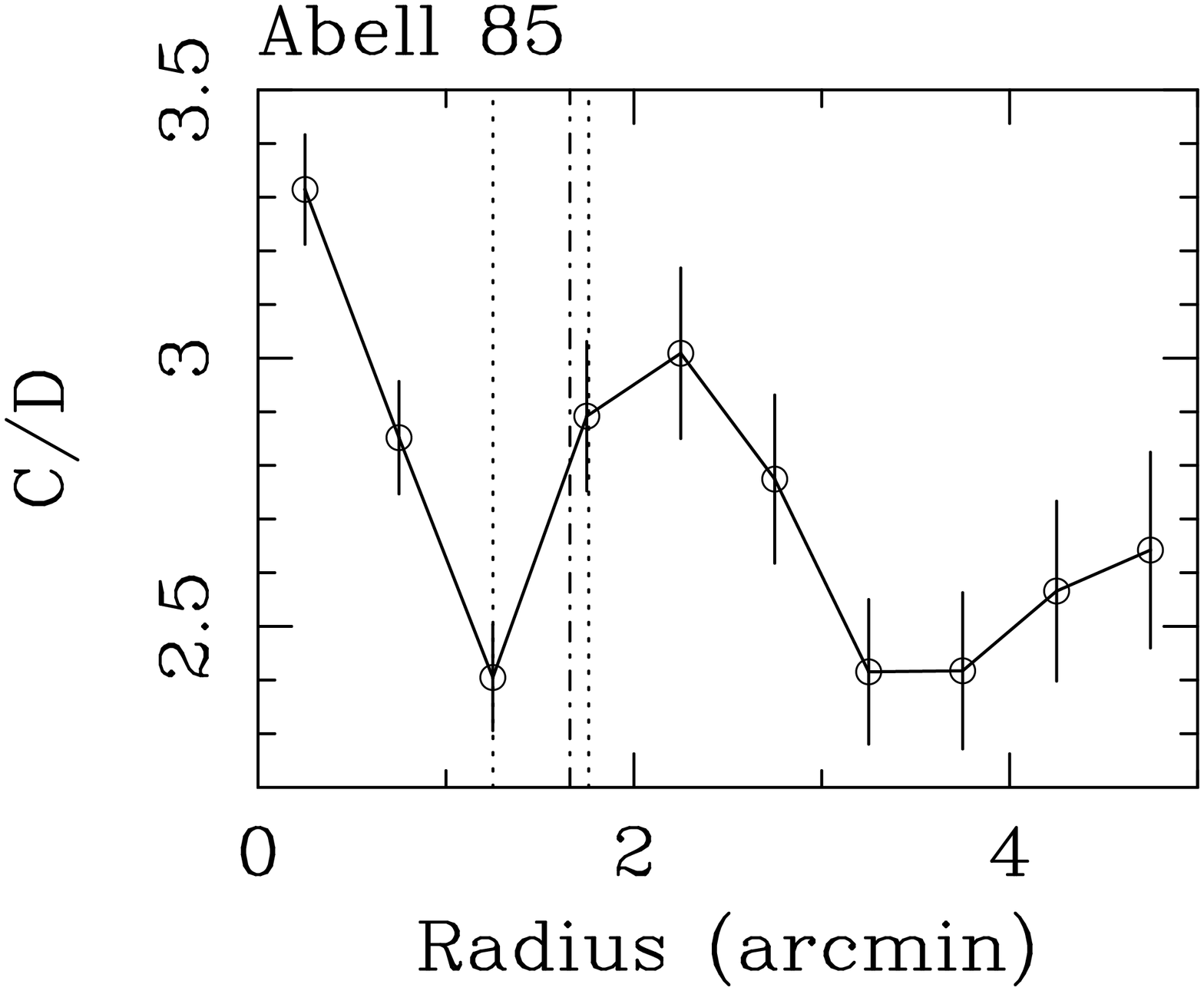,width=0.47\textwidth,angle=270}
\hspace{-0.5cm}\psfig{figure=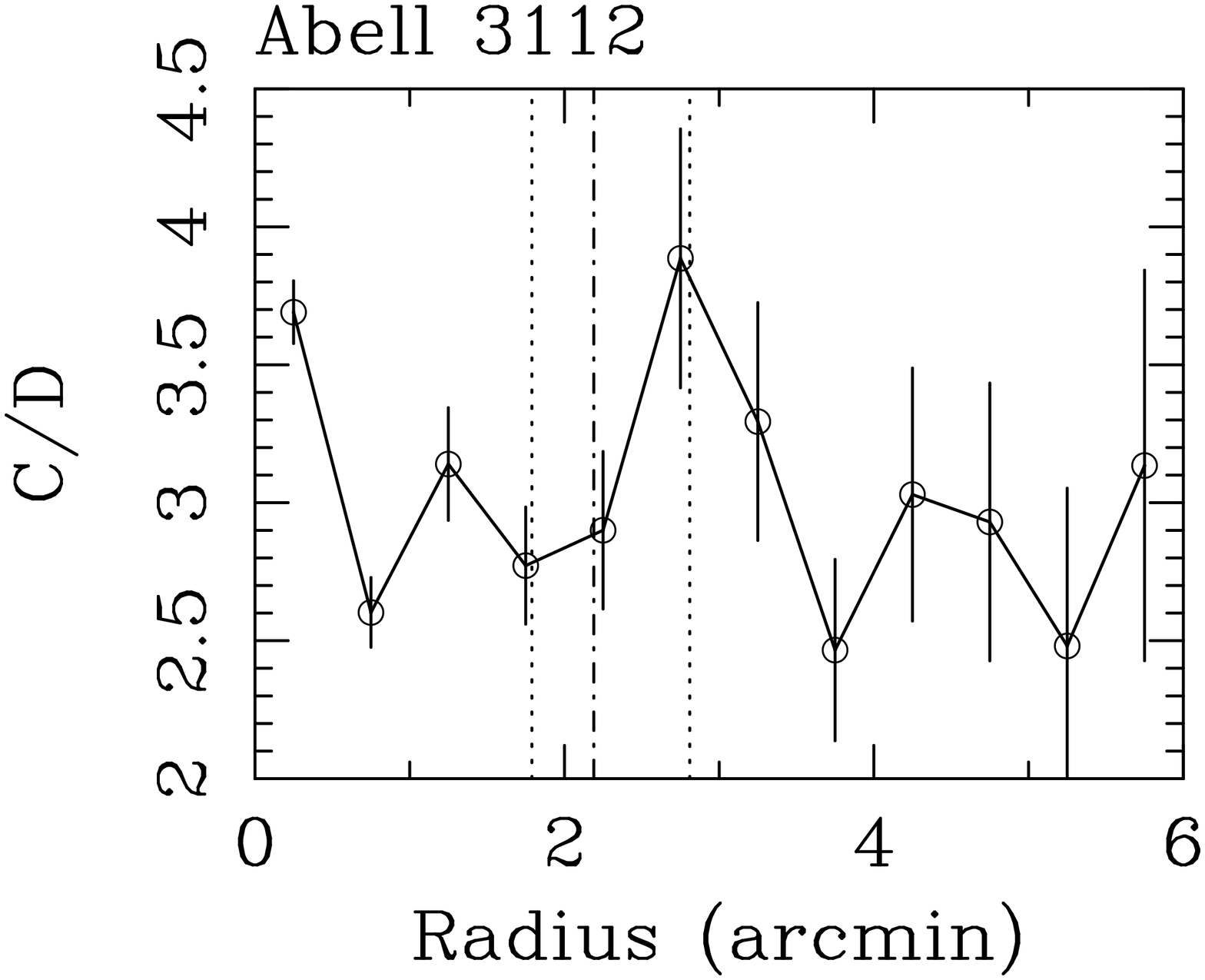,width=0.47\textwidth,angle=270}
}

\vspace{-0.2cm}

\hbox{
\hspace{1cm}\psfig{figure=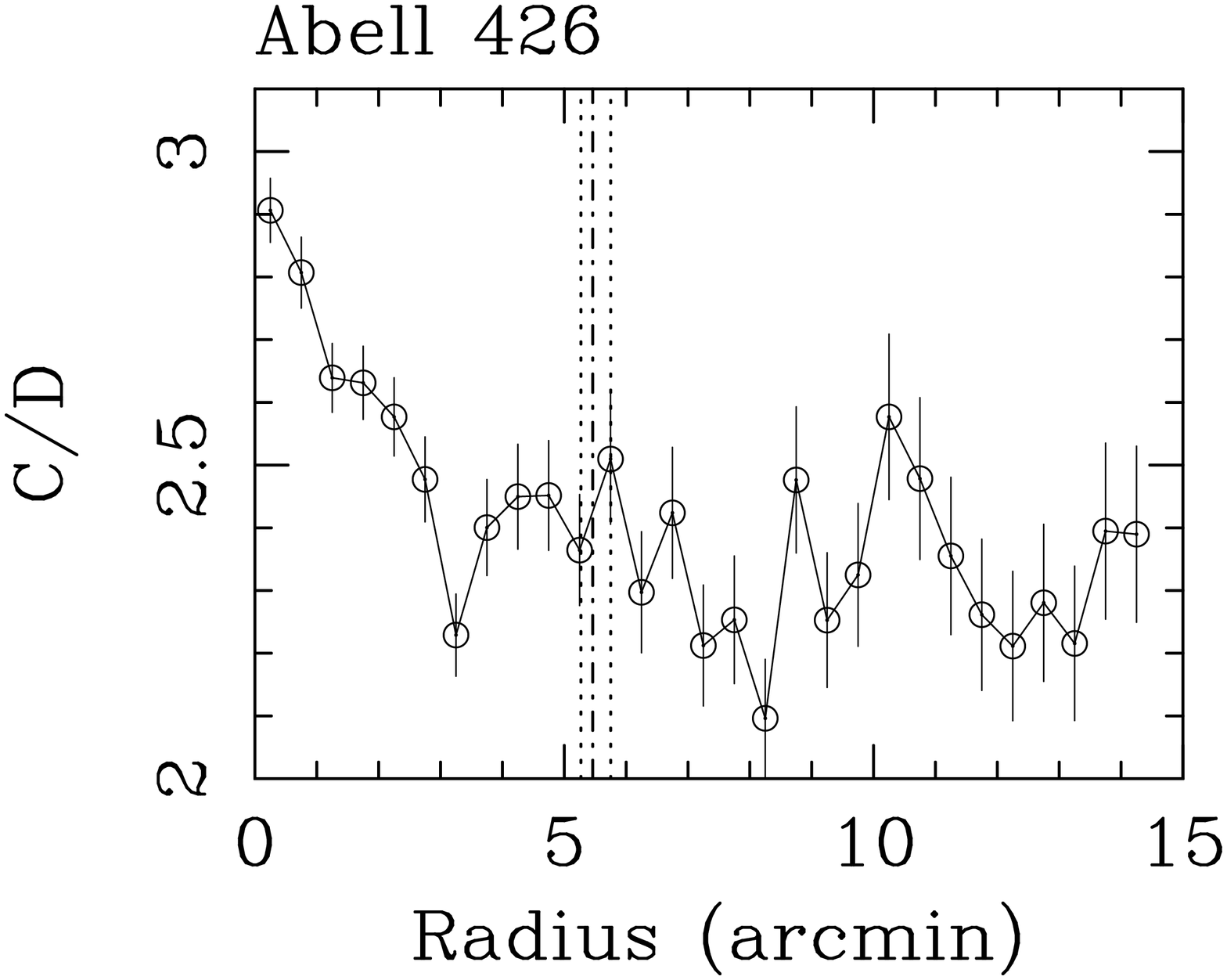,width=0.47\textwidth,angle=270}
\hspace{-0.5cm}\psfig{figure=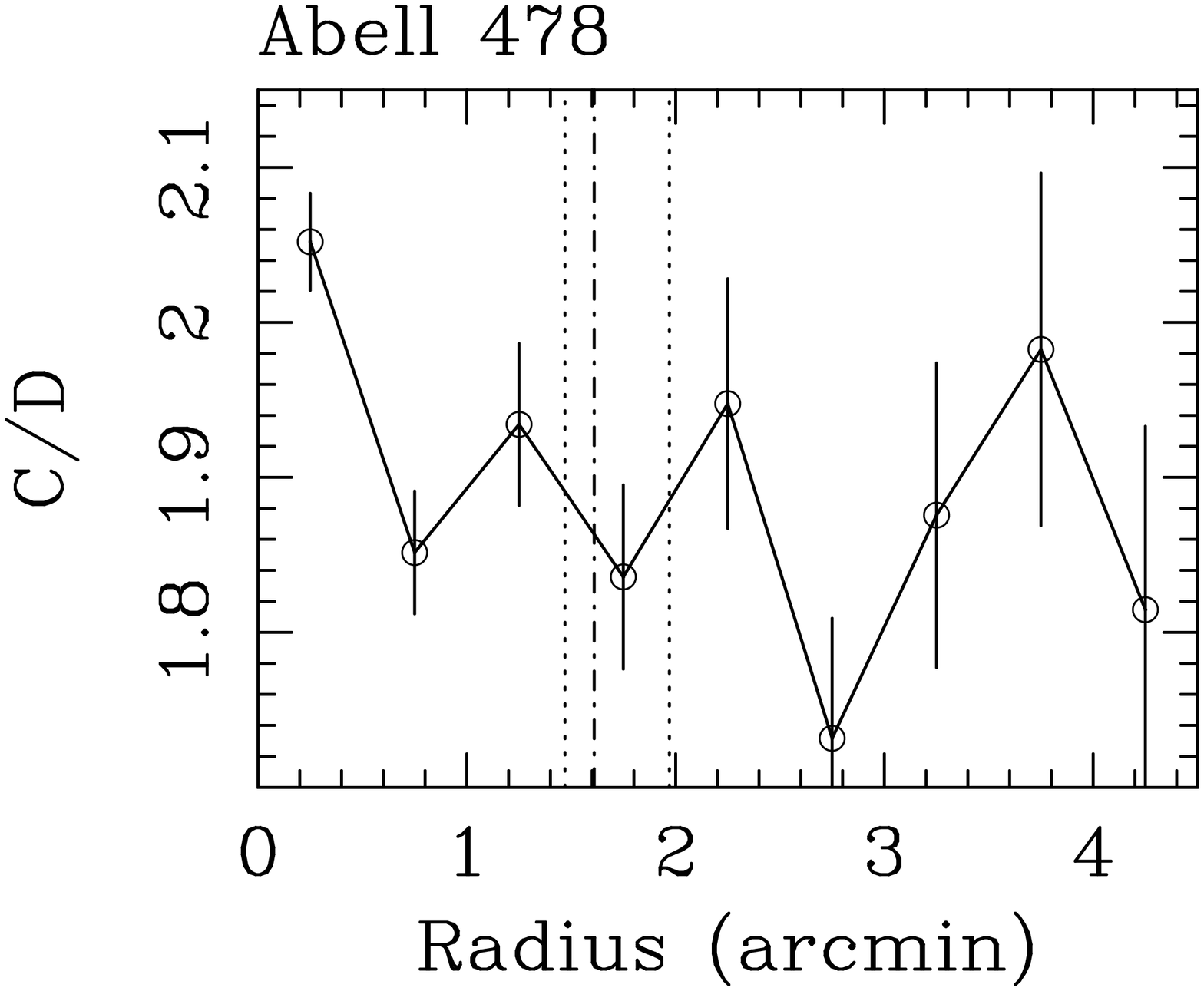,width=0.47\textwidth,angle=270}
}
\vspace{-0.2cm}

\hbox{
\hspace{1cm}\psfig{figure=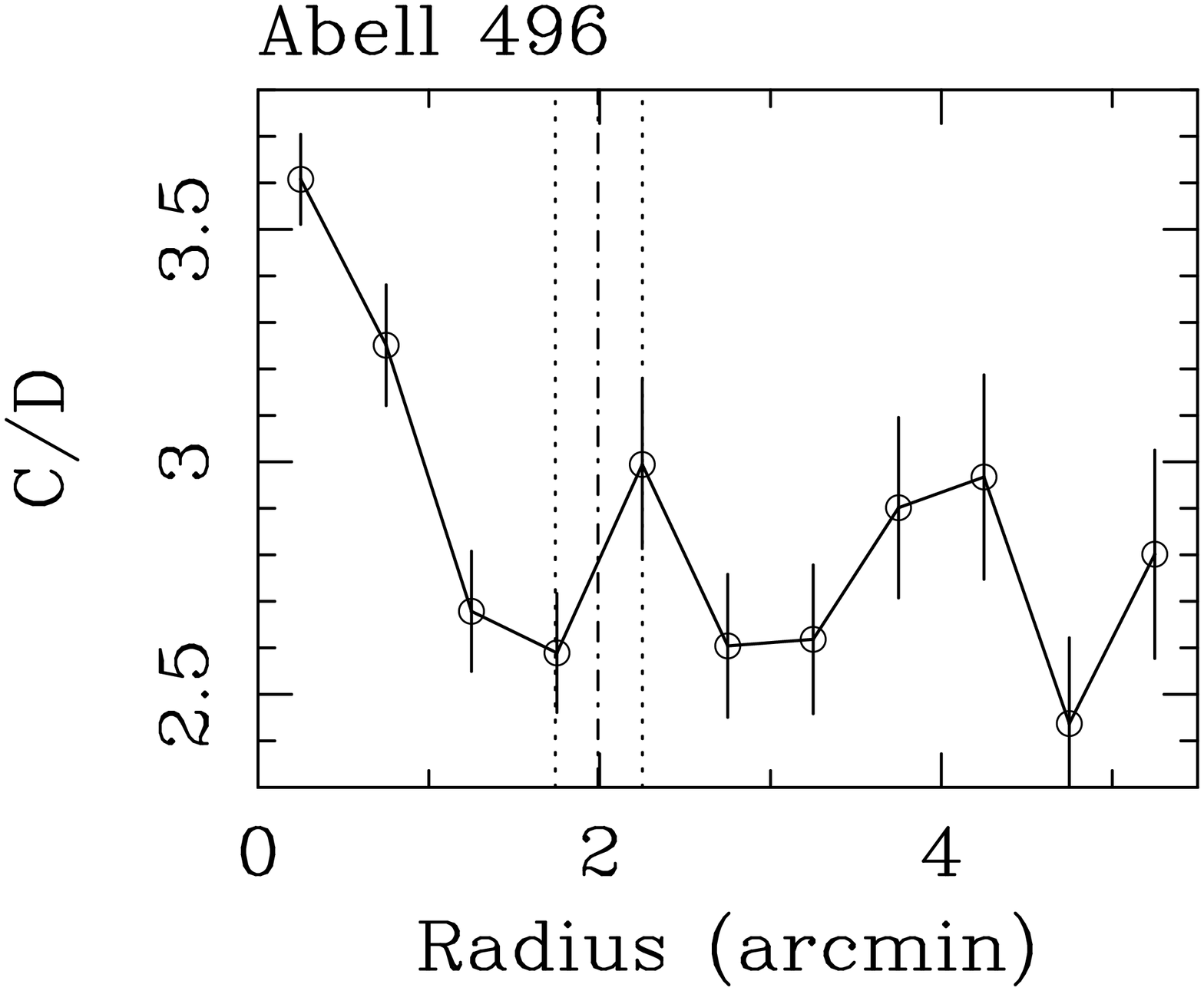,width=0.47\textwidth,angle=270}
\hspace{-0.5cm}\psfig{figure=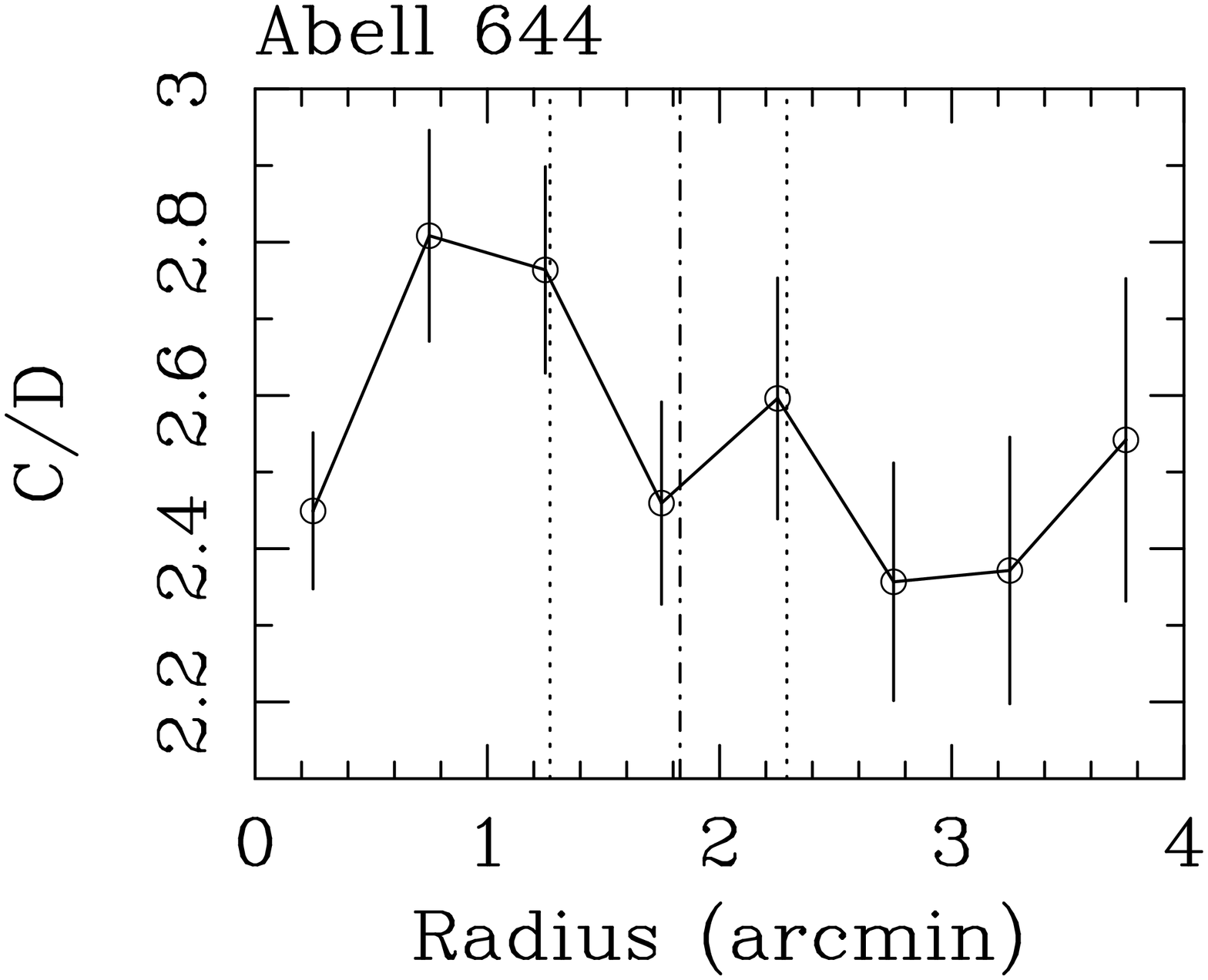,width=0.47\textwidth,angle=270}
}

\caption{The C/D ($0.80-1.39$ keV/$1.40-2.00$keV) X-ray colour profiles
(with $1\sigma$ error bars). The cooling radii determined from the 
deprojection analysis (Section 3) are
indicated by the vertical dot-dashed lines. The 10 and 90 per cent
confidence ranges on those radii are marked by the dotted lines. 
The profiles are essentially flat at large radii but show clear increases 
within $r_{\rm cool}$. 
}
\end{figure*}

\addtocounter{figure}{-1}
\begin{figure*}
\hbox{
\hspace{1cm}\psfig{figure=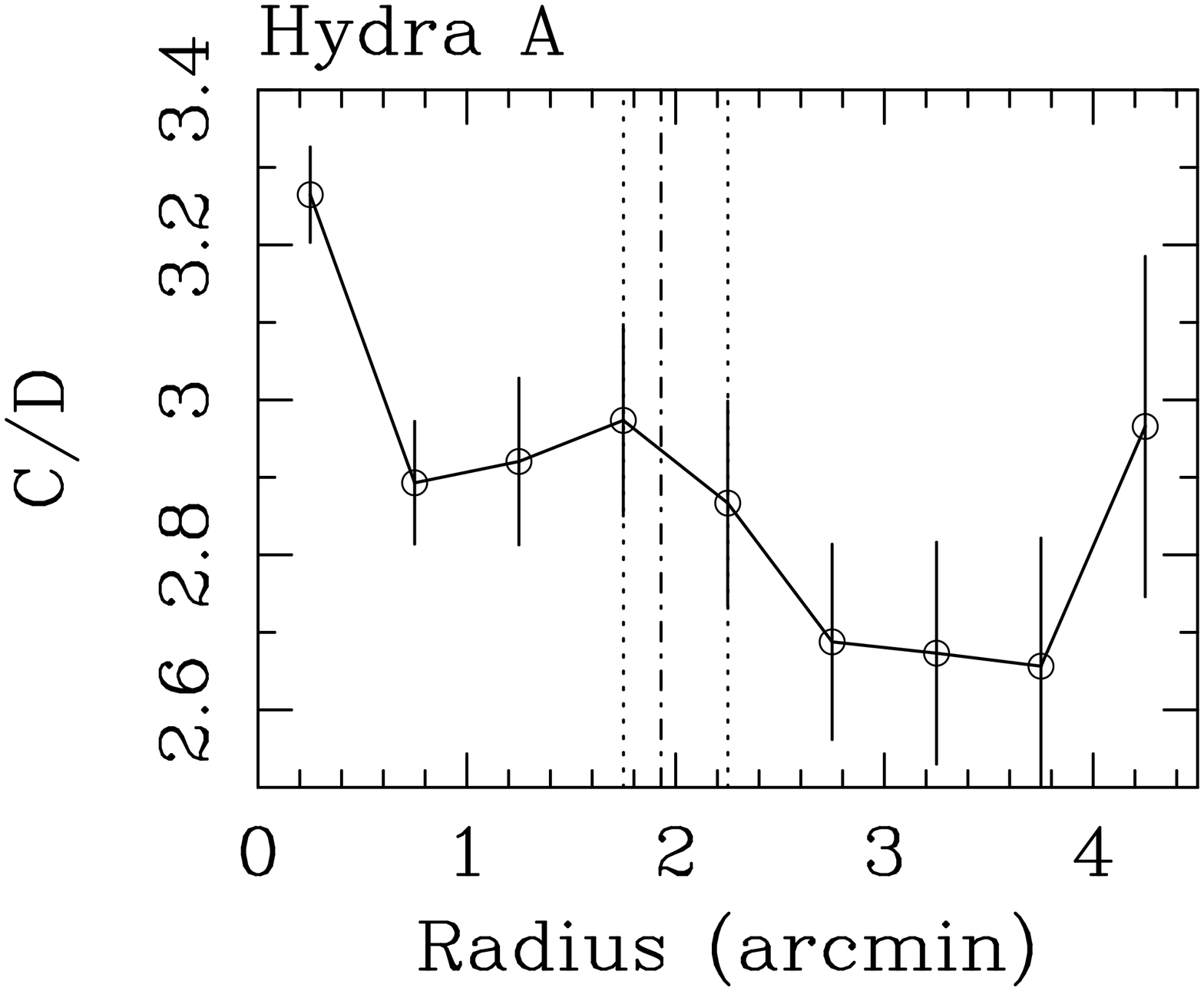,width=0.47\textwidth,angle=270}
\hspace{-0.5cm}\psfig{figure=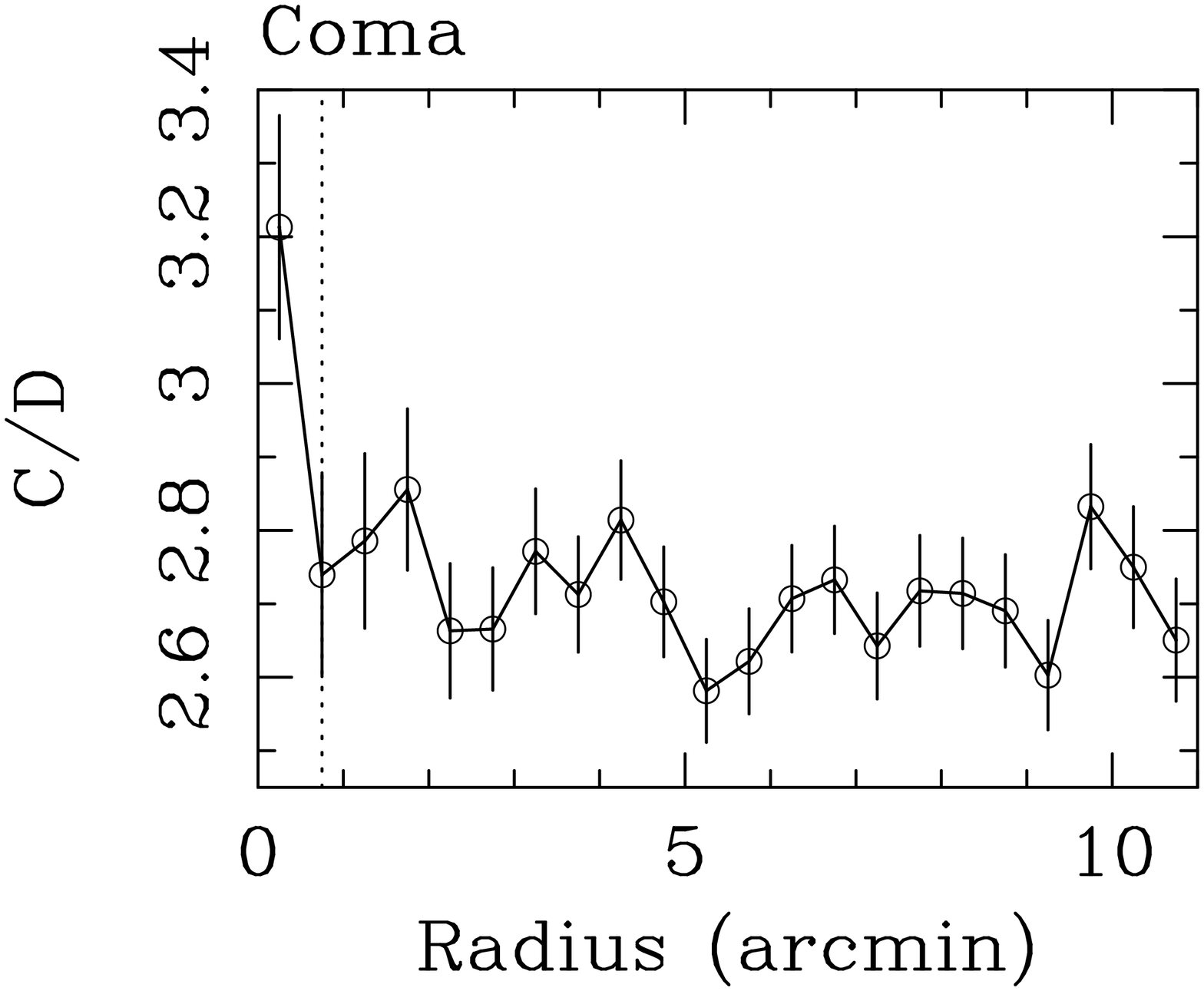,width=0.47\textwidth,angle=270}
}

\vspace{-0.2cm}

\hbox{
\hspace{1cm}\psfig{figure=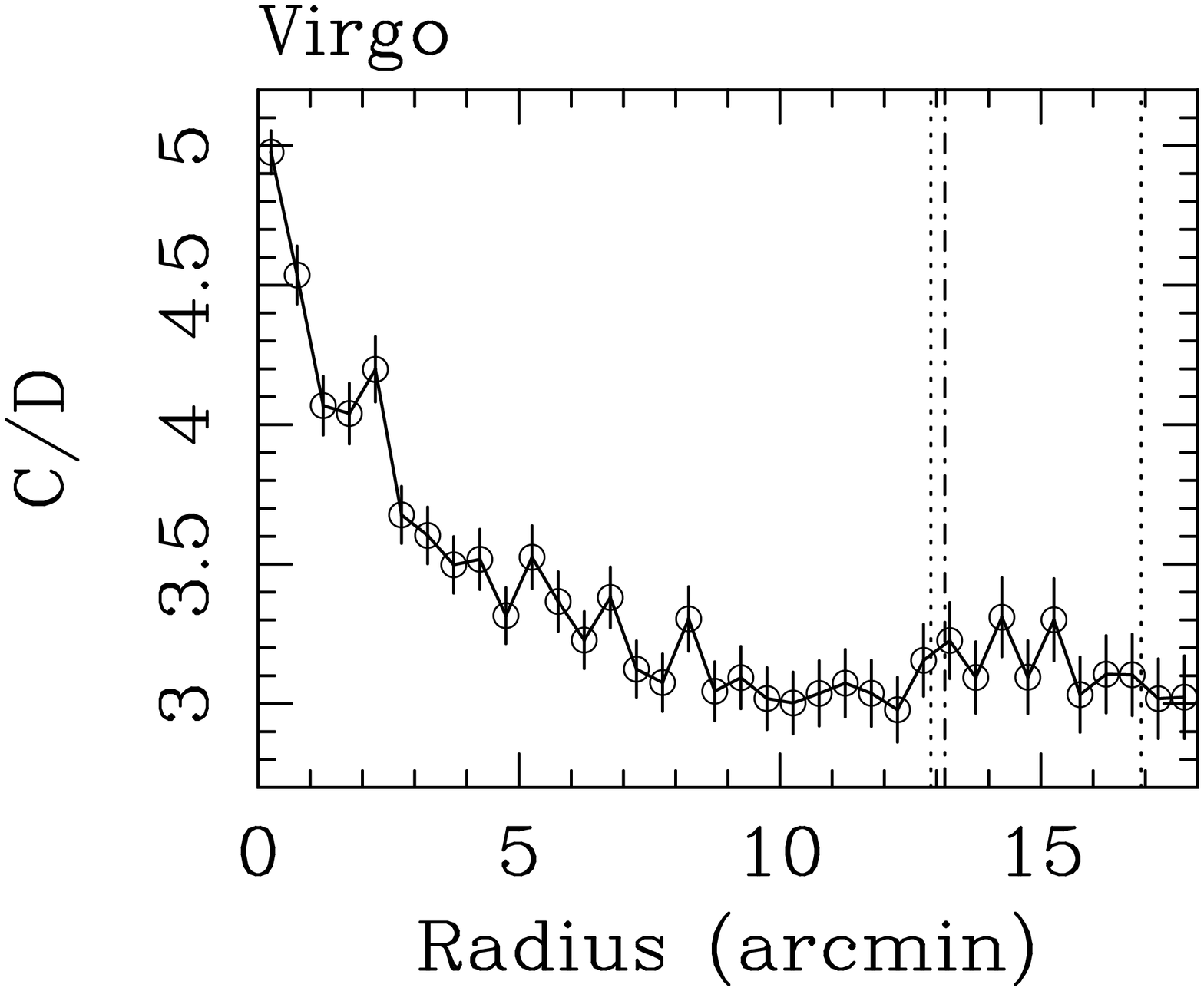,width=0.47\textwidth,angle=270}
\hspace{-0.5cm}\psfig{figure=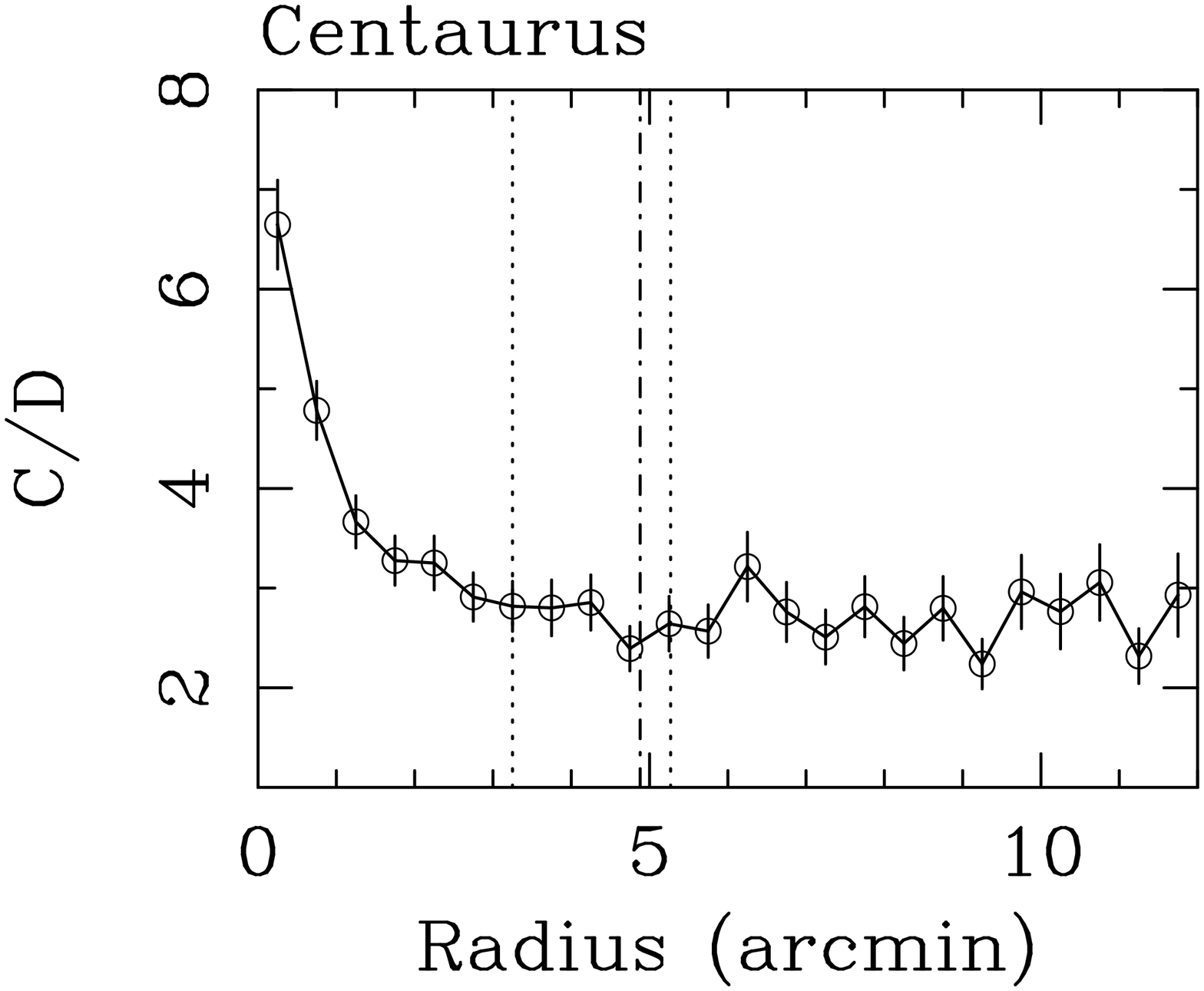,width=0.47\textwidth,angle=270}
}

\vspace{-0.2cm}

\hbox{
\hspace{1cm}\psfig{figure=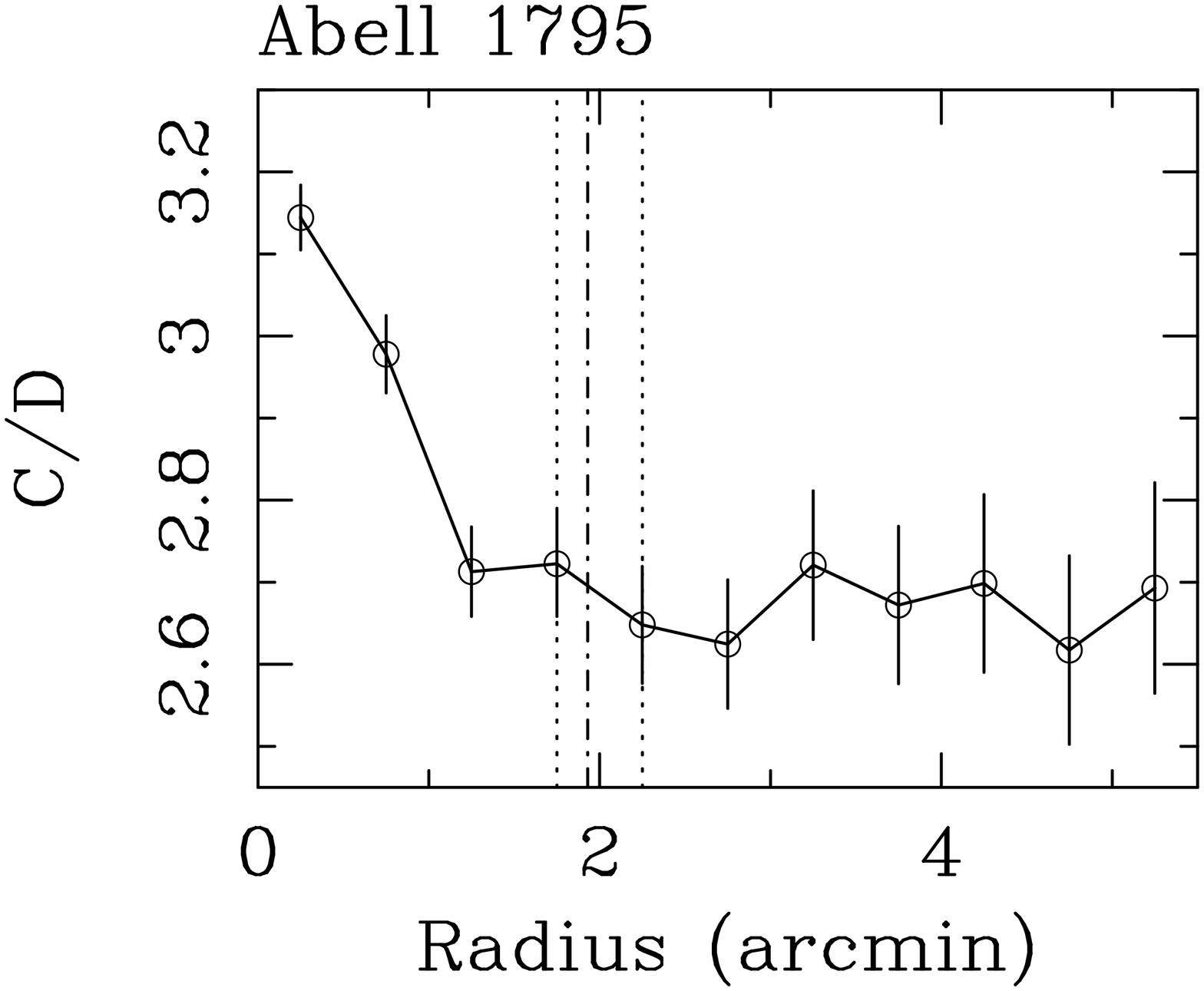,width=0.47\textwidth,angle=270}
\hspace{-0.5cm}\psfig{figure=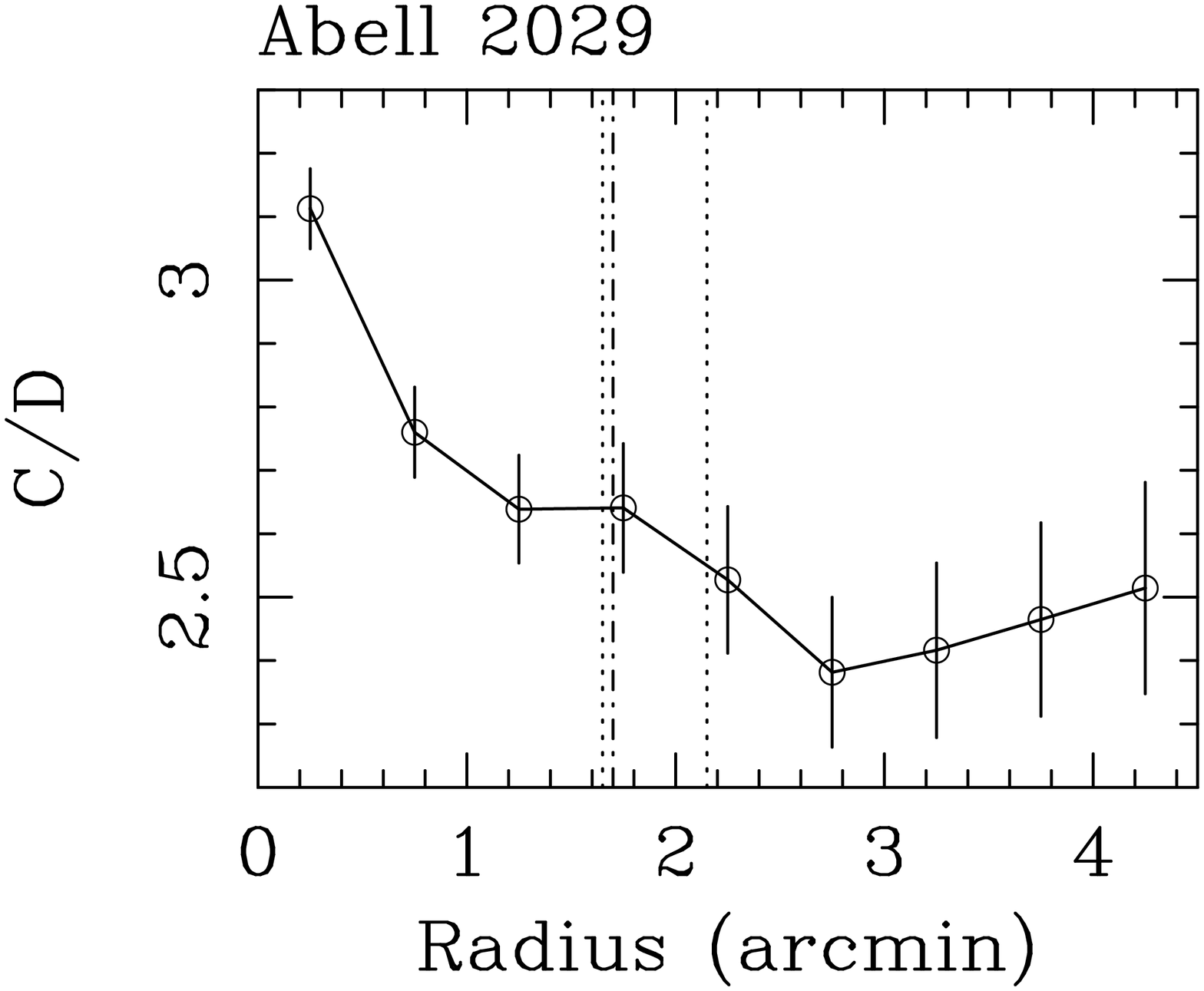,width=0.47\textwidth,angle=270}
}

\caption{ - continued}
\end{figure*}

\addtocounter{figure}{-1}
\begin{figure*}
\vspace{-0.5cm}
\hbox{
\hspace{1cm}\psfig{figure=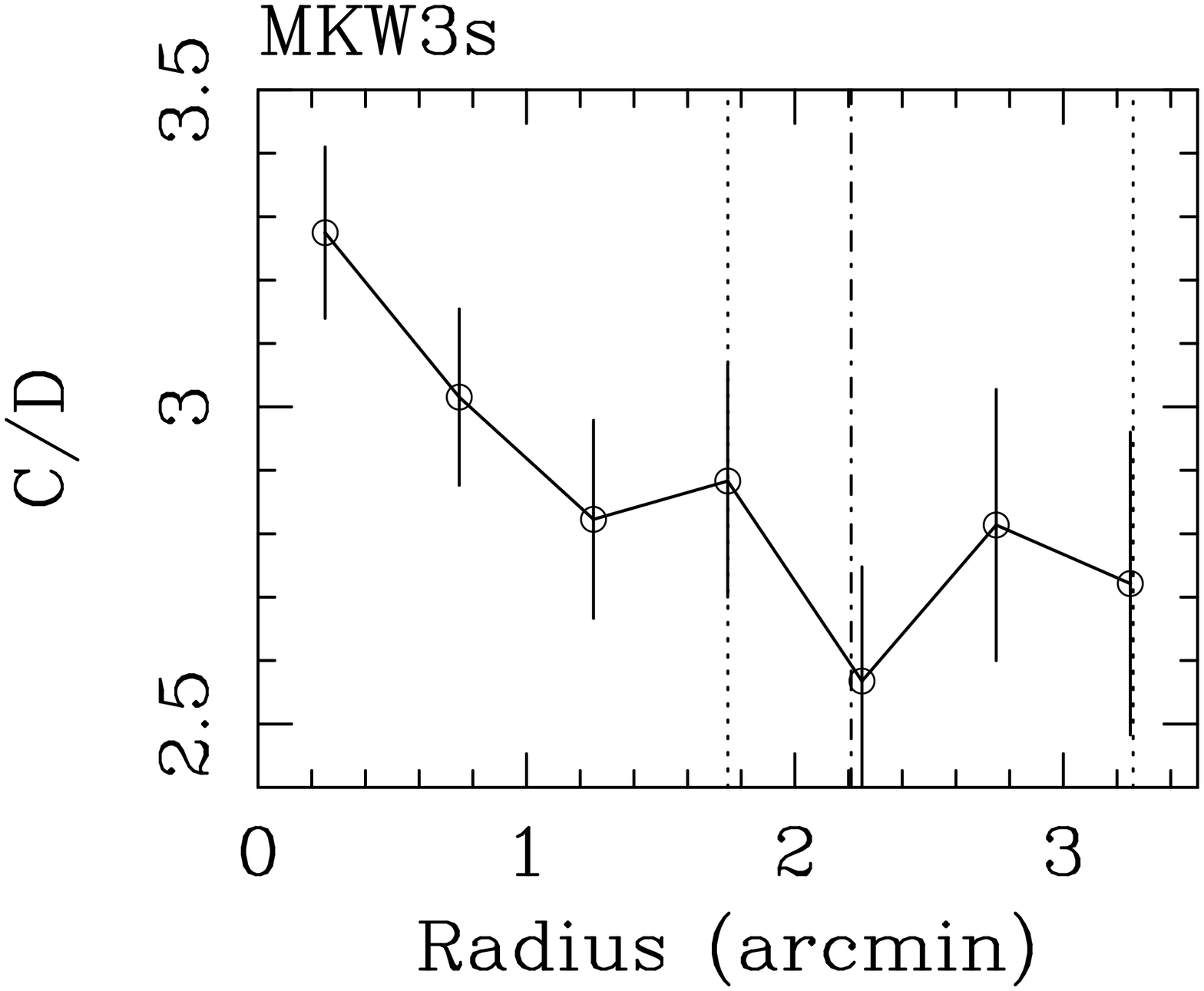,width=0.47\textwidth,angle=270}
\hspace{-0.5cm}\psfig{figure=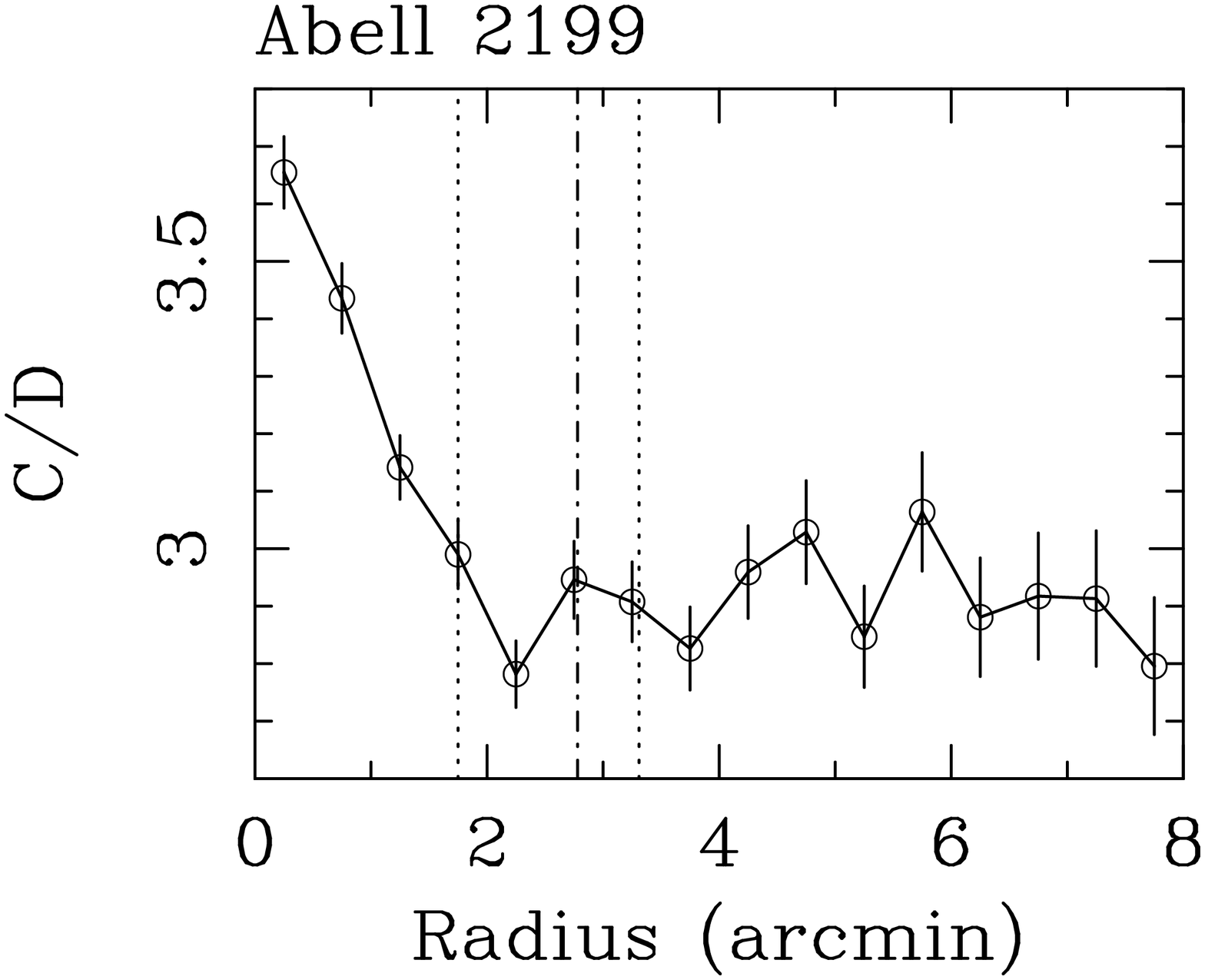,width=0.47\textwidth,angle=270}
}

\vspace{-0.2cm}

\hbox{
\hspace{1cm}\psfig{figure=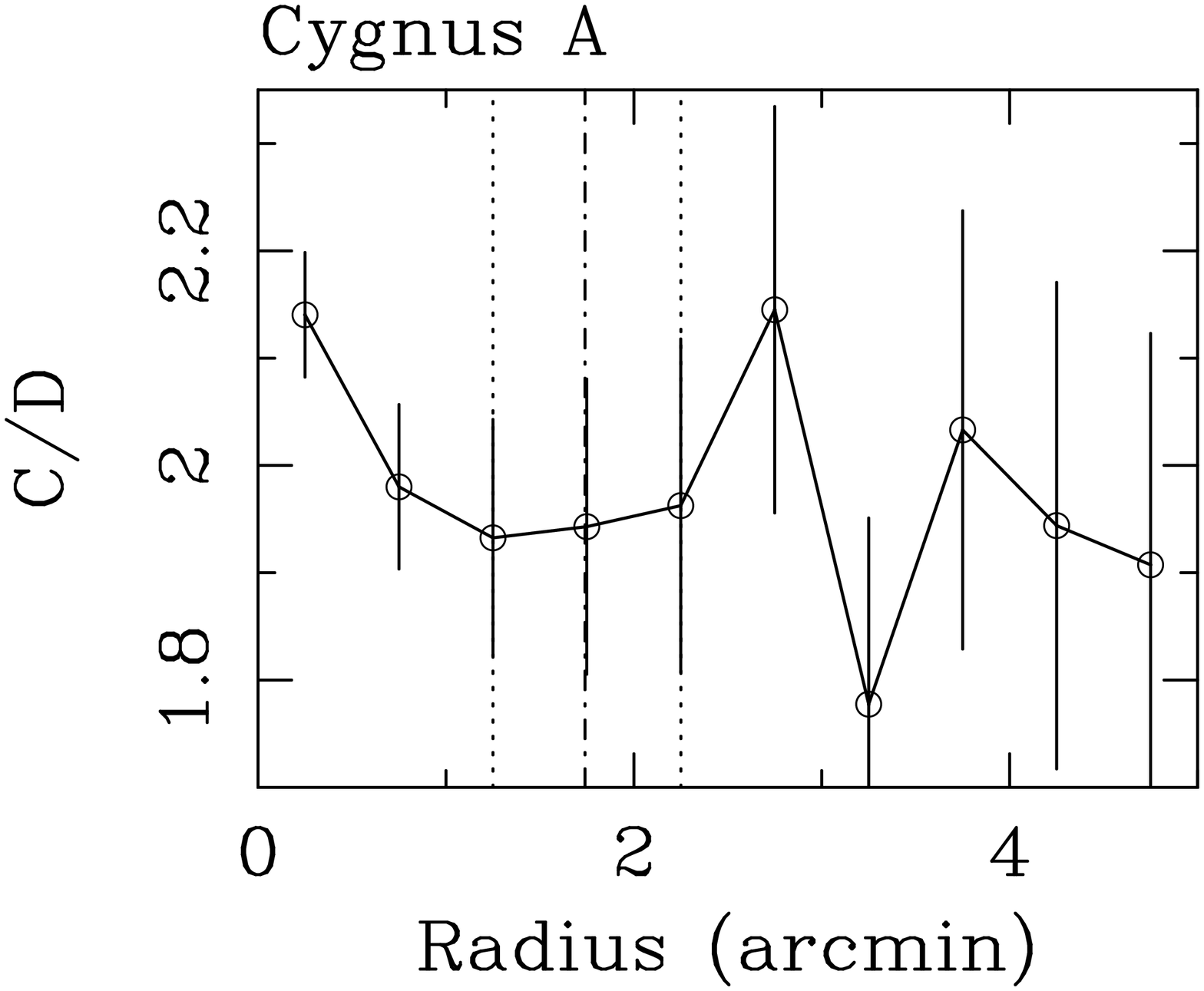,width=0.47\textwidth,angle=270}
\hspace{-0.5cm}\psfig{figure=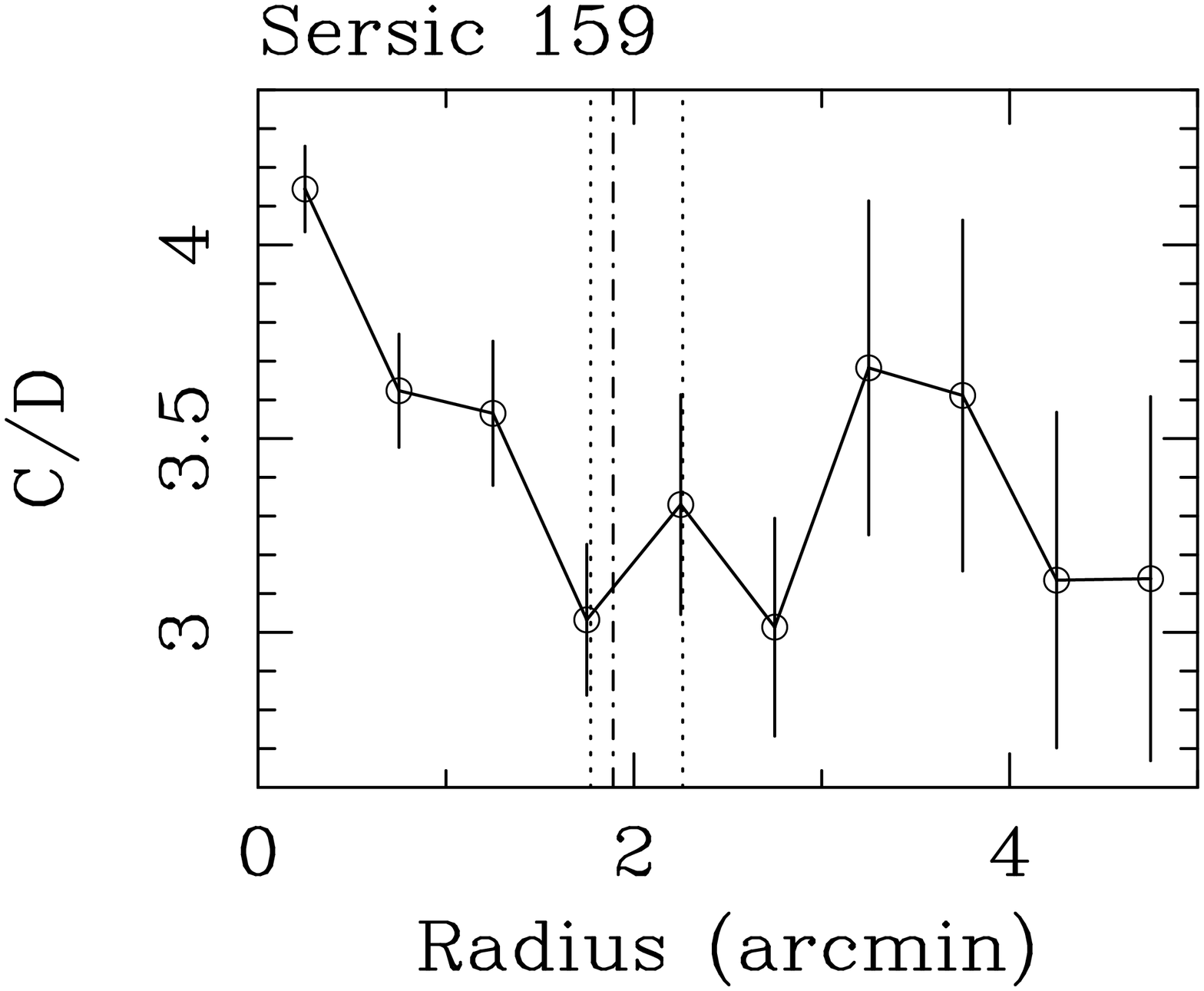,width=0.47\textwidth,angle=270}
}

\vspace{-0.2cm}

\hbox{
\hspace{1cm}\psfig{figure=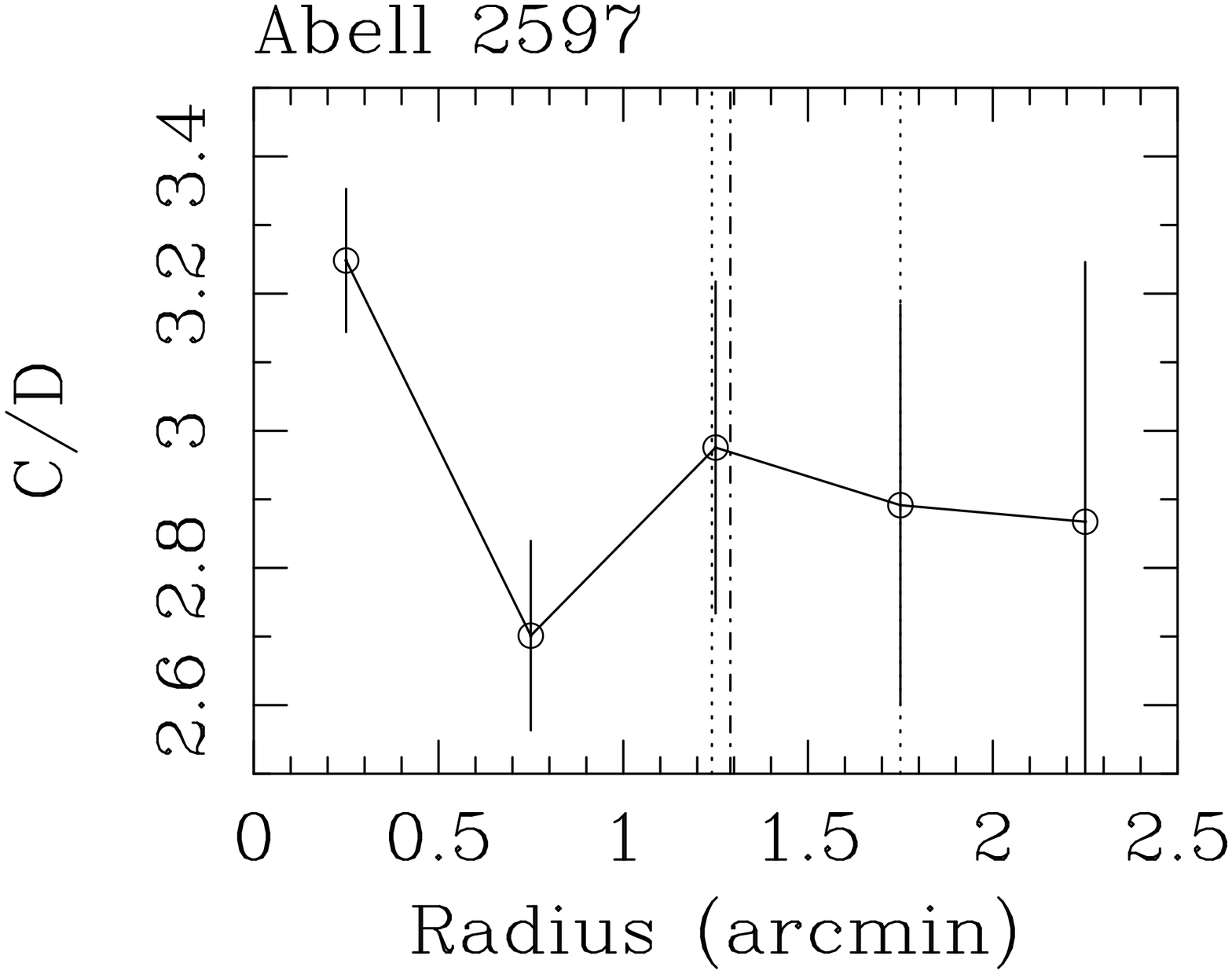,width=0.47\textwidth,angle=270}
\hspace{-0.5cm}\psfig{figure=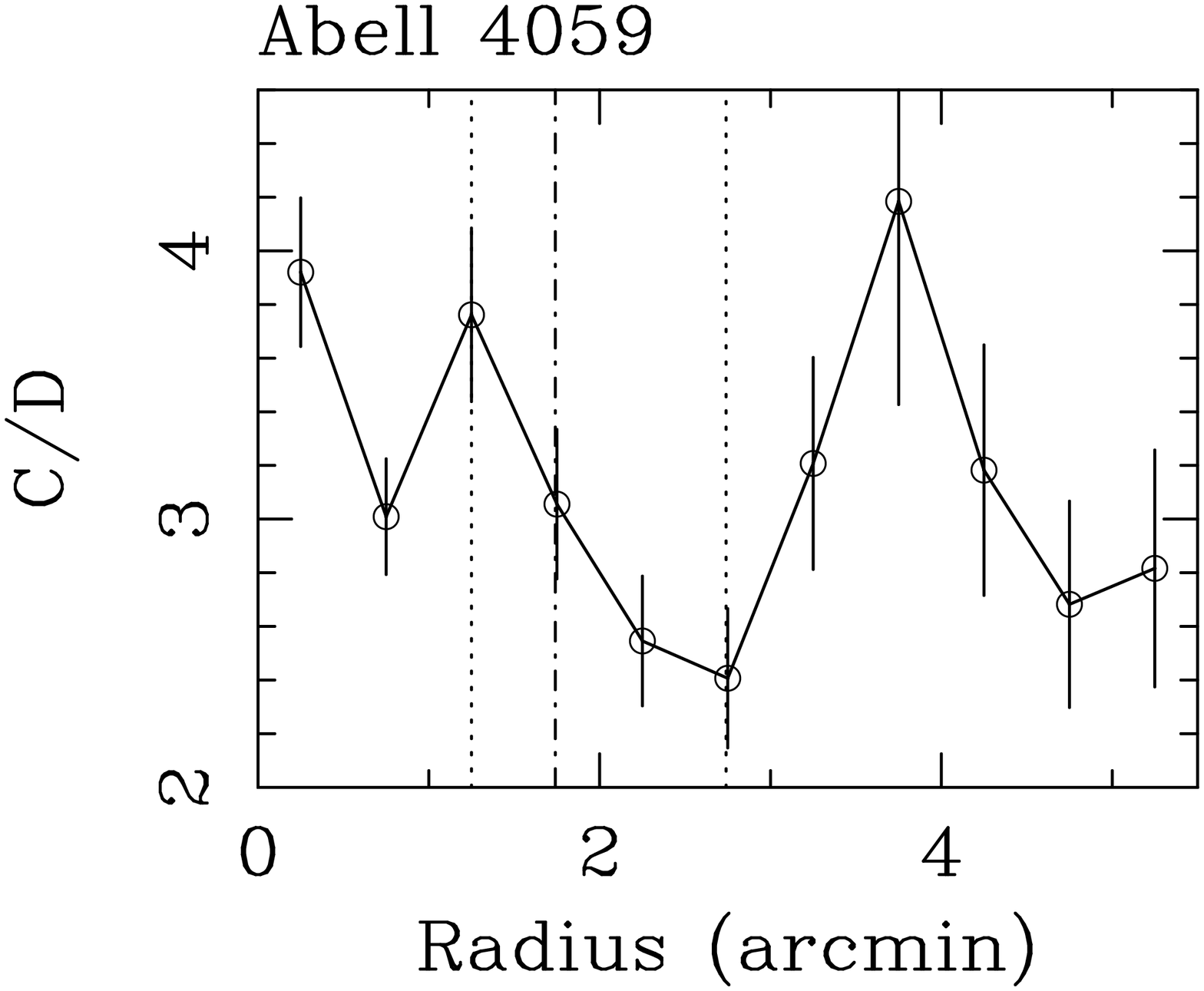,width=0.47\textwidth,angle=270}
}

\caption{ - continued}
\end{figure*}

\begin{figure*}
\hbox{
\hspace{1cm}\psfig{figure=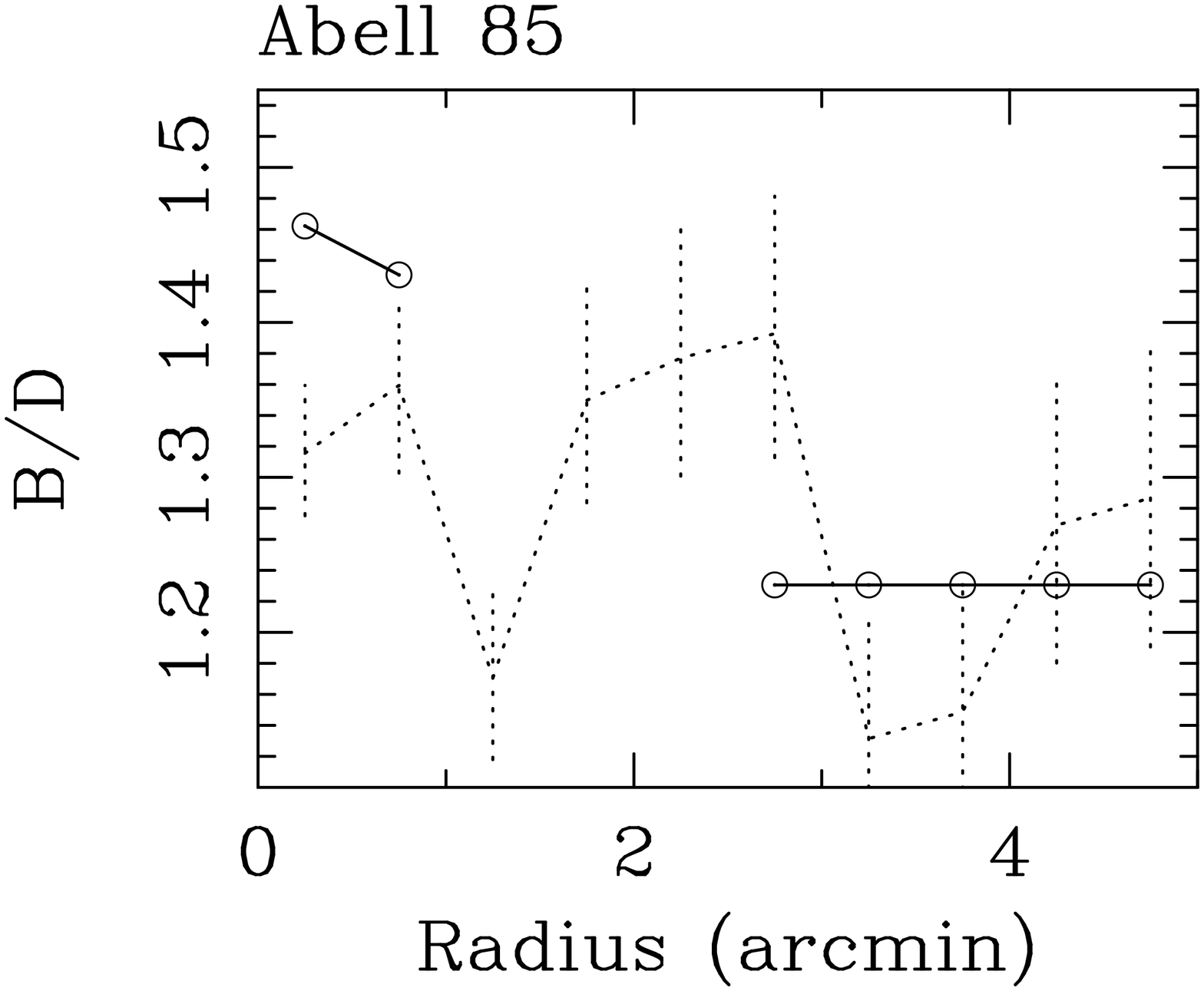,width=0.47\textwidth,angle=270}
\hspace{-0.5cm}\psfig{figure=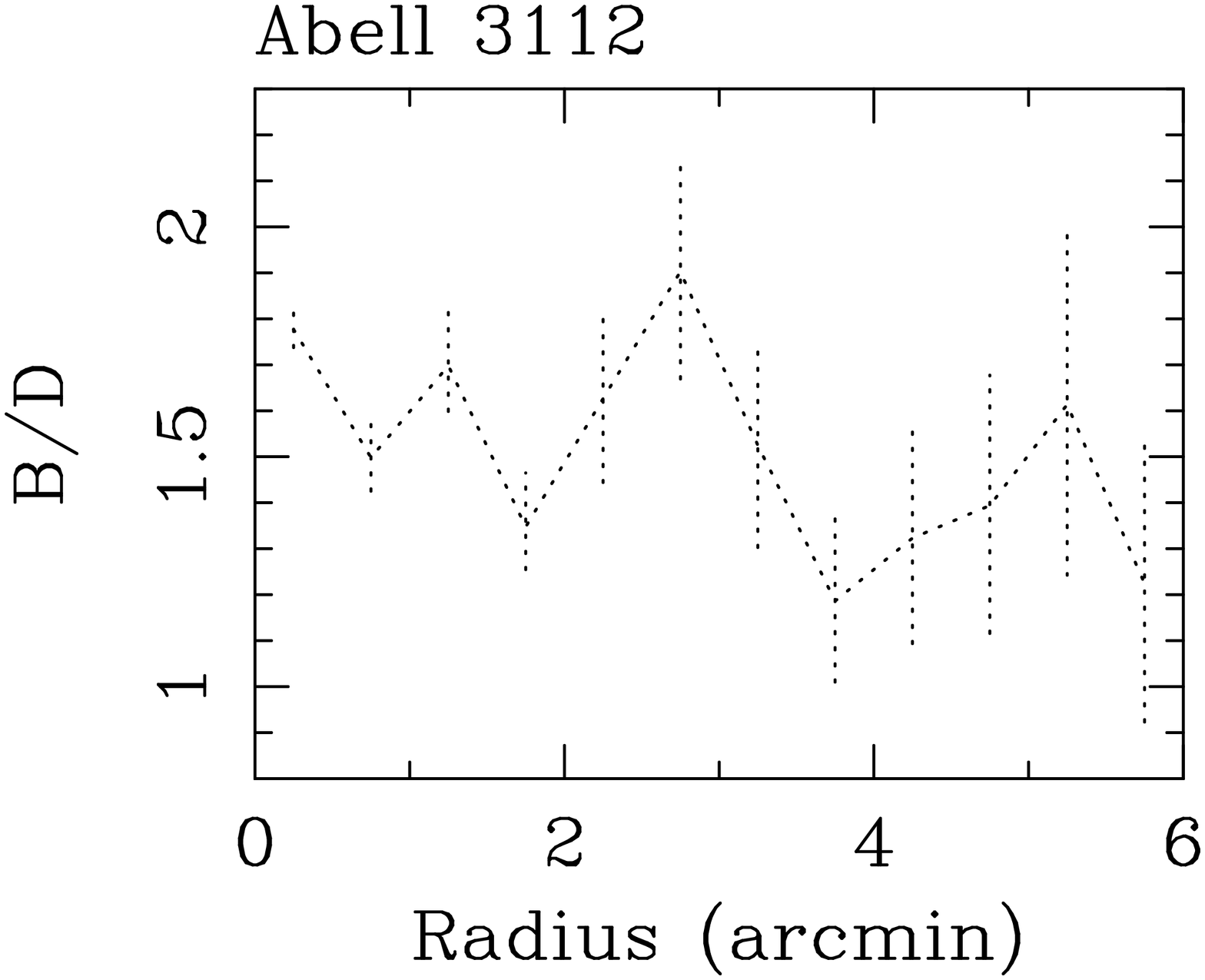,width=0.47\textwidth,angle=270}
}

\vspace{-0.2cm}

\hbox{
\hspace{1cm}\psfig{figure=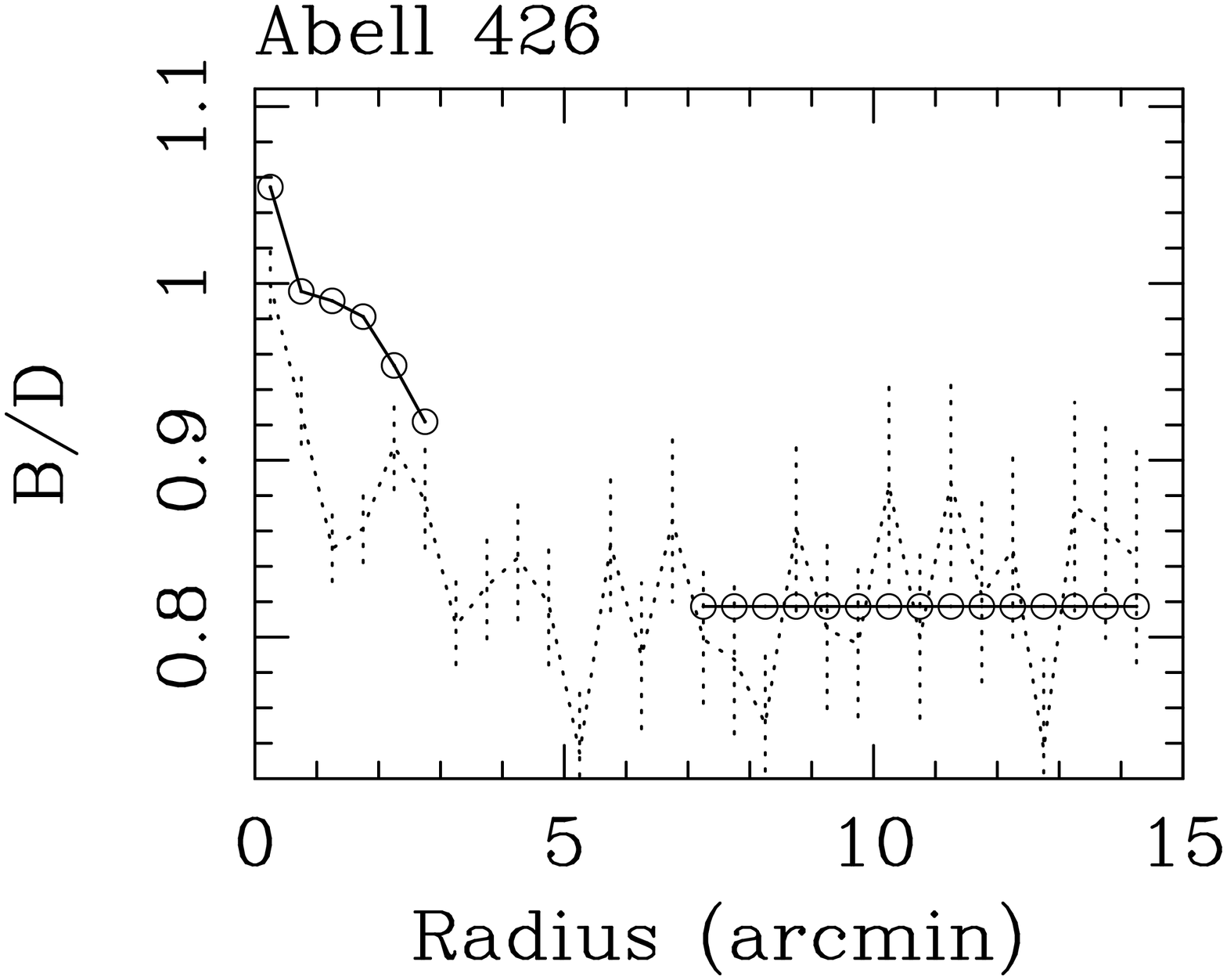,width=0.47\textwidth,angle=270}
\hspace{-0.5cm}\psfig{figure=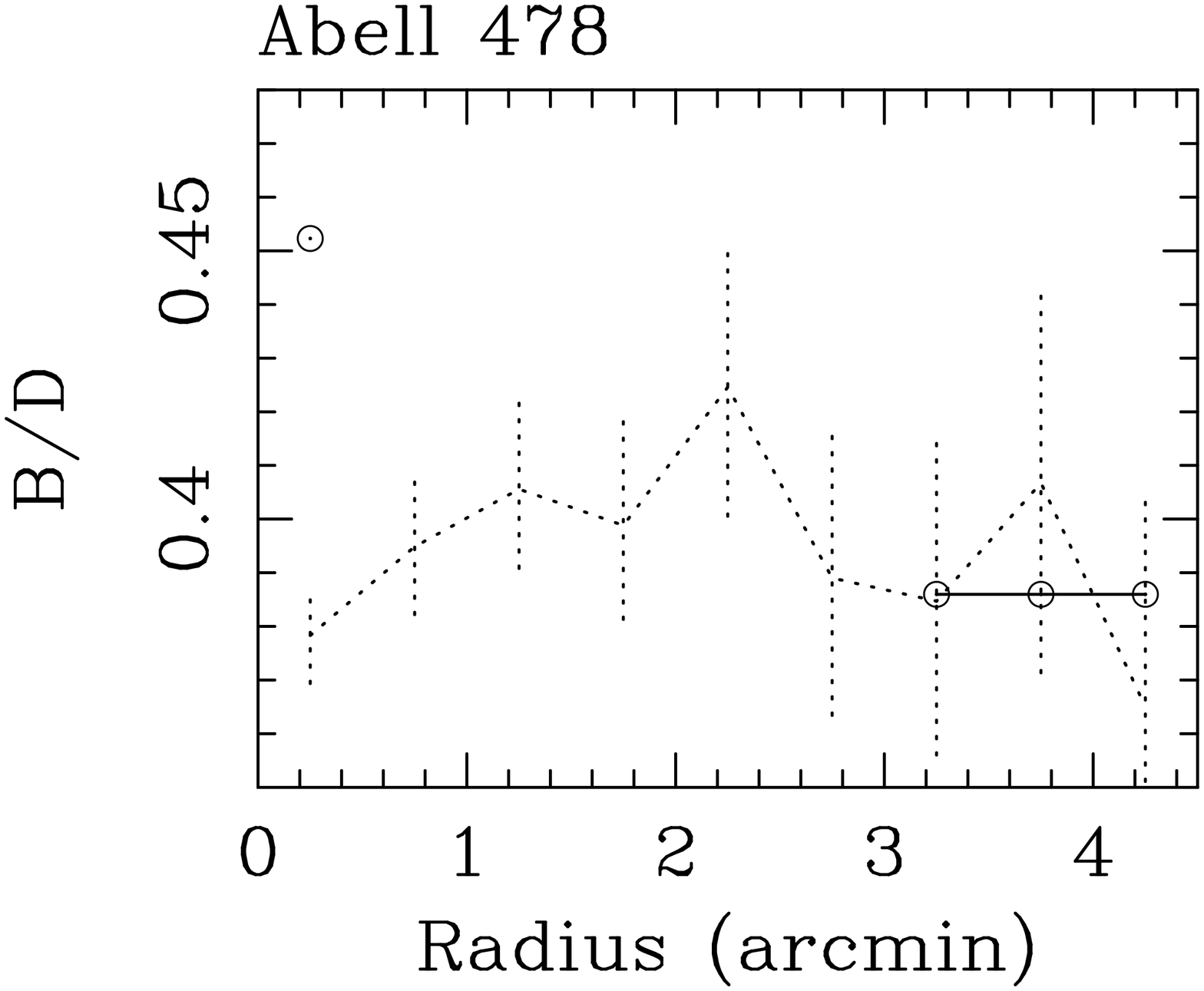,width=0.47\textwidth,angle=270}
}
\vspace{-0.2cm}

\hbox{
\hspace{1cm}\psfig{figure=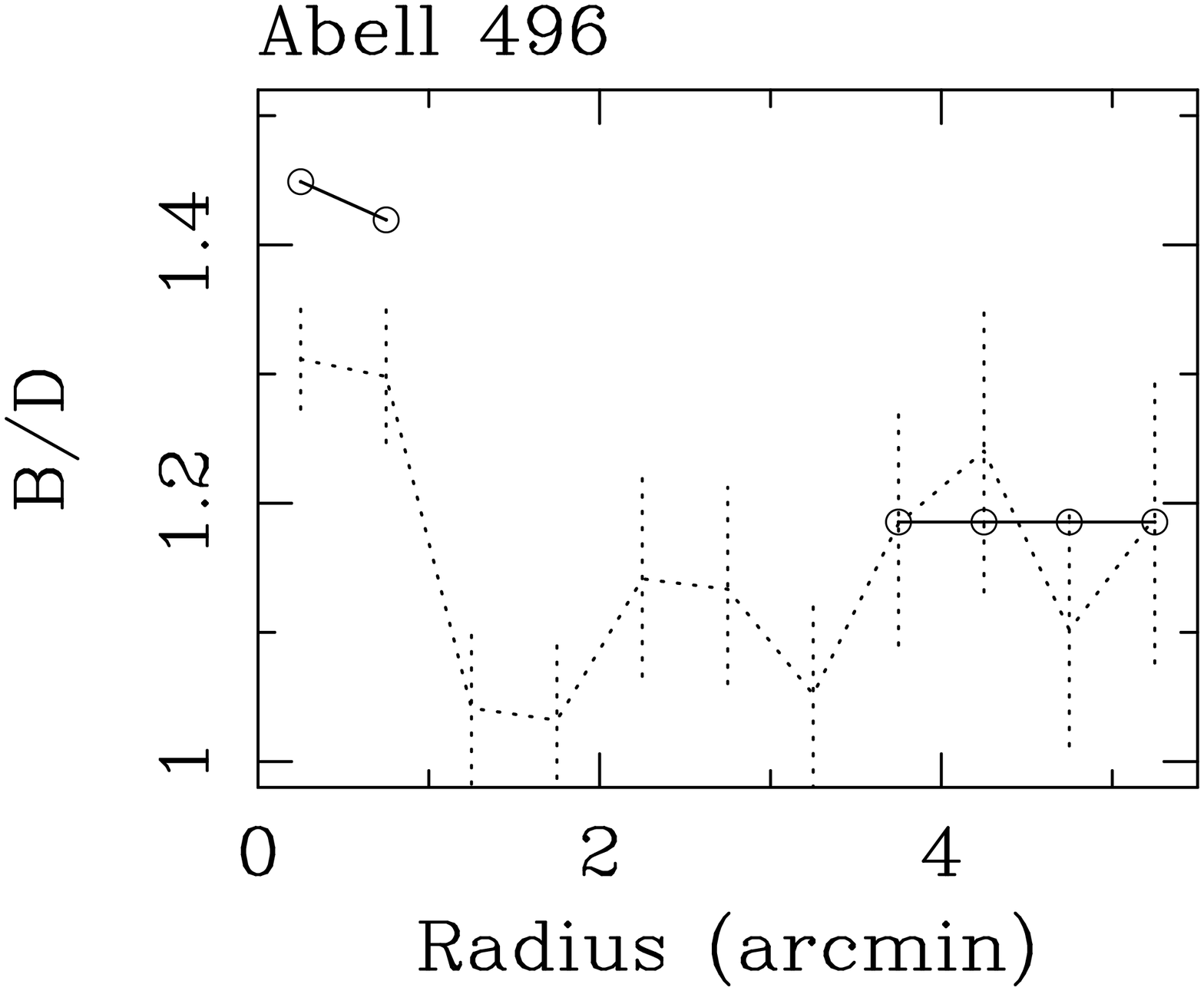,width=0.47\textwidth,angle=270}
\hspace{-0.5cm}\psfig{figure=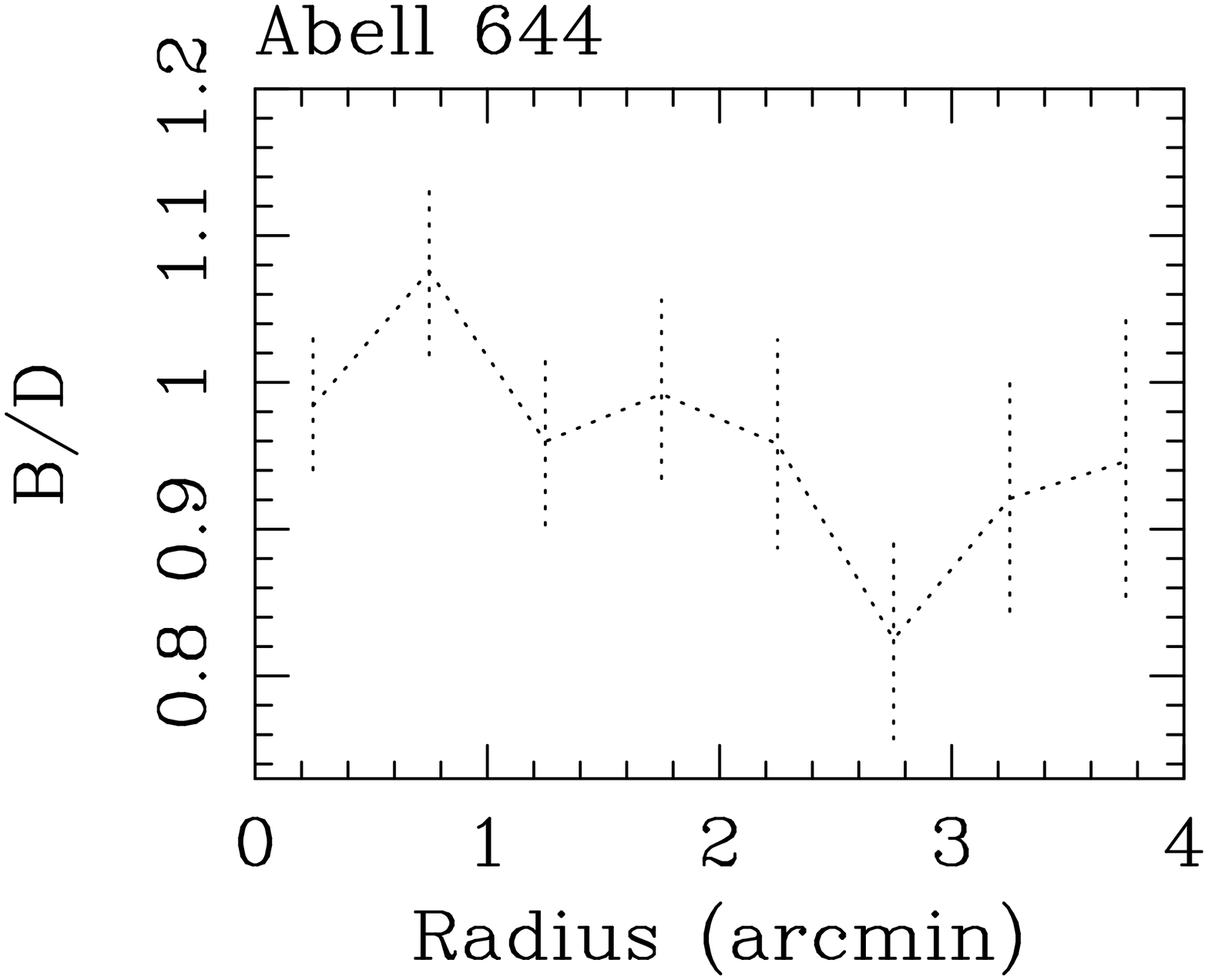,width=0.47\textwidth,angle=270}
}

\caption{The B/D ($0.40-0.79$ keV/$0.80-1.39$ keV) X-ray colour profiles
(with $1 \sigma$ error bars). Circled points mark the values
predicted from the deprojection code (Section 5.3) assuming Galactic
absorption. These predicted values are only plotted for radii where the 
cooling time of the gas is less
than $5\times 10^9$~yr. Note the clear deficit in the observed B/D values 
relative to the predictions from the deprojection code in the central
regions, which indicates the need for excess absorption. In the outer
regions the circled points mark the best-fitting constant B/D value (as
listed in Table 5).
}
\end{figure*}

\addtocounter{figure}{-1}
\begin{figure*}
\hbox{
\hspace{1cm}\psfig{figure=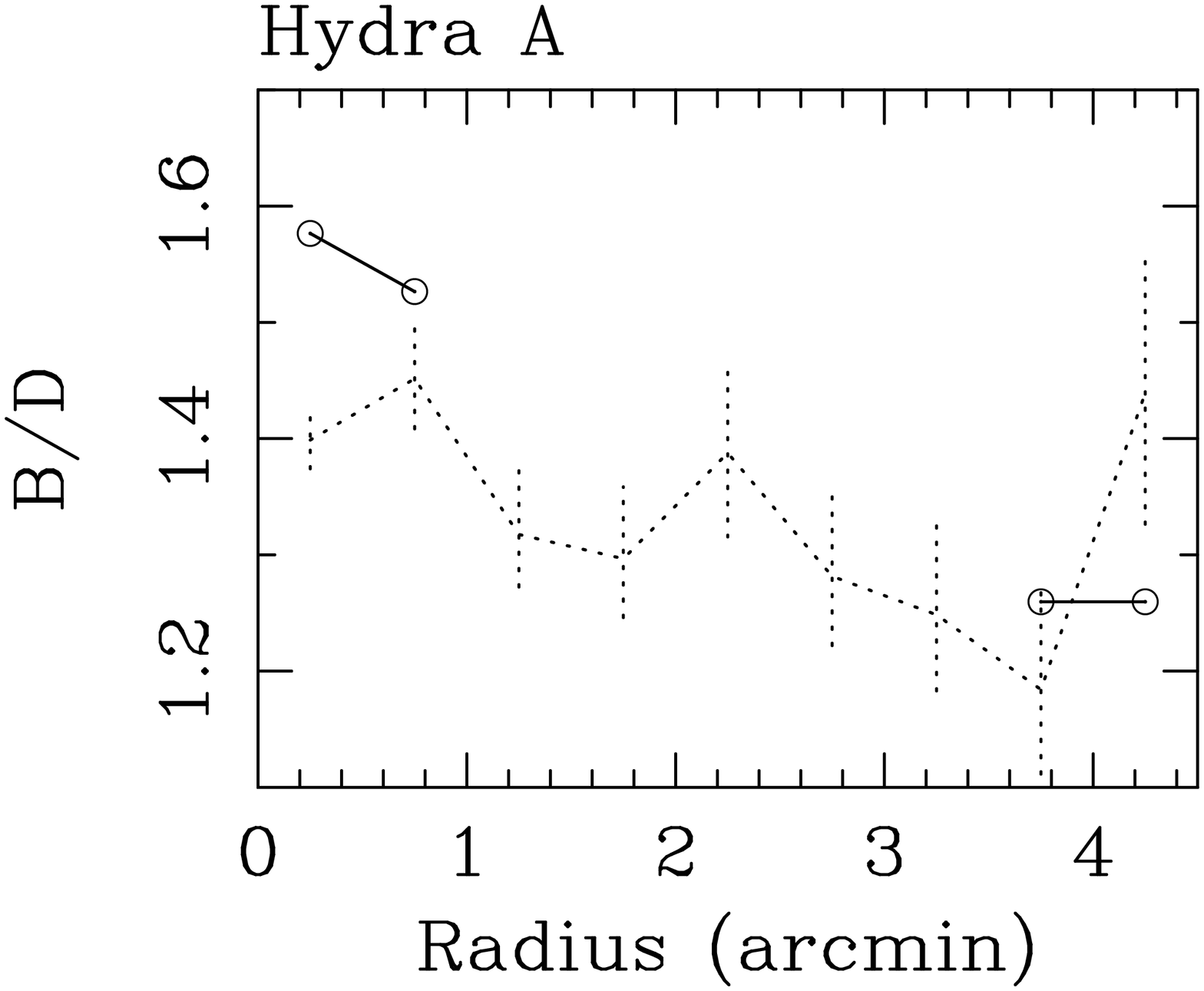,width=0.47\textwidth,angle=270}
\hspace{-0.5cm}\psfig{figure=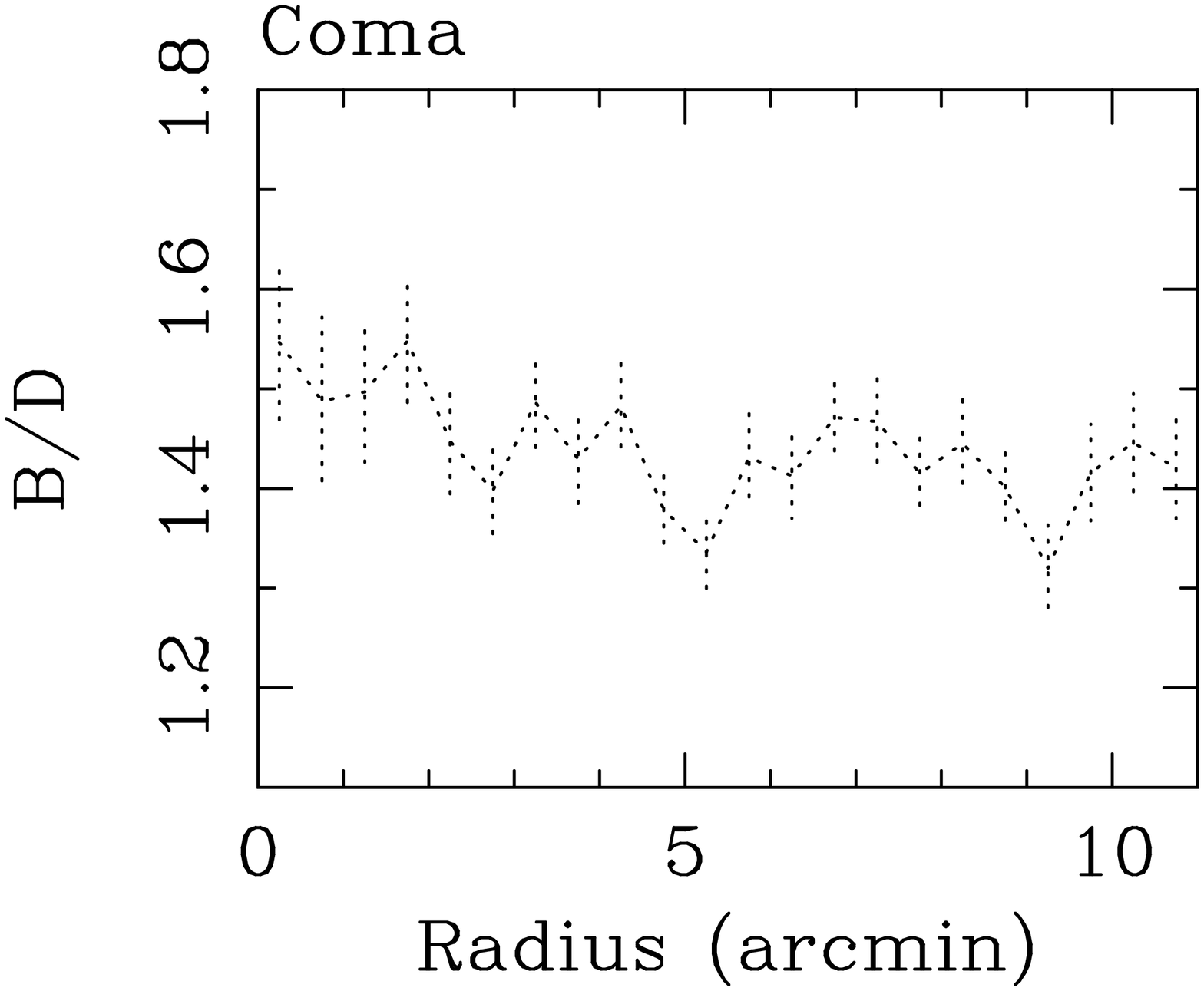,width=0.47\textwidth,angle=270}
}

\vspace{-0.2cm}

\hbox{
\hspace{1cm}\psfig{figure=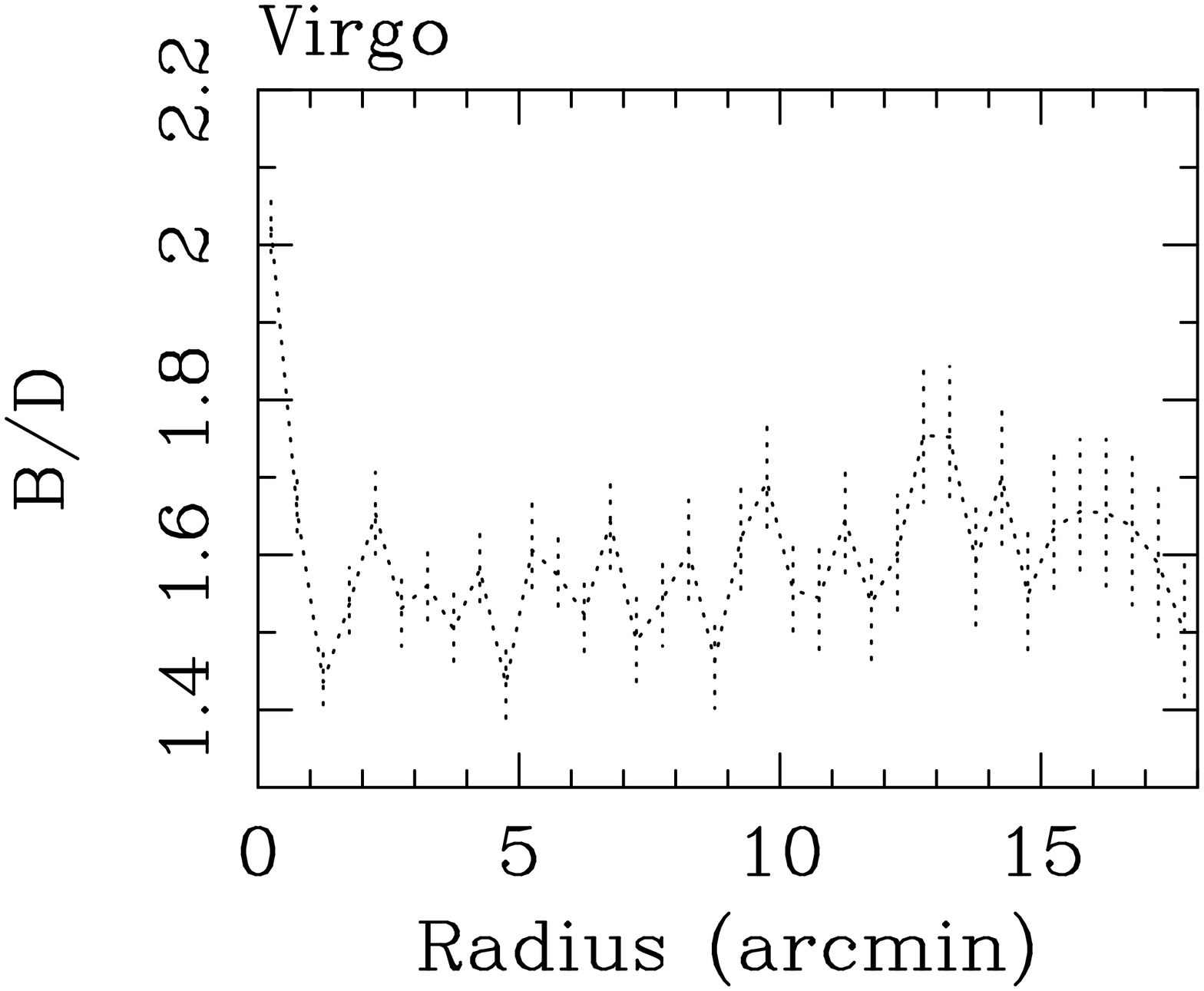,width=0.47\textwidth,angle=270}
\hspace{-0.5cm}\psfig{figure=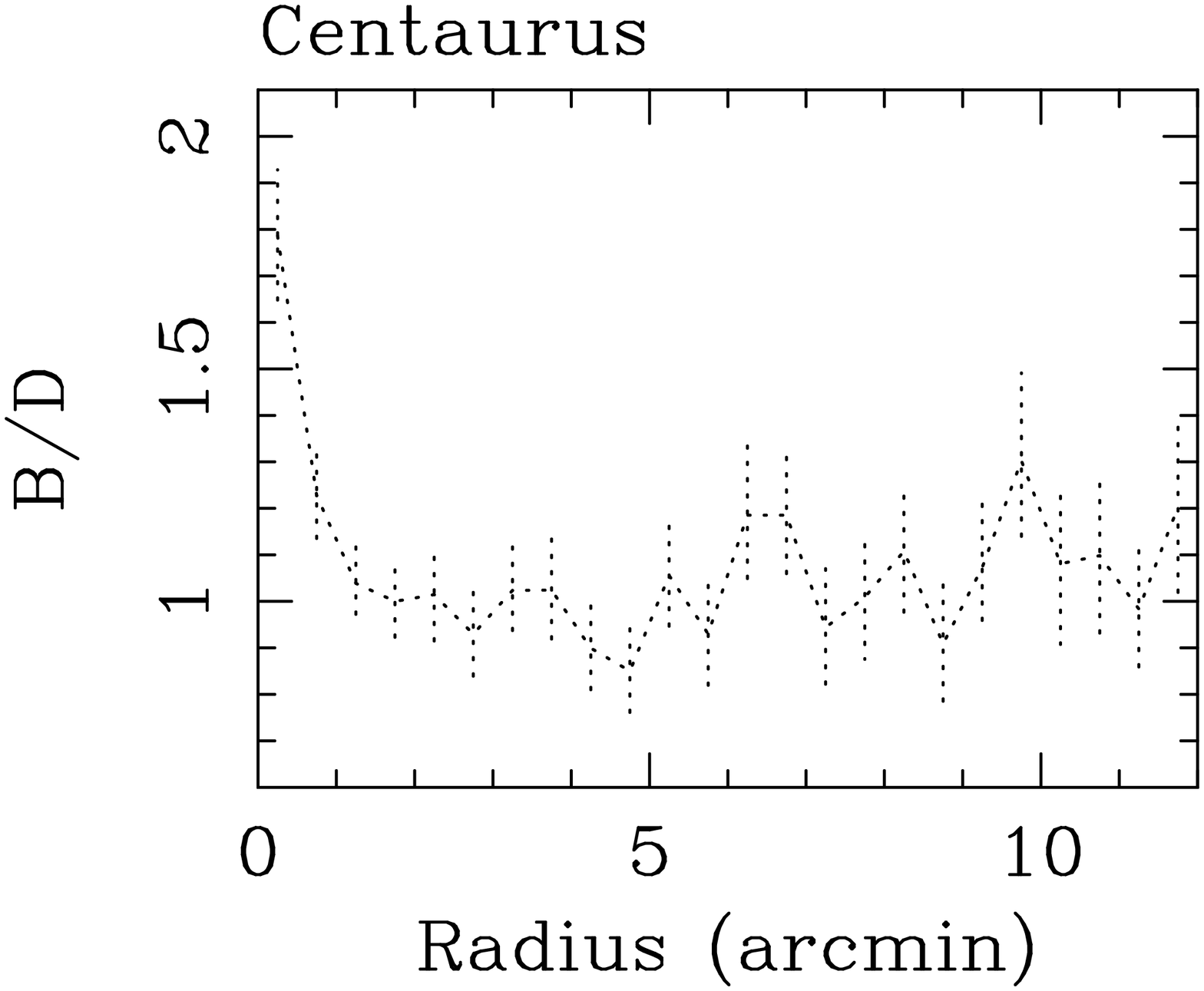,width=0.47\textwidth,angle=270}
}
\vspace{-0.2cm}

\hbox{
\hspace{1cm}\psfig{figure=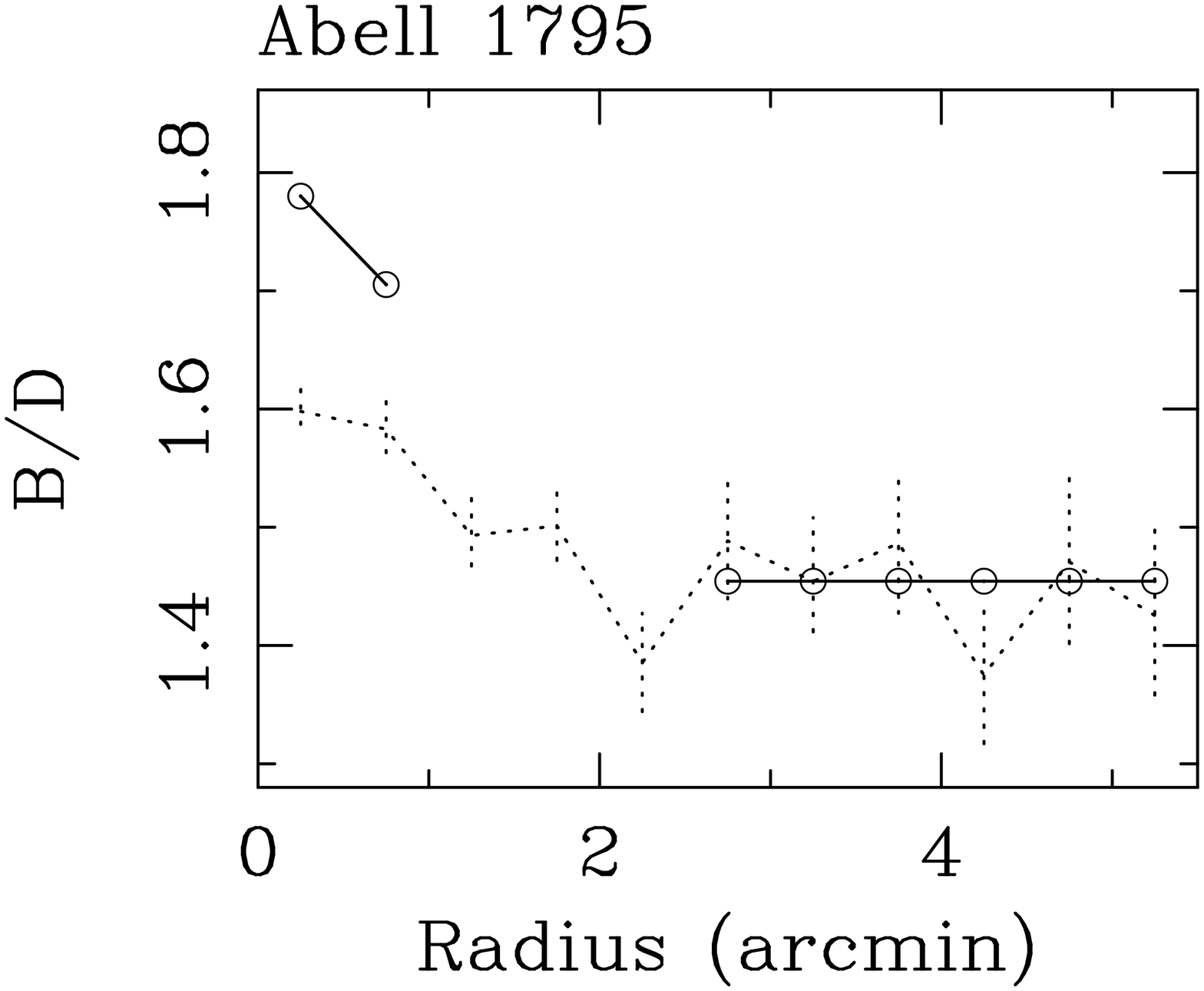,width=0.47\textwidth,angle=270}
\hspace{-0.5cm}\psfig{figure=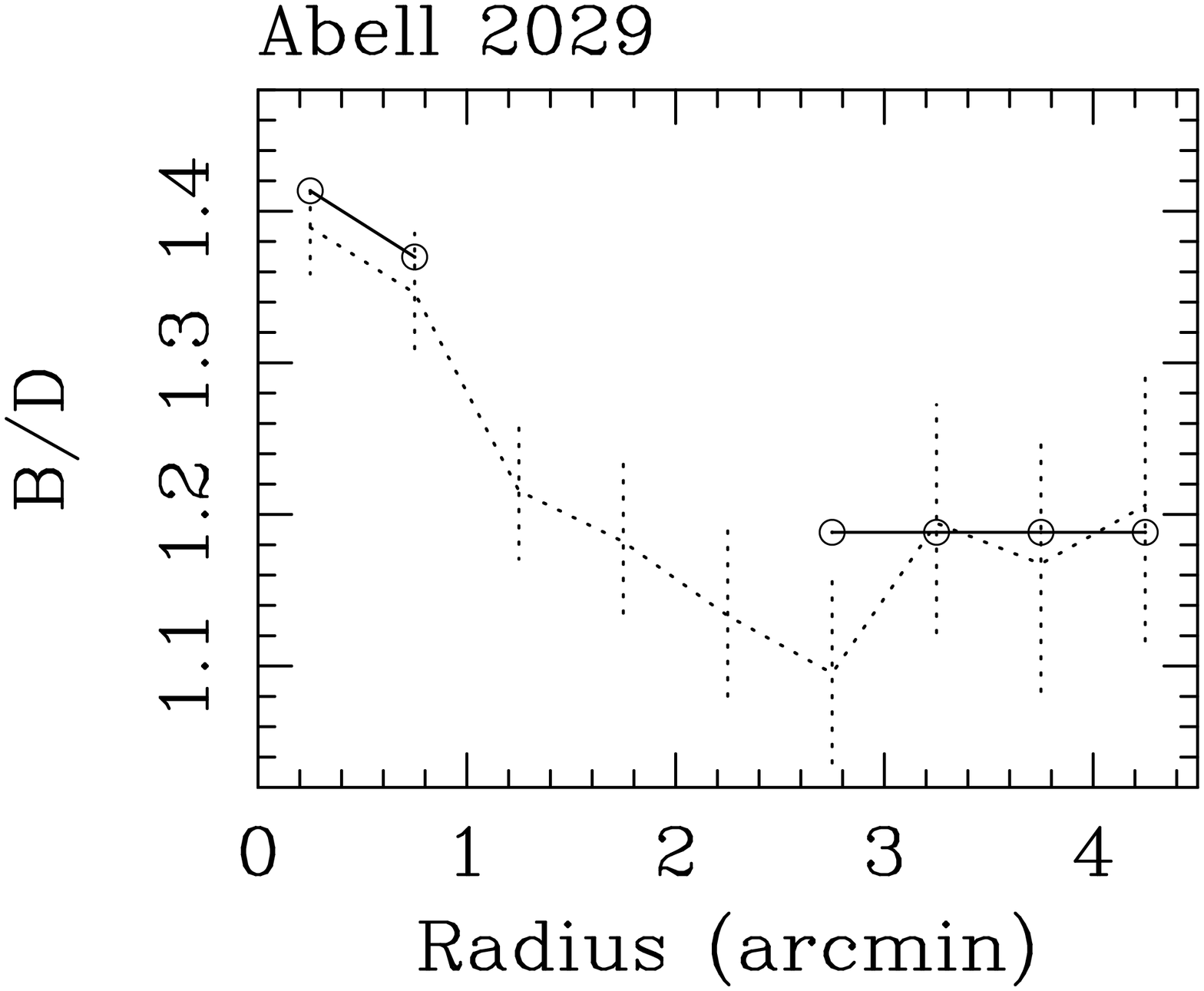,width=0.47\textwidth,angle=270}
}

\caption{ - continued}
\end{figure*}

\addtocounter{figure}{-1}
\begin{figure*}
\hbox{
\hspace{1cm}\psfig{figure=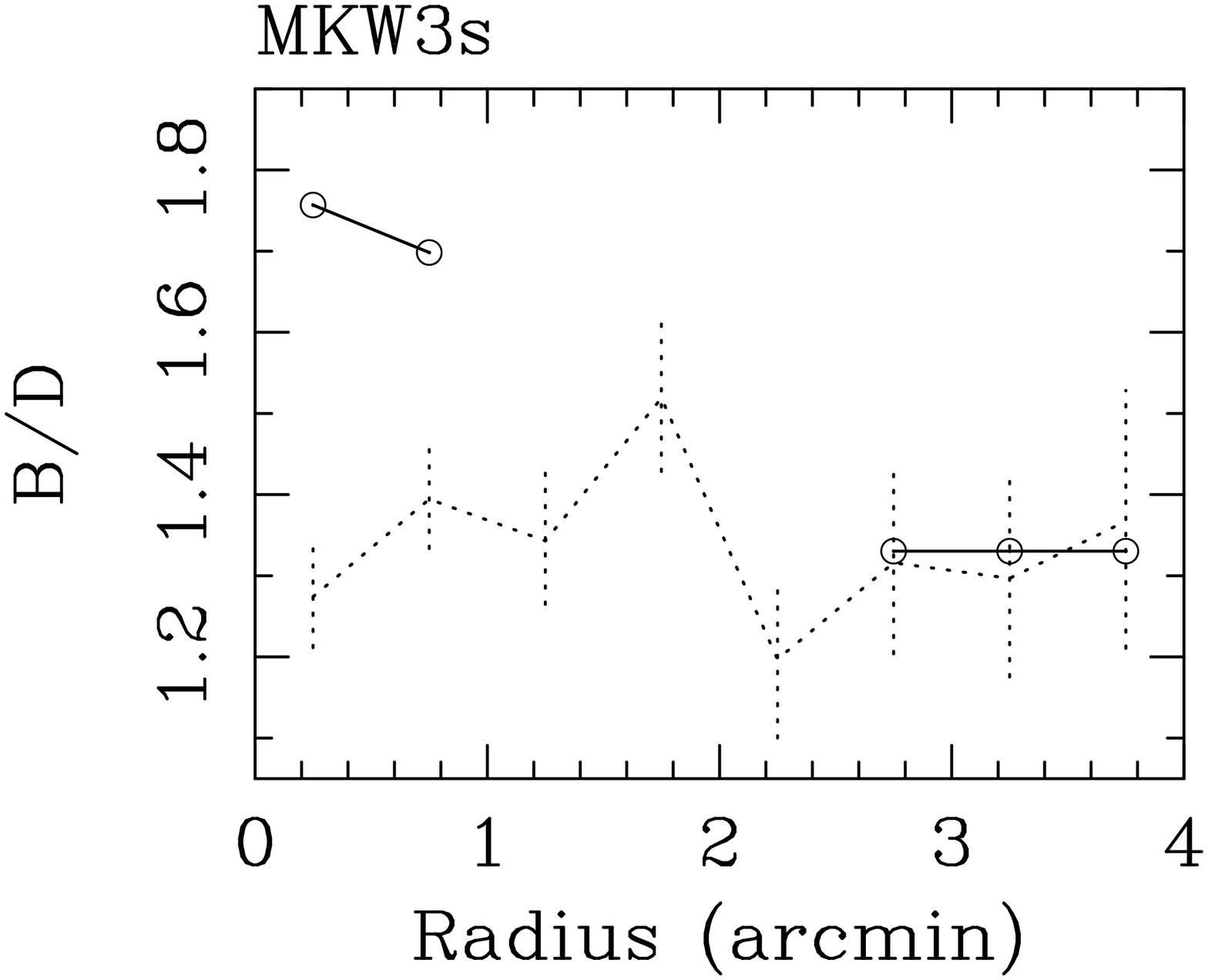,width=0.47\textwidth,angle=270}
\hspace{-0.5cm}\psfig{figure=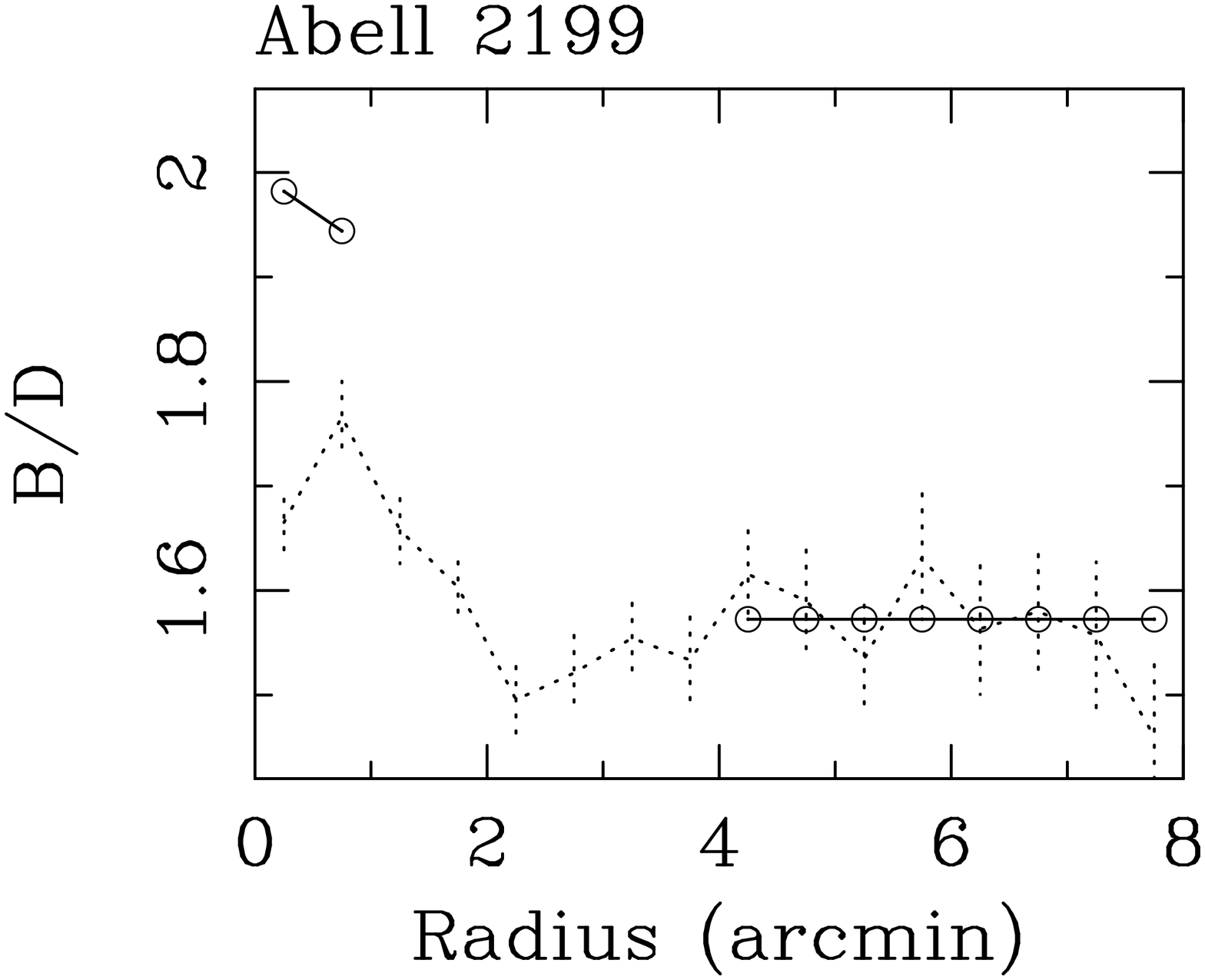,width=0.47\textwidth,angle=270}
}

\vspace{-0.2cm}

\hbox{
\hspace{1cm}\psfig{figure=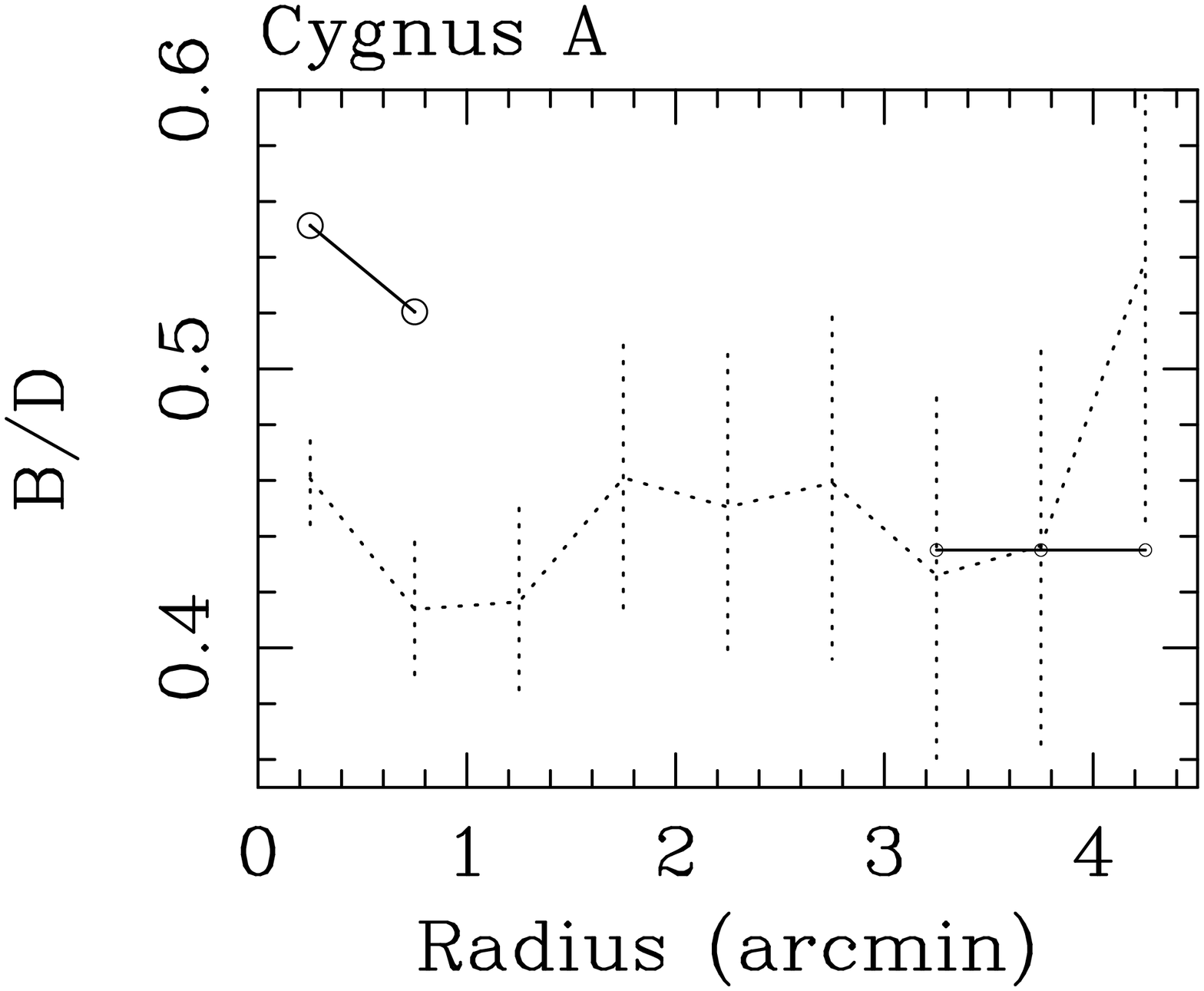,width=0.47\textwidth,angle=270}
\hspace{-0.5cm}\psfig{figure=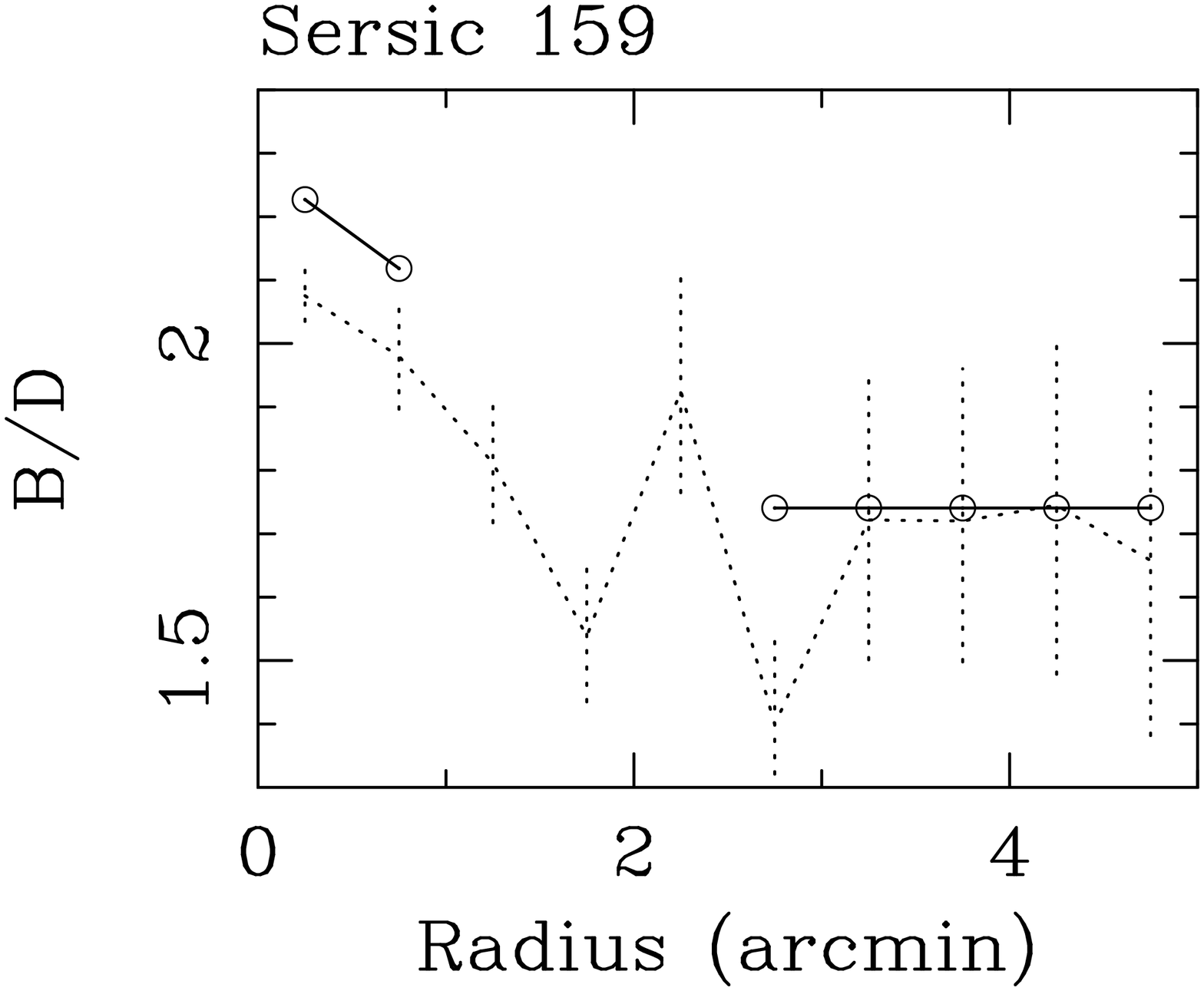,width=0.47\textwidth,angle=270}
}
\vspace{-0.2cm}

\hbox{
\hspace{1cm}\psfig{figure=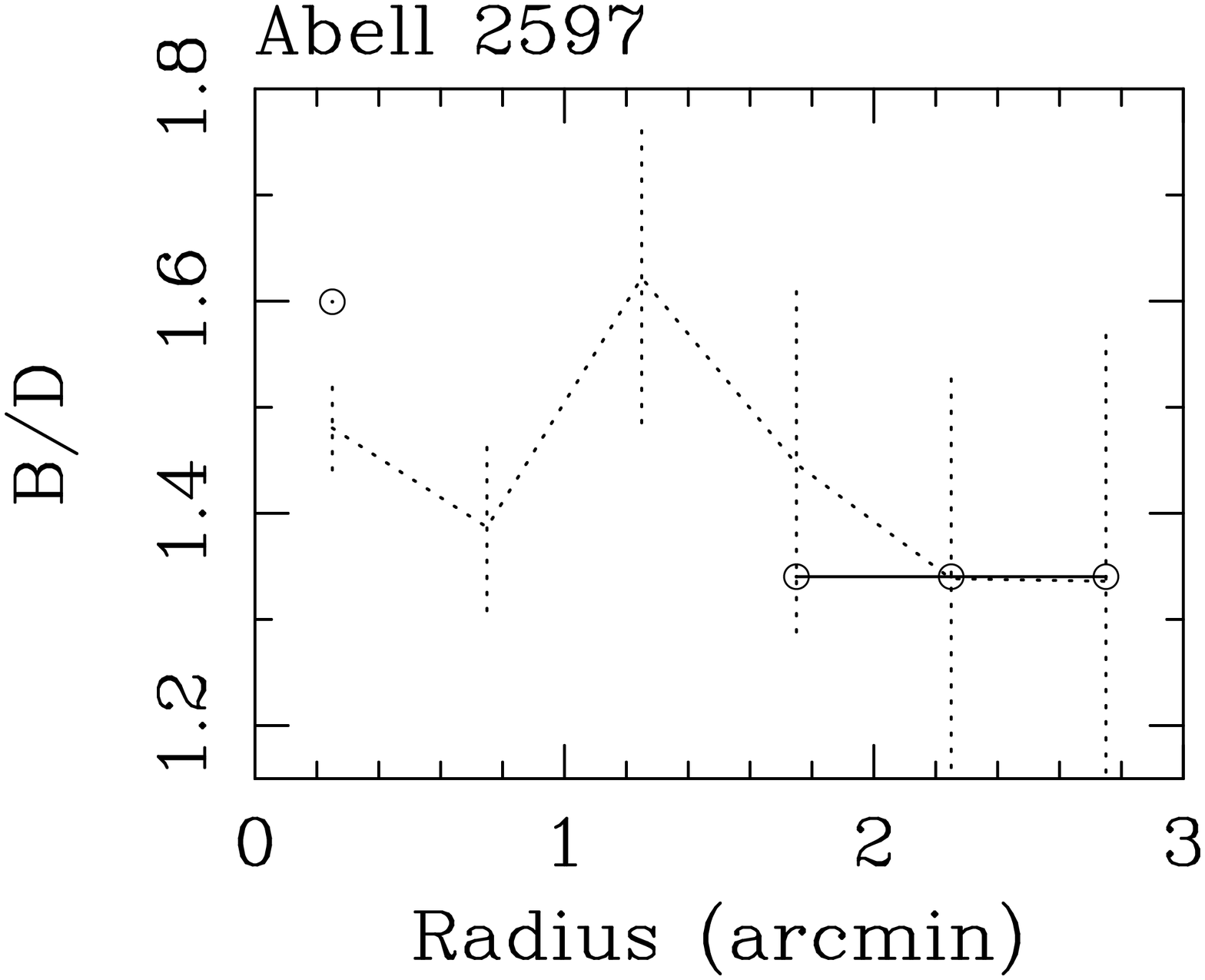,width=0.47\textwidth,angle=270}
\hspace{-0.5cm}\psfig{figure=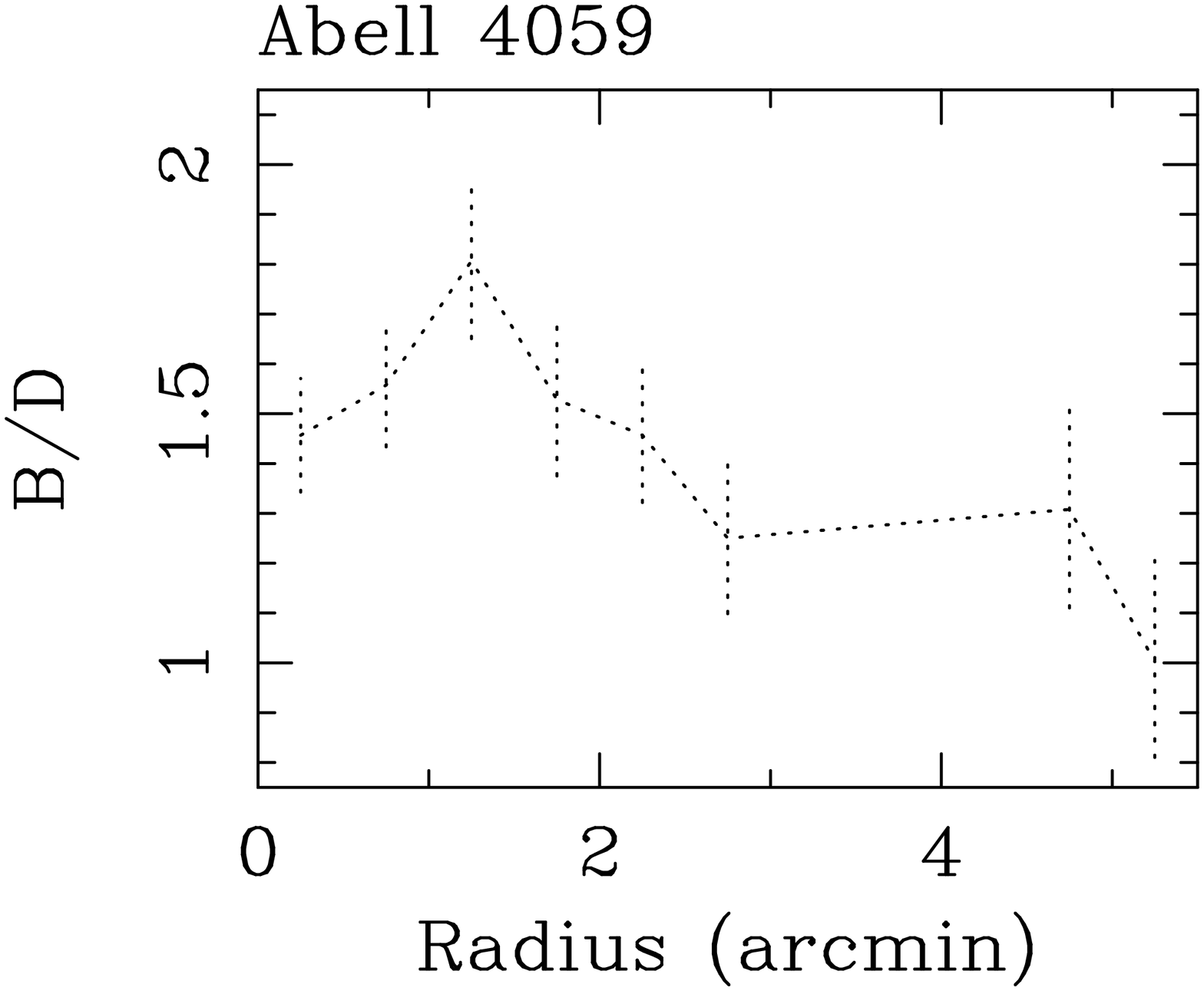,width=0.47\textwidth,angle=270}
}

\caption{ - continued}
\end{figure*}

\begin{figure*}
\hbox{
\hspace{0cm}\psfig{figure=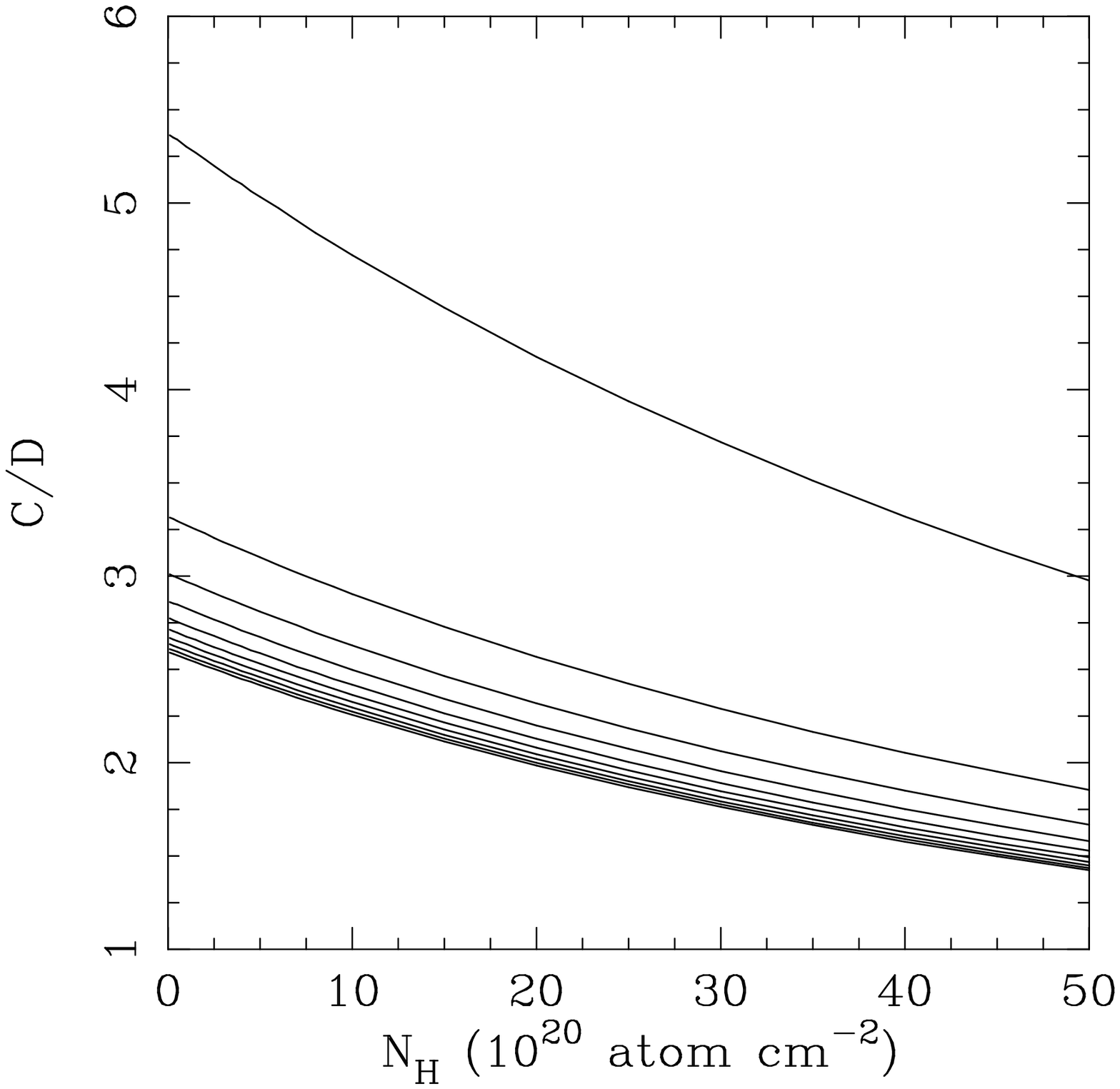,width=0.6\textwidth,angle=270}
\hspace{-2.5cm}\psfig{figure=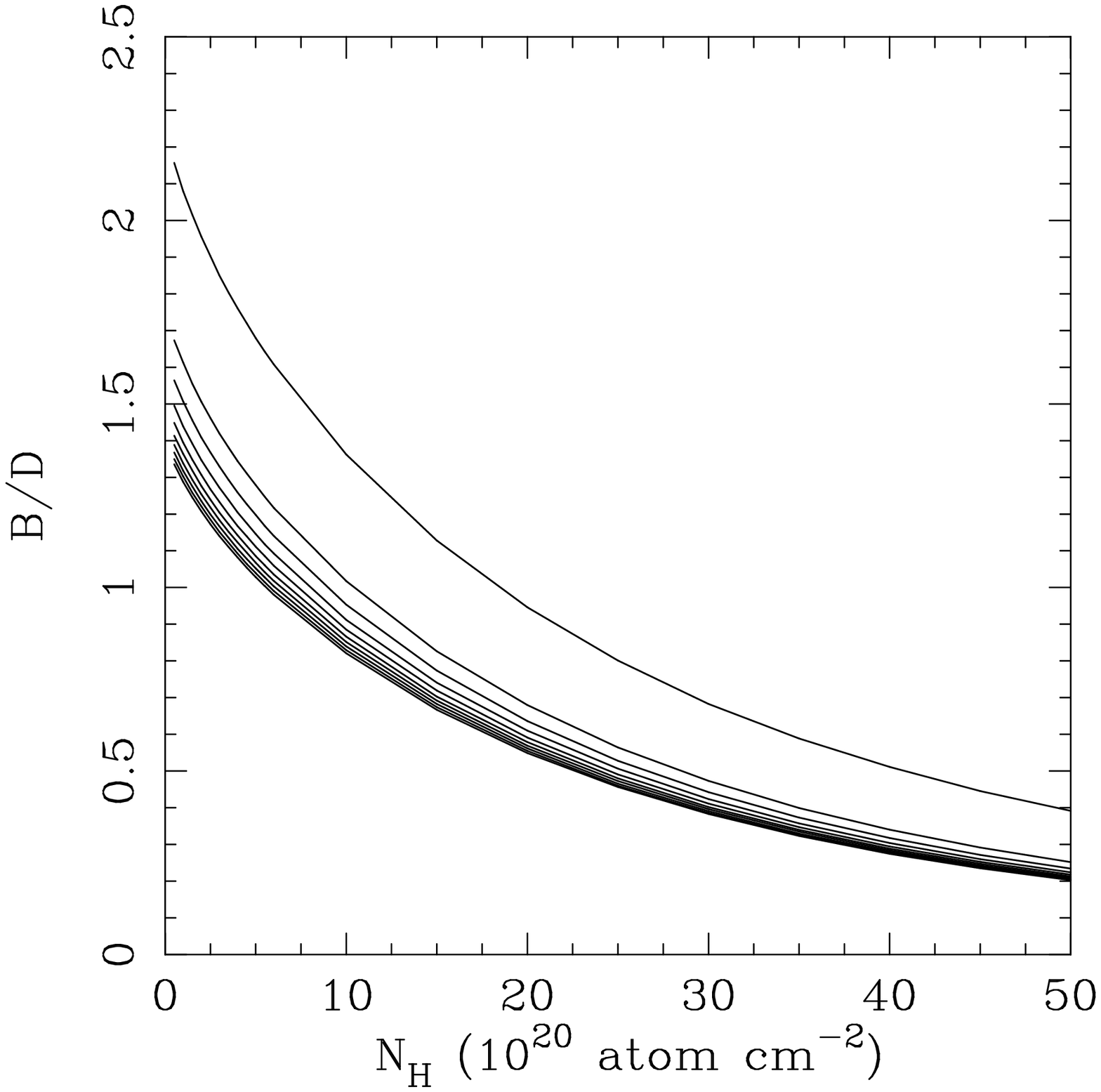,width=0.6\textwidth,angle=270}
}

\vspace{0cm}

\hbox{
\hspace{0cm}\psfig{figure=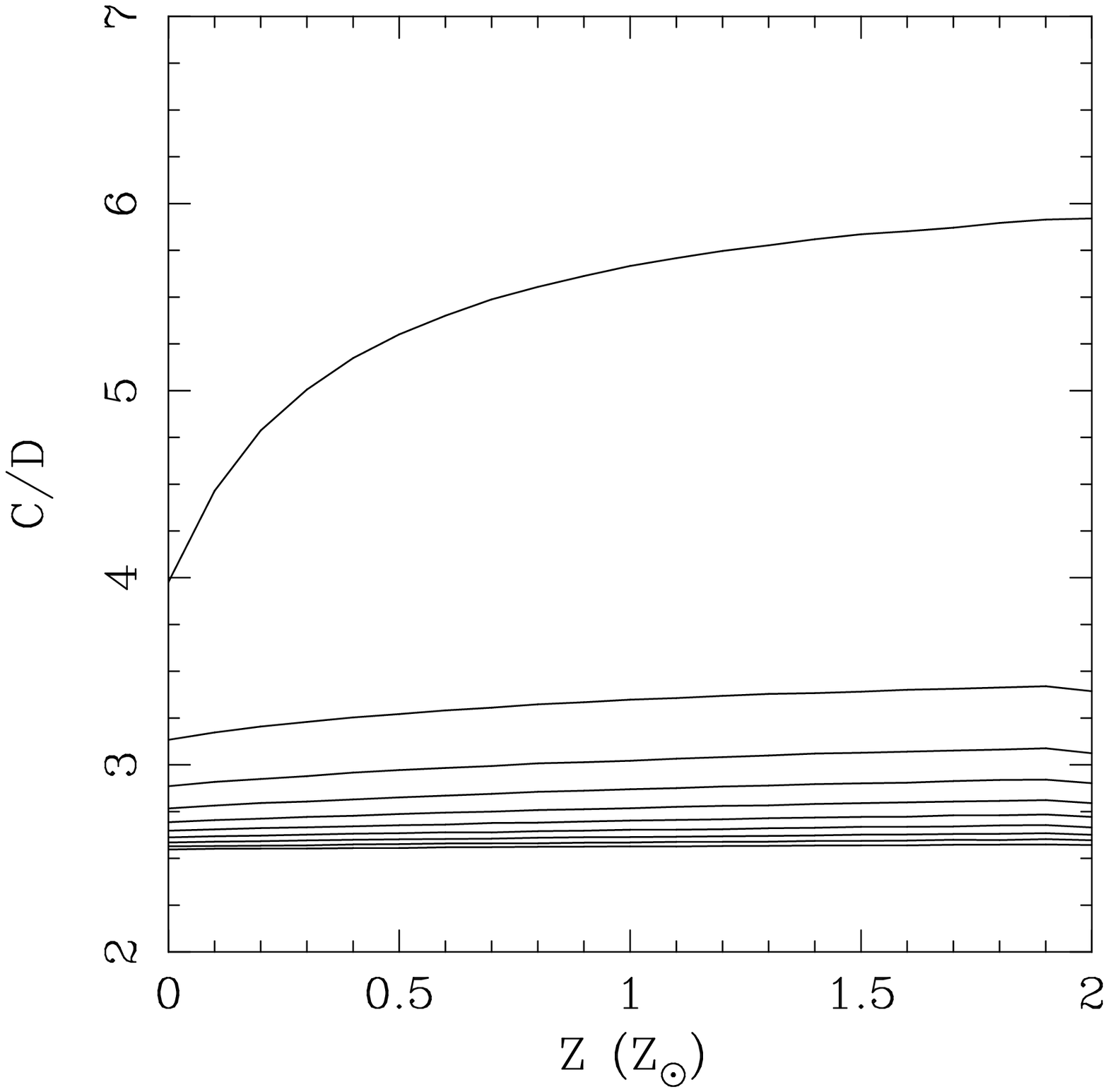,width=0.6\textwidth,angle=270}
\hspace{-2.5cm}\psfig{figure=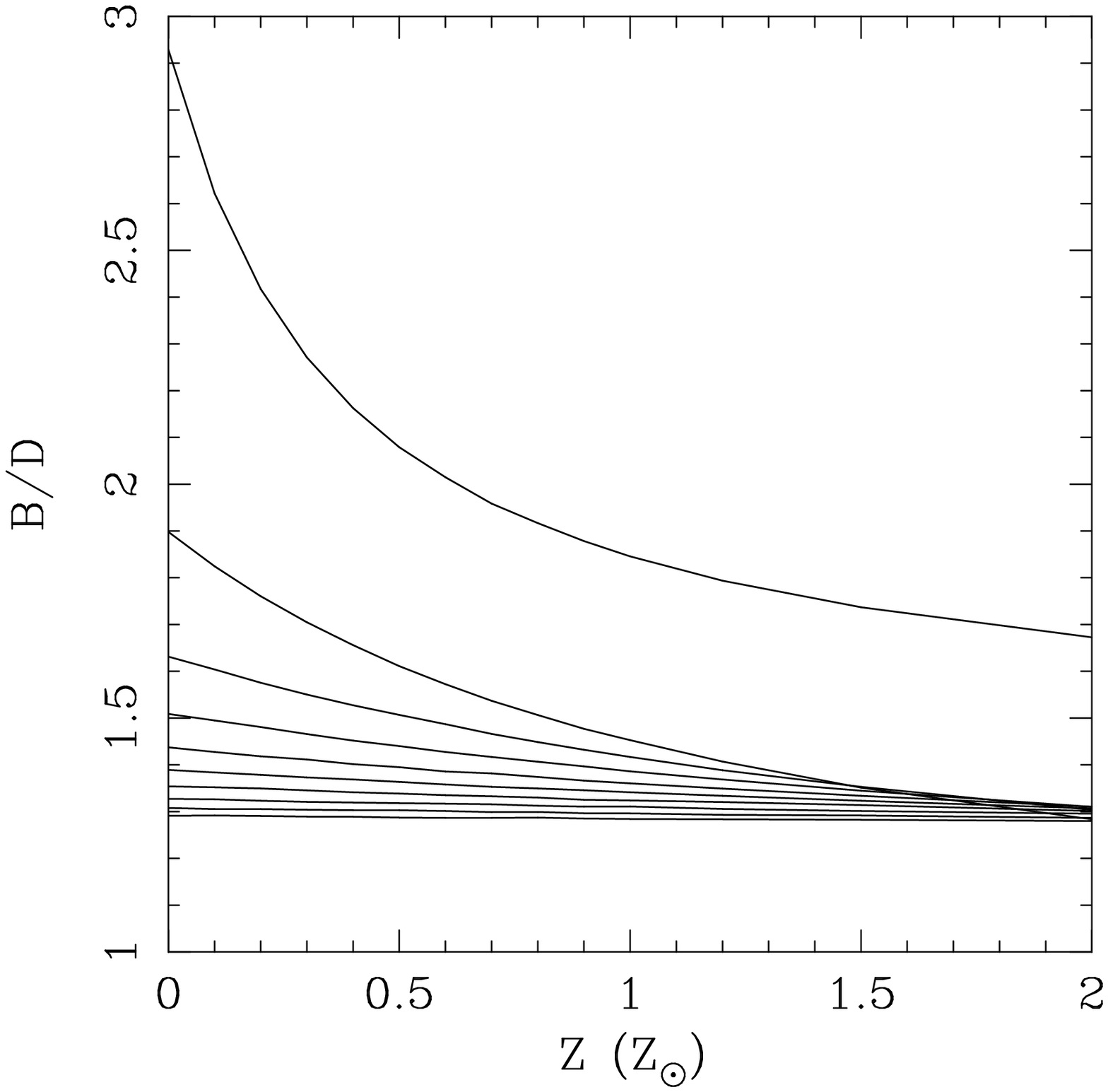,width=0.6\textwidth,angle=270}
}

\caption{Theoretical predictions for the C/D and B/D ratios for isothermal gas 
at temperatures of 1-10 keV. (a) C/D as a function of column density 
($Z = 0.5Z_\odot$ fixed). 
(b) B/D as a function of column density ($Z = 0.5Z_\odot$ fixed). 
(c) C/D as a function of
metallicity ($N_{\rm H} = 10^{20}$ \apc~fixed). 
(d) B/D as a function of metallicity ($N_{\rm H} = 10^{20}$ \apc~fixed).
In all cases the temperature of the gas increases  
with decreasing C/D and B/D along the y axis. 
}
\end{figure*}

\begin{figure*}
\hbox{
\hspace{0cm}\psfig{figure=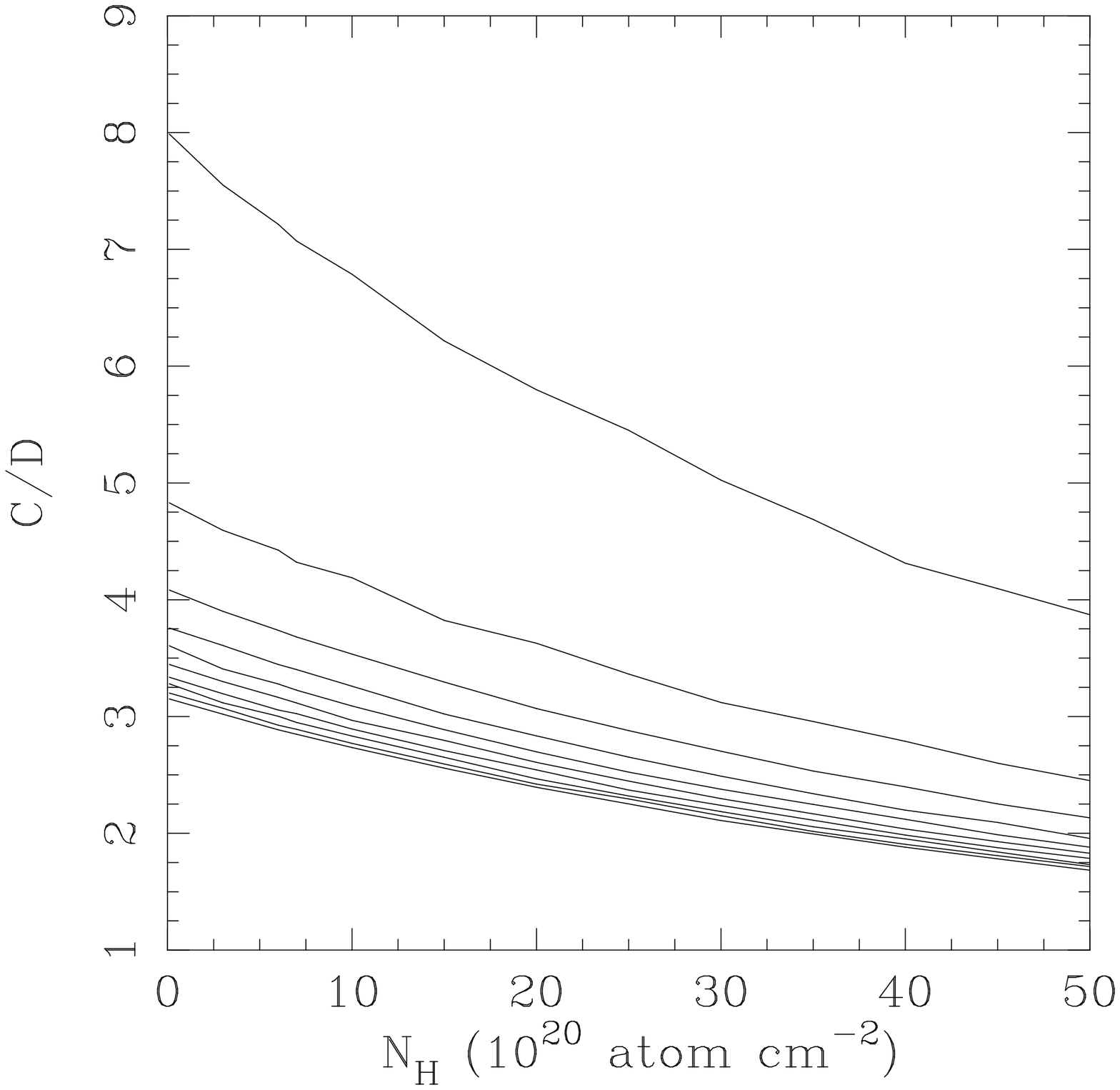,width=0.6\textwidth,angle=270}
\hspace{-2.5cm}\psfig{figure=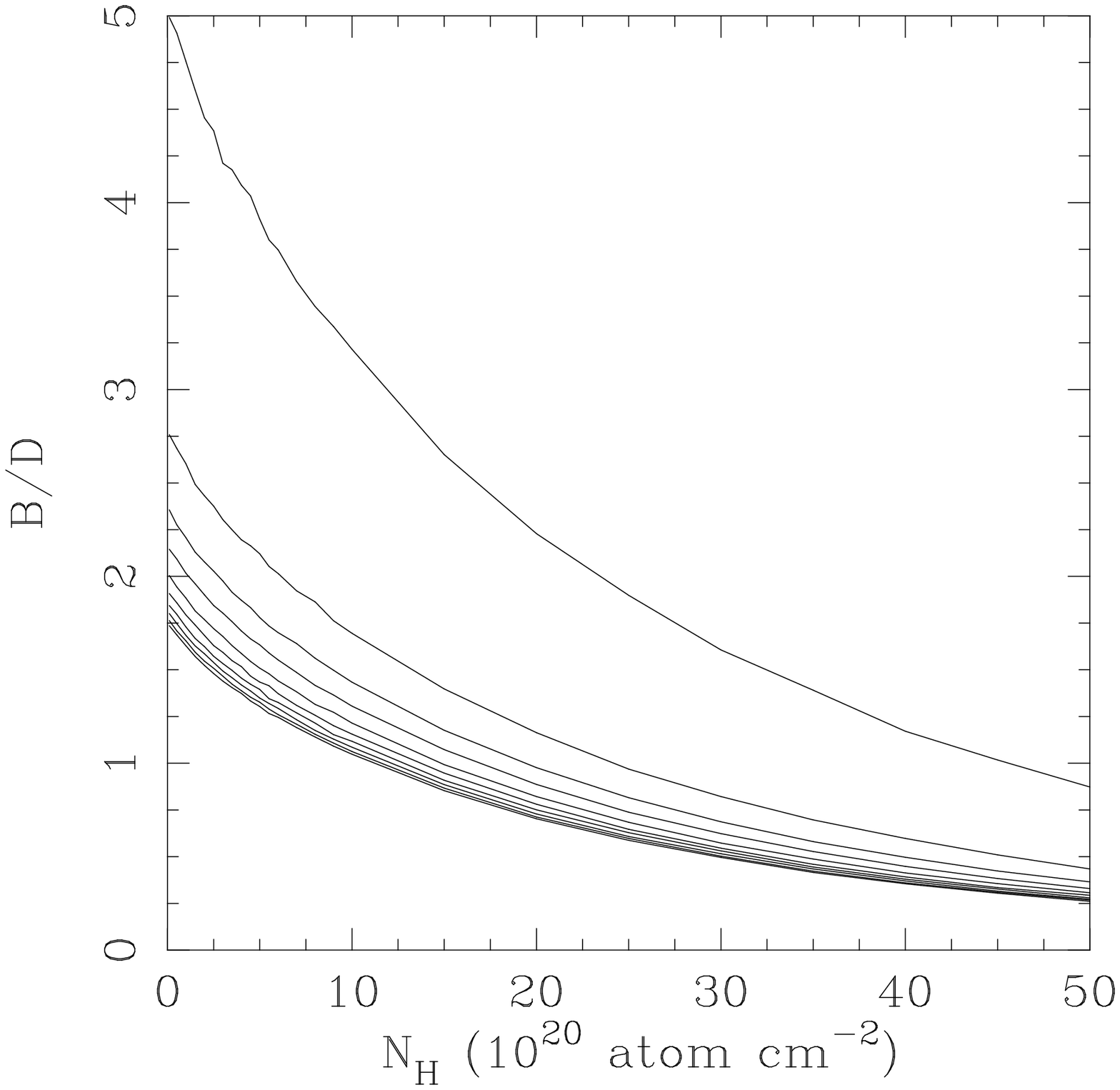,width=0.6\textwidth,angle=270}
}

\vspace{0cm}

\hbox{
\hspace{0cm}\psfig{figure=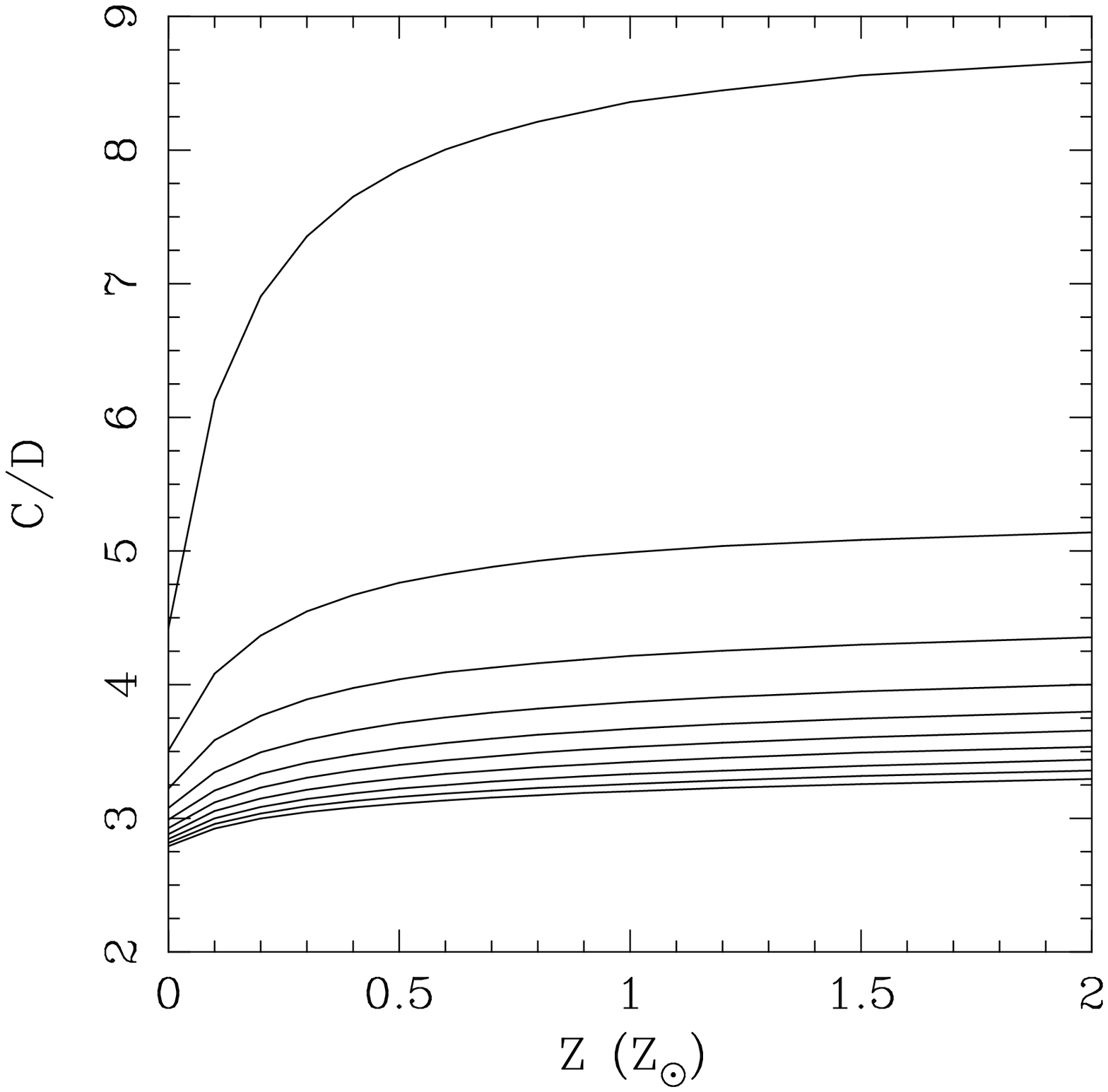,width=0.6\textwidth,angle=270}
\hspace{-2.5cm}\psfig{figure=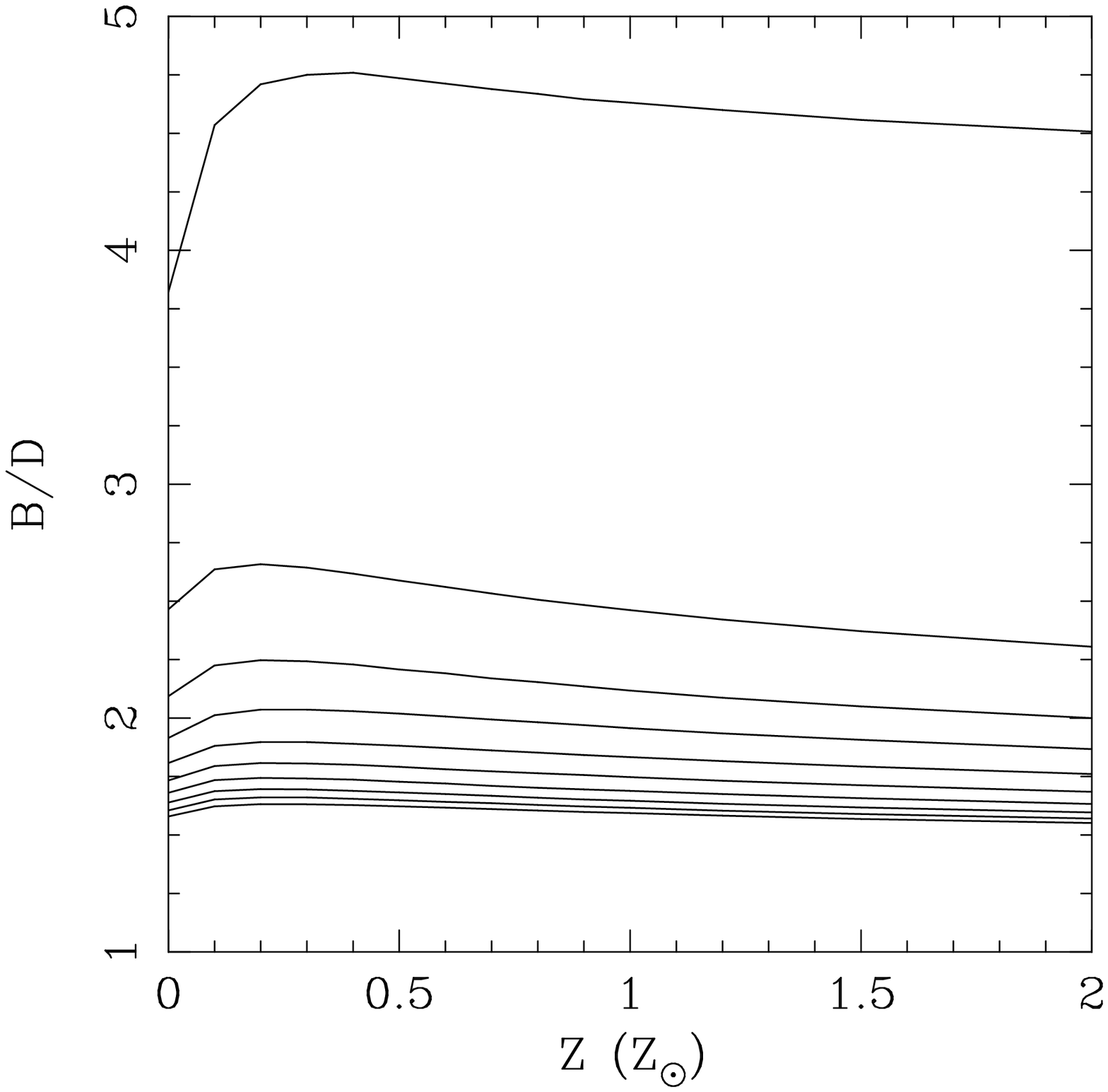,width=0.6\textwidth,angle=270}
}

\caption{Theoretical predictions for the C/D and B/D ratios for 
gas cooling at constant pressure from and upper temperature of 
1-10 keV to a lower
temperature of .001 keV. (a) C/D as a function of column density 
($Z = 0.5Z_\odot$ fixed). 
(b) B/D as a function of column density ($Z = 0.5Z_\odot$ fixed). 
(c) C/D as a function of
metallicity ($N_{\rm H} = 10^{20}$ \apc~fixed). 
(d) B/D as a function of metallicity ($N_{\rm H} = 10^{20}$ \apc~fixed).
In all cases the temperature of the gas increases  
with decreasing C/D and B/D along the y axis. 
}
\end{figure*}

\begin{table}
\vskip 0.2truein
\begin{center}
\caption{A comparison of isothermal and cooling gas}
\vskip 0.2truein
\begin{tabular}{ c c c c c c c }
\multicolumn{1}{c}{} &
\multicolumn{1}{c}{} &
\multicolumn{2}{c}{Isothermal} &
\multicolumn{1}{c}{} &
\multicolumn{2}{c}{C. Flow} \\
 $kT$ (keV)   & ~ &   C/D & B/D   & ~ &  C/D  & B/D    \\
\hline                   
&&&&&& \\                
  1           & ~ &  5.30 &  2.08 & ~ &  7.76 &  4.76 \\
  2           & ~ &  3.27 &  1.61 & ~ &  4.72 &  2.60 \\
  3           & ~ &  2.97 &  1.51 & ~ &  4.02 &  2.21 \\
  4           & ~ &  2.83 &  1.44 & ~ &  3.71 &  2.02 \\
  5           & ~ &  2.74 &  1.39 & ~ &  3.53 &  1.88 \\
  6           & ~ &  2.68 &  1.36 & ~ &  3.40 &  1.79 \\
  7           & ~ &  2.64 &  1.34 & ~ &  3.30 &  1.73 \\
  8           & ~ &  2.60 &  1.32 & ~ &  3.22 &  1.68 \\
  9           & ~ &  2.58 &  1.30 & ~ &  3.15 &  1.66 \\
  10          & ~ &  2.56 &  1.29 & ~ &  3.11 &  1.63 \\
&&&&&& \\
\end{tabular}
\end{center}

\parbox {3.2in}
{Notes: A comparison of the theoretical C/D and B/D ratios for isothermal gas and
a constant pressure cooling flow. For the cooling flow data, the $kT$
value is the ambient cluster temperature from which the gas cools. 
A Galactic column density of $10^{20}$ \apc~
and a metallicity of $0.5 Z_\odot$ are assumed.}
\end{table}

As discussed in Section 1, our method of analysis makes use of X-ray colour
profiles, rather than the more standard approach of taking annular or regional spectra, 
to study the distributions of cool gas and intrinsic absorbing
material in the clusters. Such a technique allows us to make the best use
of the spatial resolution offered by the ROSAT PSPC.

We have constructed radial surface brightness profiles for the clusters 
in each the energy bands, B, C and D 
(Table 2). The surface brightness profiles were binned to a spatial
scale of 30 arcsec pixel$^{-1}$ (slightly larger than the 
FWHM resolution of the PSPC) which provides a good signal-to-noise
ratio in the pixels over the regions of interest. 
Figs. 2 and 3 summarize the results on the X-ray colour profiles.
Fig. 2 shows the C/D profiles, which are
approximately constant at large radii but increase in
their central regions. The B/D profiles, shown in Fig. 3, are generally
flatter than the C/D data.

Figs. 4 and 5 illustrate how the X-ray colour profiles may be used to
probe the physical properties of the clusters. 
These figures show the behaviour of the C/D and B/D ratios in simulated
cluster spectra as a function of temperature, column density and
metallicity. Results are shown both for isothermal gas and for 
gas cooling at constant pressure, from the ambient cluster temperature 
to a temperature below the ROSAT band \ie a constant pressure cooling flow. 
The simulated spectra were constructed using the plasma code
of Kaastra \& Mewe (1993) and the photoelectric absorption models of
Morrison \& McCammon (1983). We adopt the definition of solar abundance
from Anders \& Grevesse (1989). In all cases the 
absorbing column density is modelled as a uniform screen, with solar
metallicity, at zero redshift. A cluster redshift of $z = 0.01$ is
assumed (although the results do not vary significantly for any
redshift $\leq 0.1$). 

The theoretical curves for an isothermal plasma are shown in Figs.
4(a)--(d). At high temperatures ($kT \approxgt 3$ keV) both the C/D and
B/D ratios are weak functions of temperature. At lower temperatures,
however, the C/D ratio (in particular) increases sharply as 
the temperature drops, due primarily to
Fe-L emission lines in the C band produced by gas at $\sim 1$ keV. 
Both the B/D and C/D ratios decrease with increasing column density,
although the variation of the B/D ratio is steeper (particularly at low
column densities). Both
ratios also exhibit some variation with metallicity. In general, as the
metallicity rises the C/D ratio increases (slightly) and the B/D ratio drops,
although for $kT \approxgt 3$ keV the functions are very weak.  

The behaviour for cooling gas, shown in Figs. 5(a)--(d), is
qualitatively much like that for isothermal emission, except that
the C/D and B/D ratios generally give higher values for the same
temperature, column density and metallicity. This is illustrated by the
data presented in Table 4 where we list the C/D and B/D values for
isothermal and cooling plasmas as a function of temperature, for a fixed
column density of $10^{20}$ \apc~and a metallicity of 0.5Z$_\odot$. The
variation of B/D with metallicity, Fig. 5(d), is flatter for cooling
gas than for an isothermal plasma, and exhibits a downwards trend at low
metallicities not seen in the isothermal data.

In what follows, we use the C/D profiles to demonstrate the presence of
distributed cooling gas in the central regions of the cooling flow clusters, and the 
B/D ratio to demonstrate the presence of intrinsic X-ray absorbing gas.

\section{Results from the Colour Profiles}

\subsection{C/D and cooling radii} 

\begin{table*}
\vskip 0.2truein
\begin{center}
\caption{C/D and B/D values at large radii}
\vskip 0.2truein
\begin{tabular}{ c c c c c c c c c }
\multicolumn{1}{c}{} &
\multicolumn{1}{c}{} &
\multicolumn{1}{c}{} &
\multicolumn{1}{c}{} &
\multicolumn{2}{c}{C/D} &
\multicolumn{1}{c}{} &
\multicolumn{2}{c}{B/D} \\
 cluster     & ~ &  radii      & ~ &  observed    &    predicted    & ~ &  observed    &  predicted \\  
\hline                                                                                                                                   
&&&&&&&& \\                                                                                                                             
Abell 85     & ~ & $3.0-6.0$   & ~ & $2.58\pm0.08$  &  2.58     & ~ &  $1.23\pm0.05$   &  1.20  \\  
Abell 3112   & ~ & $3.0-6.0$   & ~ & $2.83\pm0.19$  &  2.80     & ~ &  $1.34\pm0.11$   &  1.32  \\  
Abell 426    & ~ & $6.0-15.0$  & ~ & $2.31\pm0.03$  &  2.24     & ~ &  $0.82\pm0.01$   &  0.74  \\  
Abell 478    & ~ & $2.5-4.5$   & ~ & $1.83\pm0.05$  &  2.07     & ~ &  $0.39\pm0.02$   &  0.58  \\
Abell 496    & ~ & $3.6-9.0$   & ~ & $2.78\pm0.07$  &  2.64     & ~ &  $1.18\pm0.04$   &  1.17  \\
Abell 644    & ~ & $3.0-7.2$   & ~ & $2.40\pm0.09$  &  2.35     & ~ &  $0.89\pm0.04$   &  0.90  \\  
Hydra A      & ~ & $3.0-4.5$   & ~ & $2.72\pm0.10$  &  2.77     & ~ &  $1.26\pm0.05$   &  1.20  \\  
Coma         & ~ & $2.0-12.0$  & ~ & $2.69\pm0.02$  &  2.61     & ~ &  $1.42\pm0.01$   &  1.33  \\  
Virgo        & ~ & $10.0-12.0$ & ~ & $3.04\pm0.05$  &  3.07     & ~ &  $1.57\pm0.04$   &  1.48  \\  
Centaurus    & ~ & $6.0-12.0$  & ~ & $2.66\pm0.09$  &  2.58     & ~ &  $1.07\pm0.04$   &  0.97  \\  
Abell 1795   & ~ & $2.5-5.5$   & ~ & $2.67\pm0.04$  &  2.76     & ~ &  $1.46\pm0.03$   &  1.40  \\  
Abell 2029   & ~ & $3.0-4.5$   & ~ & $2.46\pm0.09$  &  2.58     & ~ &  $1.19\pm0.05$   &  1.19  \\  
MKW3s        & ~ & $3.0-4.2$   & ~ & $2.85\pm0.19$  &  2.99     & ~ &  $1.33\pm0.11$   &  1.38  \\  
Abell 2199   & ~ & $4.0-8.0$   & ~ & $2.93\pm0.04$  &  2.78     & ~ &  $1.57\pm0.02$   &  1.43  \\
Cyg A        & ~ & $1.5-3.0$   & ~ & $2.01\pm0.10$  &  1.91     & ~ &  $0.44\pm0.04$   &  0.39  \\  
Sersic 159   & ~ & $3.0-7.8$   & ~ & $3.15\pm0.17$  &  3.02     & ~ &  $1.74\pm0.11$   &  1.48  \\  
Abell 2597   & ~ & $1.8-3.0$   & ~ & $2.79\pm0.21$  &  2.83     & ~ &  $1.34\pm0.16$   &  1.35  \\  
Abell 4059   & ~ & $4.2-8.4$   & ~ & $2.79\pm0.17$  &  2.91     & ~ &  $1.26\pm0.09$   &  1.44  \\  
                           
&&&&&&&& \\                
\end{tabular}
\end{center}
\parbox {7in}
{Notes: The observed and predicted C/D and B/D ratios in the outer
regions of the clusters. Column 2 lists the range of radii (in arcmin) 
that define the `outer regions'. 
Columns 3 and 5 summarize the observed C/D and B/D values and 1$\sigma$
errors. Columns 4 and 6 list the predicted ratios for an 
isothermal plasma, at the temperature and redshift 
of the cluster. Galactic column densities (Table 3) and a metallicity of 
$0.5Z_\odot$ are assumed. 
The general agreement between the observed and predicted C/D and B/D ratios 
demonstrates the validity of the isothermal assumption in these regions, 
and the absence of excess absorbing material at these radii (see 
discussion in Section 5.3).
}
\end{table*}

The C/D colour profiles shown in Fig. 2 exhibit a number of striking
features. Firstly, we see that at large radii ($r > r_{\rm cool}$) the
profiles are essentially flat. The observed values are generally in good
agreement with the theoretical predictions for isothermal gas at
the temperatures listed in Table 3, absorbed by the appropriate
Galactic column density (although significant
deviations are observed for Abell 478, the Coma cluster, 
and Abell 2199, the  
origins of which are discussed in Section 5.3.) 
The observed and predicted C/D and B/D ratios
at large radii are summarized in Table 5. 

Within radii $r \approxlt r_{\rm cool}$, however, the cooling-flow 
clusters exhibit 
clear increases in their C/D ratios. 
From comparison with the theoretical curves in Figs. 4 and 5, 
we see that these increases can only plausibly be explained by the presence
of cooler gas. The X-ray colour profiles
thus provide firm evidence for distributed cool(ing)
 gas in the central regions of the
cooling flow clusters and show that this gas is 
confined within the cooling radii 
inferred from the deprojection analyses. 
For the one non-cooling flow cluster in the sample, the Coma cluster, no
significant enhancement in the C/D ratio (outside the central 30 arcsec
\ie 20 kpc) is observed.

\subsection{The spatial distribution of cooling gas} 

\begin{figure*}
\hbox{
\hspace{1cm}\psfig{figure=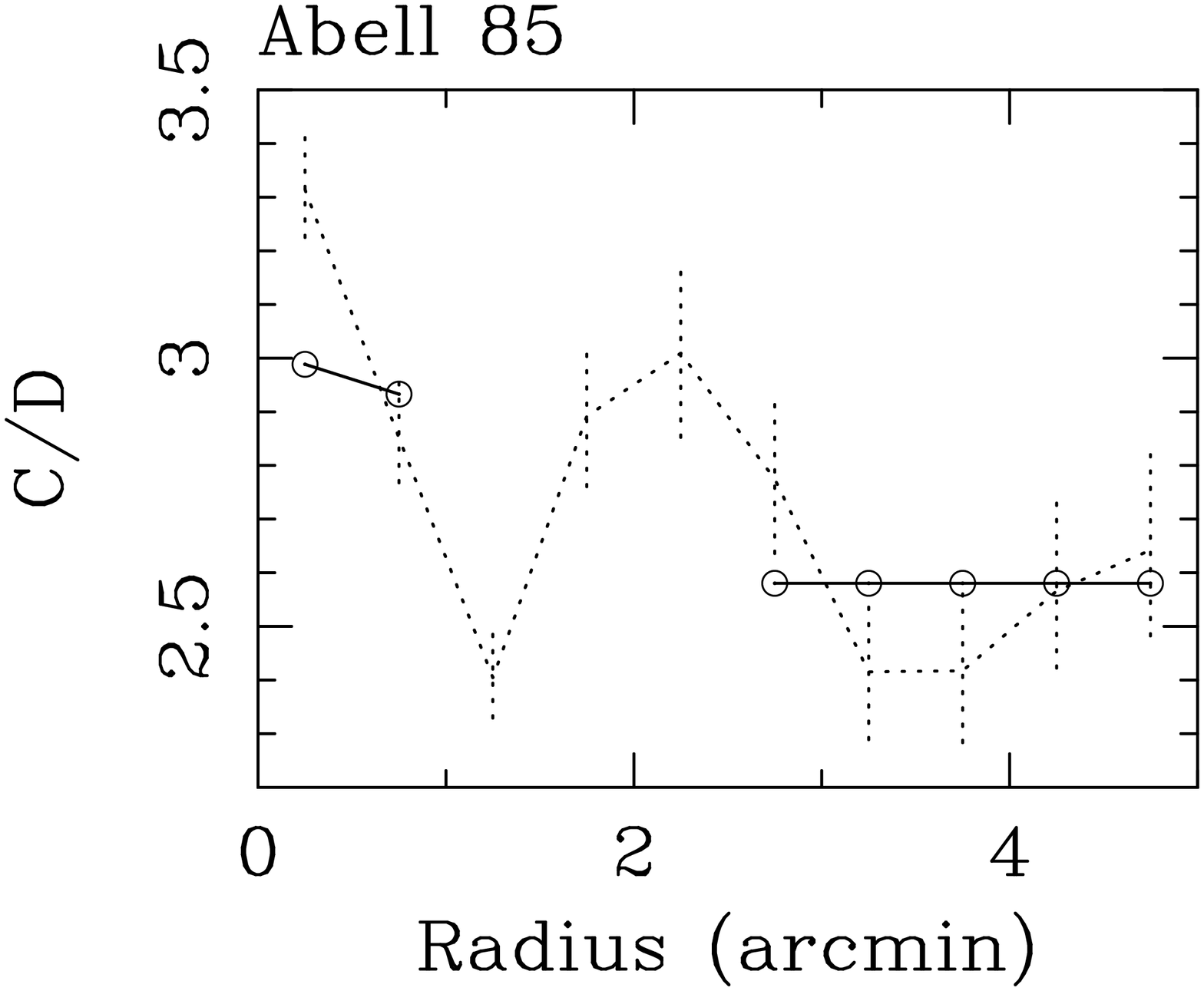,width=0.47\textwidth,angle=270}
\hspace{-0.5cm}\psfig{figure=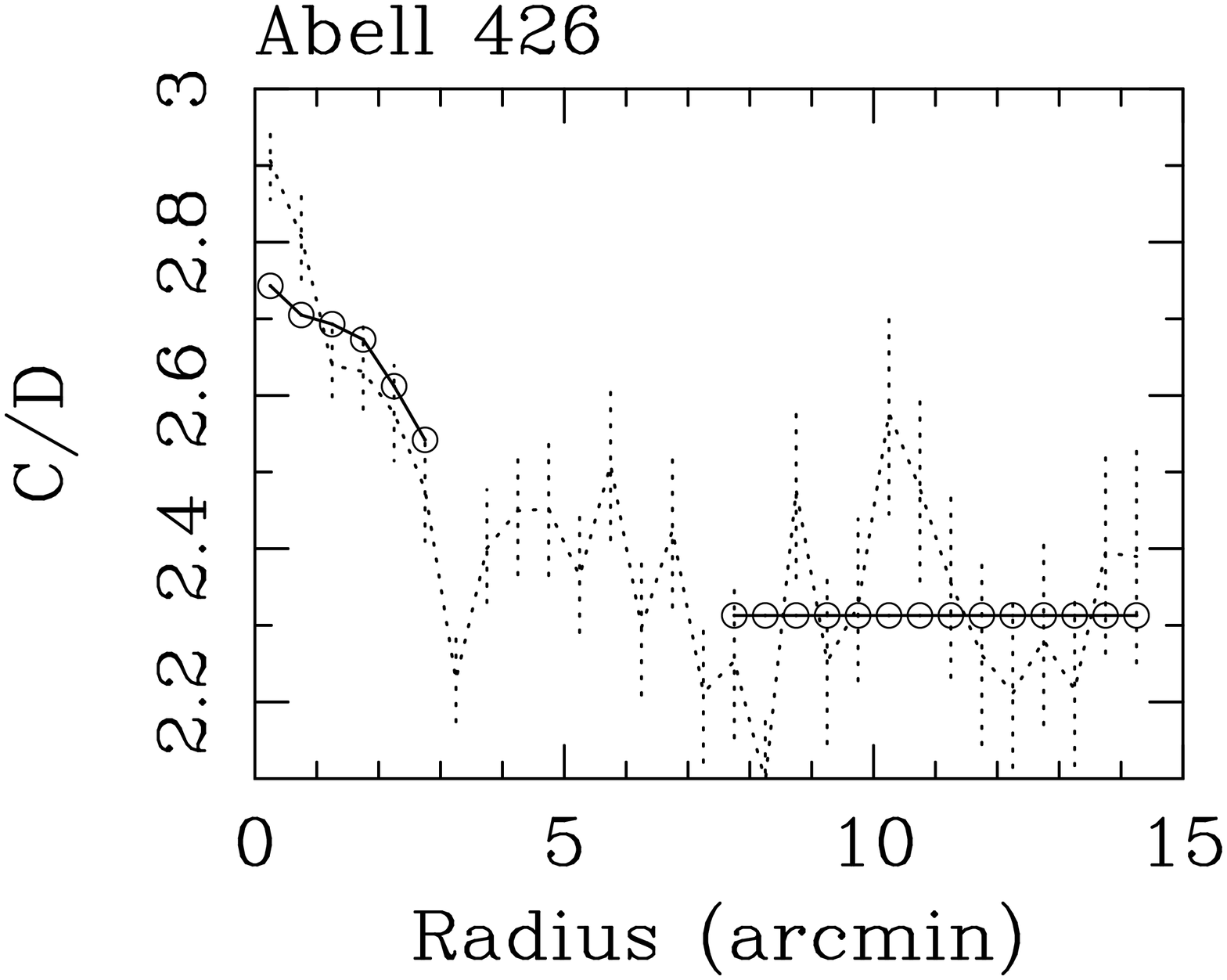,width=0.47\textwidth,angle=270}
}

\vspace{-0.2cm}

\hbox{
\hspace{1cm}\psfig{figure=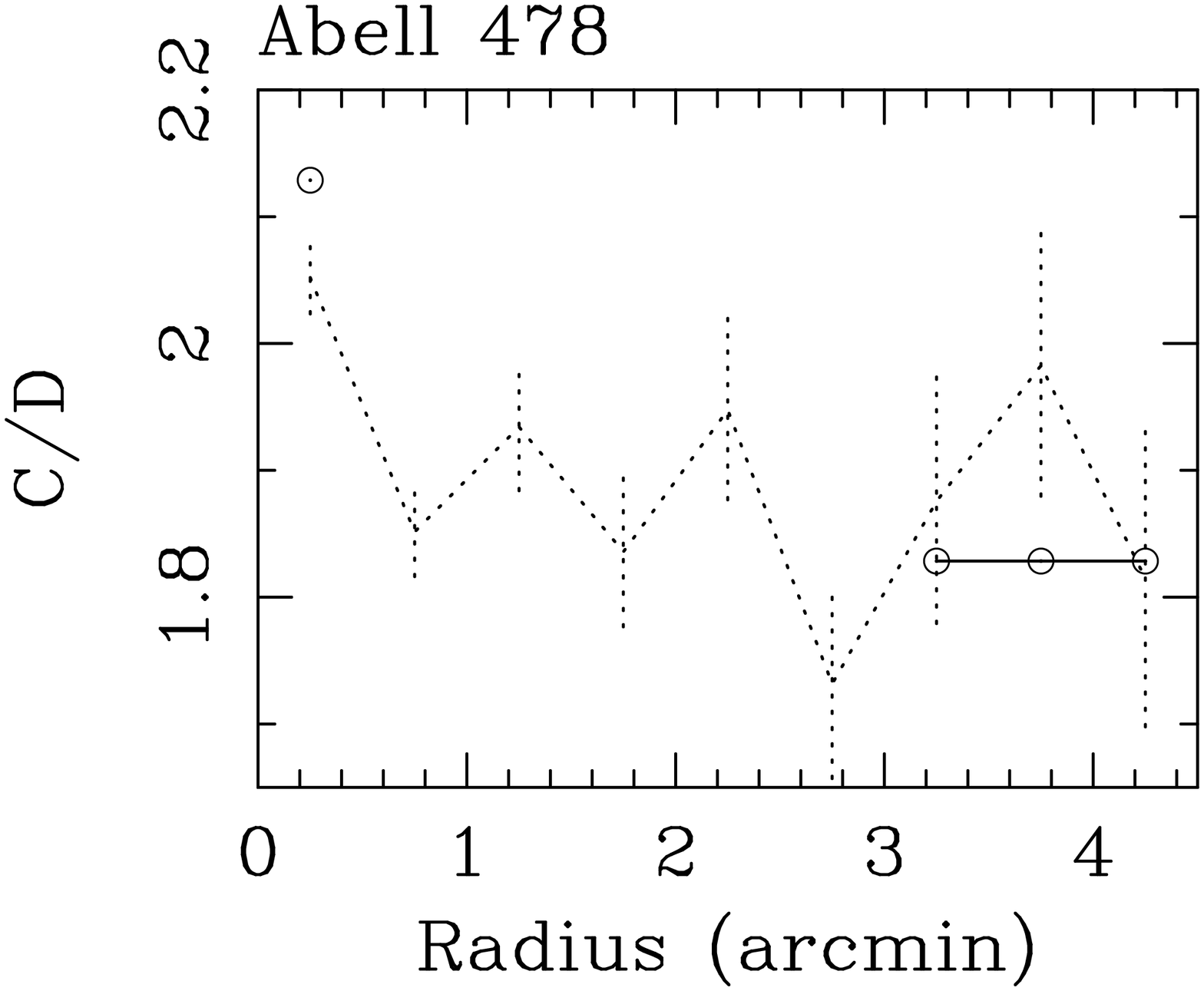,width=0.47\textwidth,angle=270}
\hspace{-0.5cm}\psfig{figure=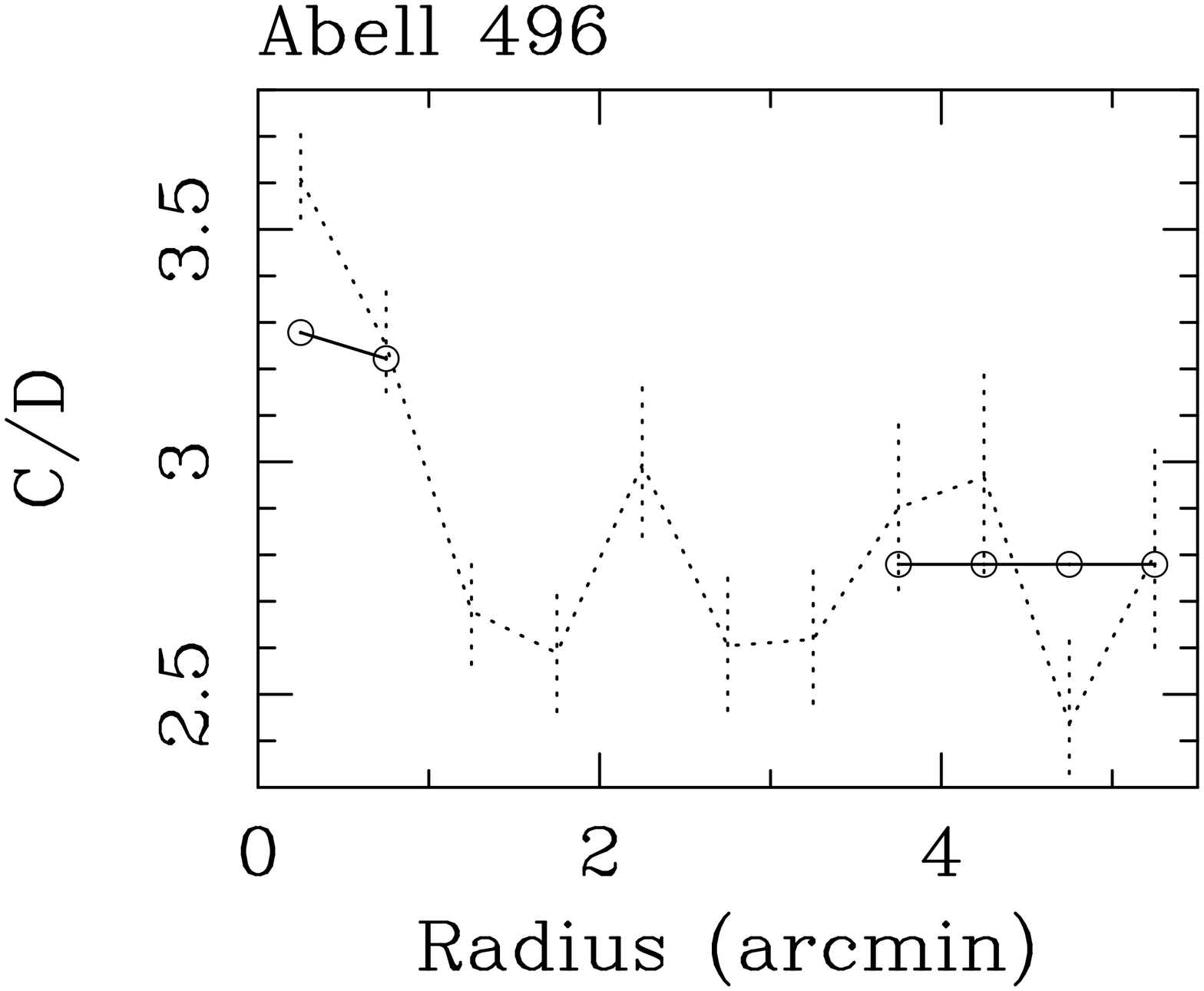,width=0.47\textwidth,angle=270}
}

\vspace{-0.2cm}

\hbox{
\hspace{1cm}\psfig{figure=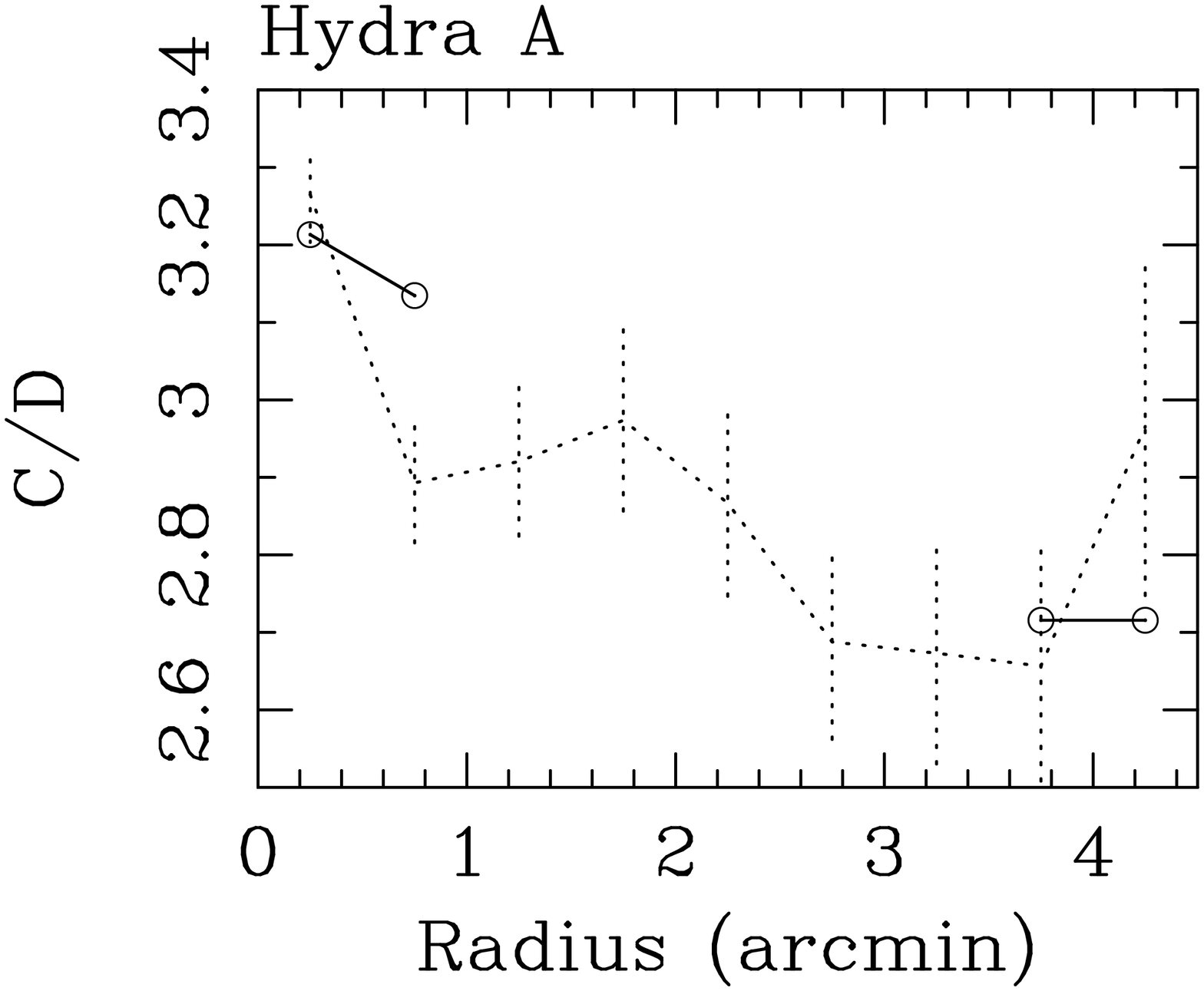,width=0.47\textwidth,angle=270}
\hspace{-0.5cm}\psfig{figure=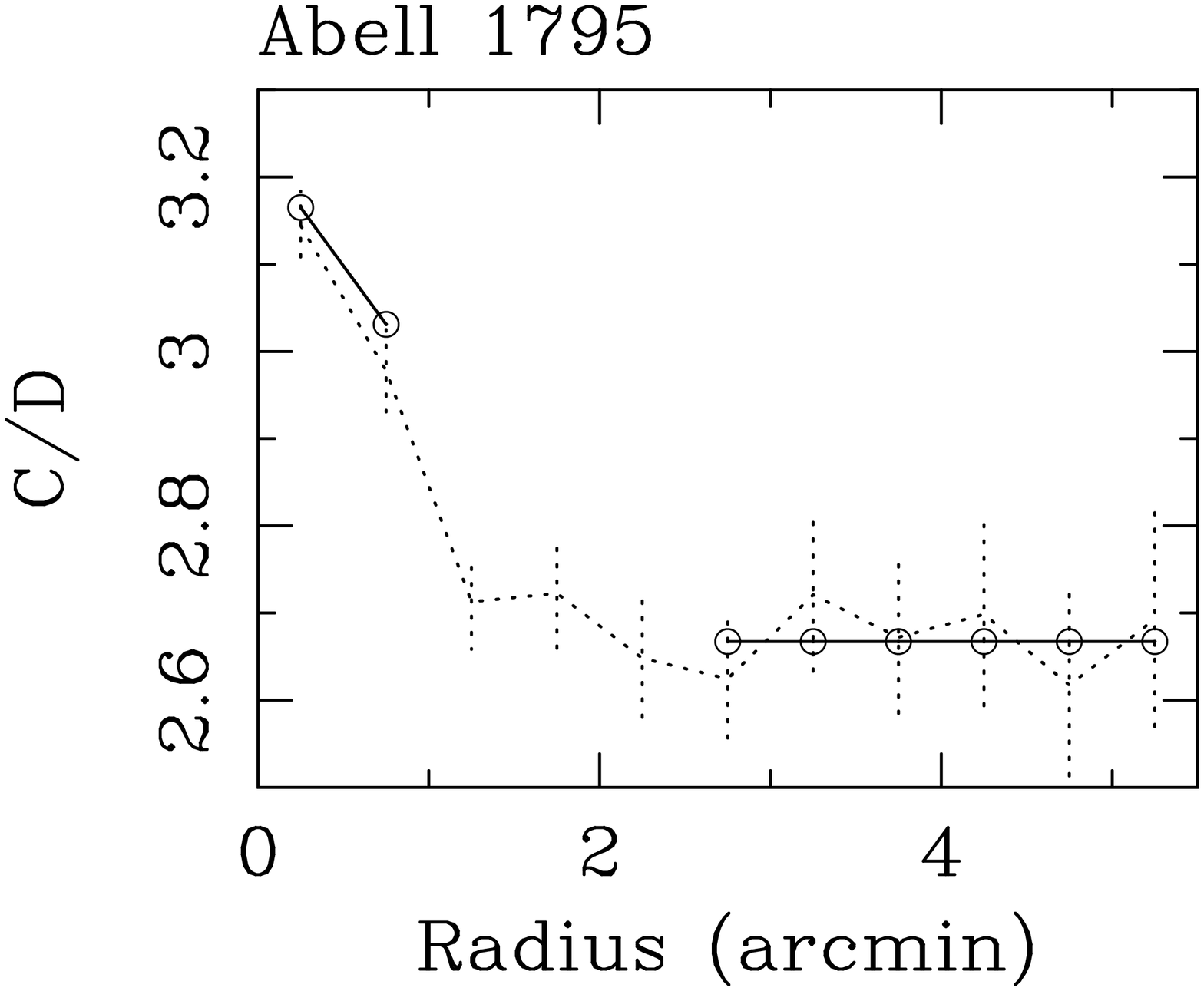,width=0.47\textwidth,angle=270}
}

\caption{A comparison of the observed C/D profiles (as shown in Fig. 2) 
with the ratios predicted from
the deprojection code (Section 5.2).}
\end{figure*}

\addtocounter{figure}{-1}
\begin{figure*}
\hbox{
\hspace{1cm}\psfig{figure=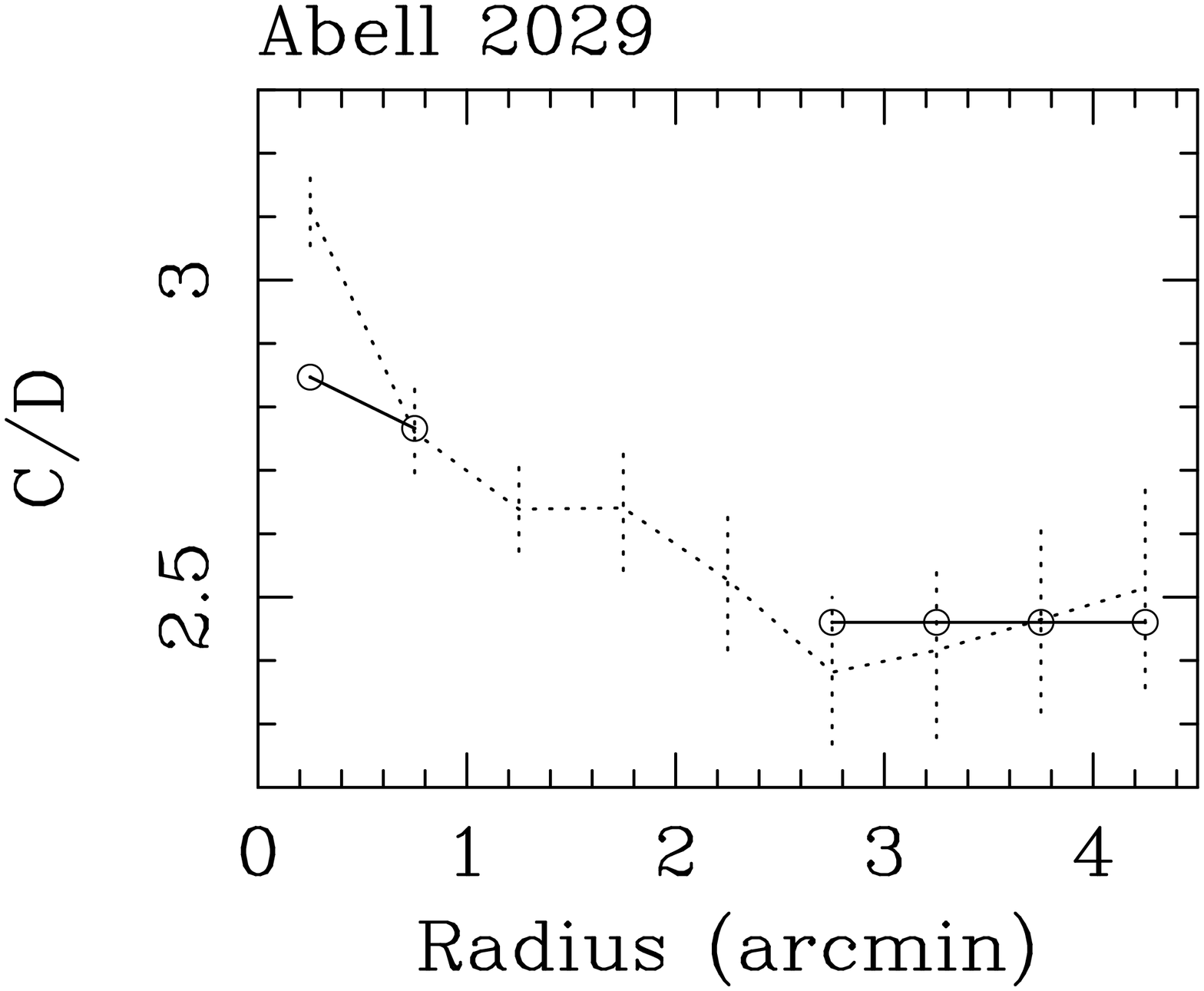,width=0.47\textwidth,angle=270}
\hspace{-0.5cm}\psfig{figure=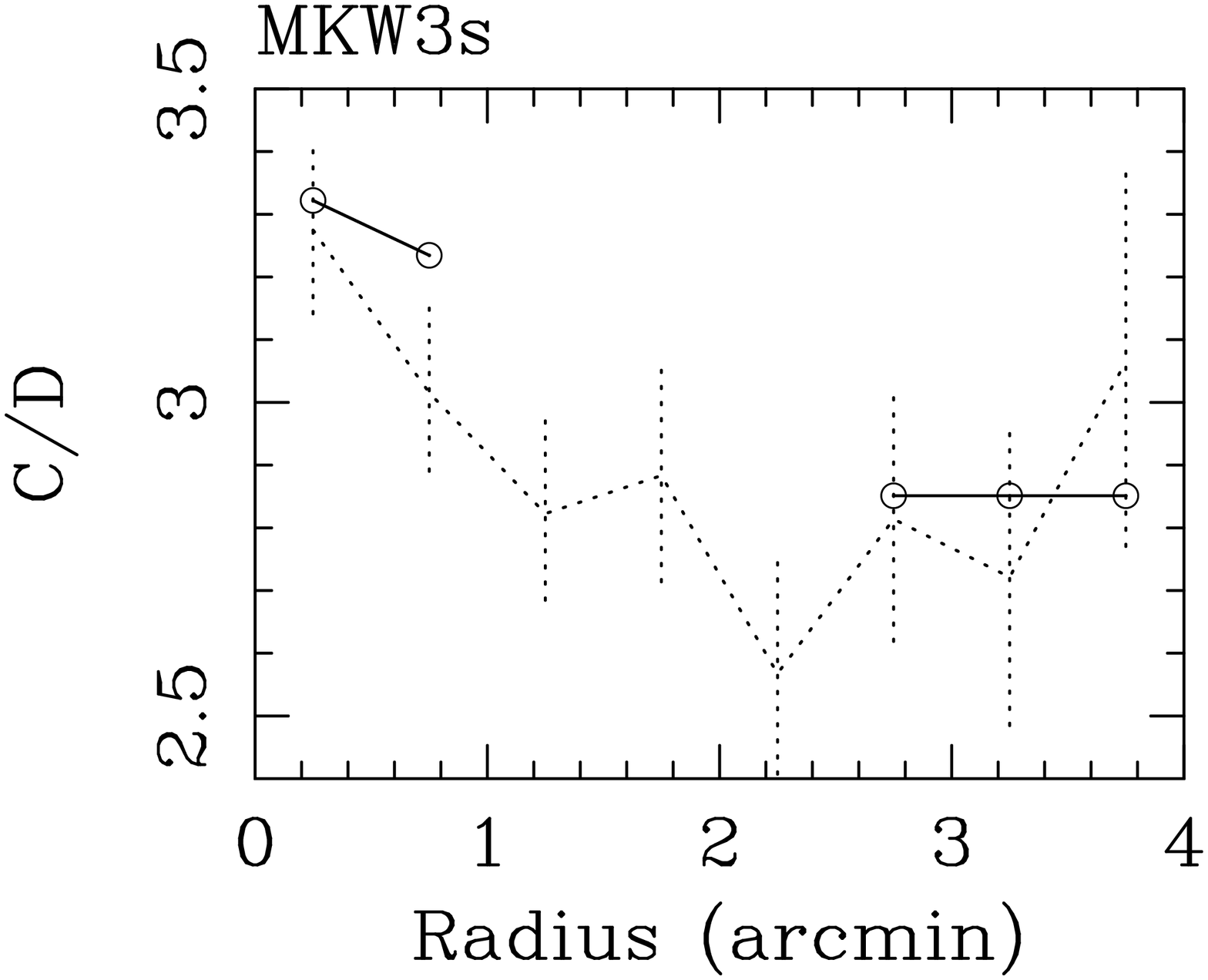,width=0.47\textwidth,angle=270}
}

\vspace{-0.2cm}

\hbox{
\hspace{1cm}\psfig{figure=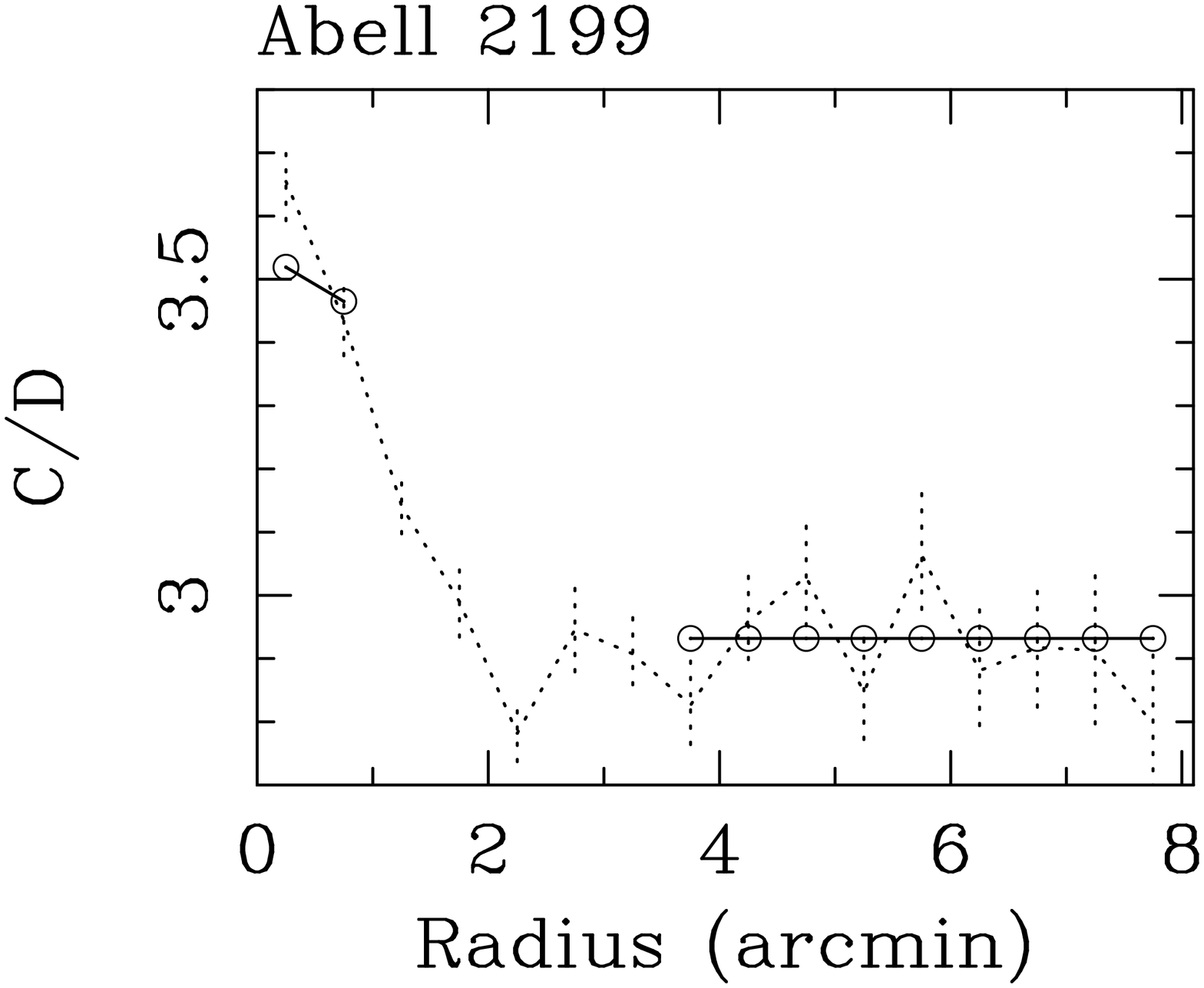,width=0.47\textwidth,angle=270}
\hspace{-0.5cm}\psfig{figure=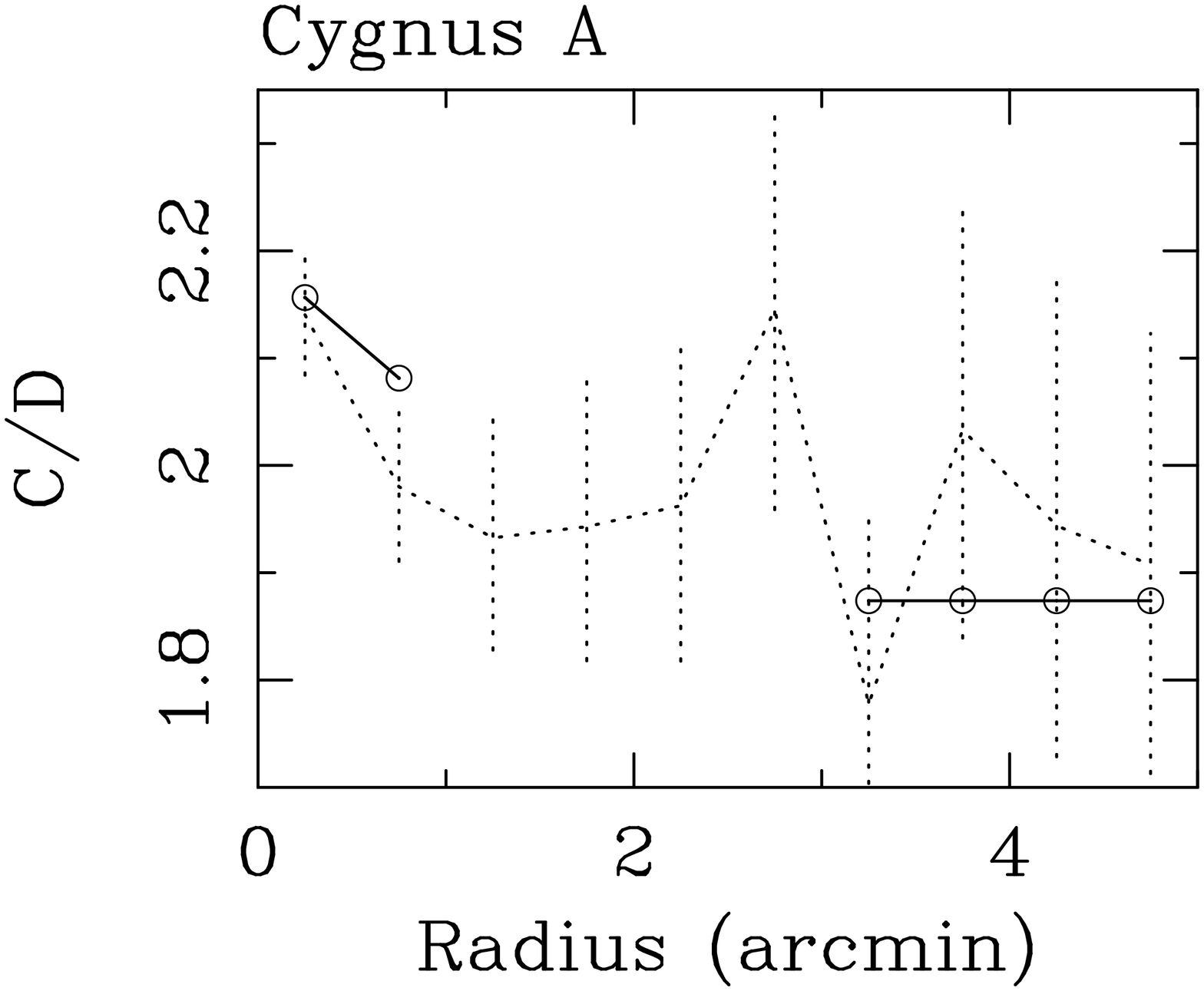,width=0.47\textwidth,angle=270}
}

\vspace{-0.2cm}

\hbox{
\hspace{1cm}\psfig{figure=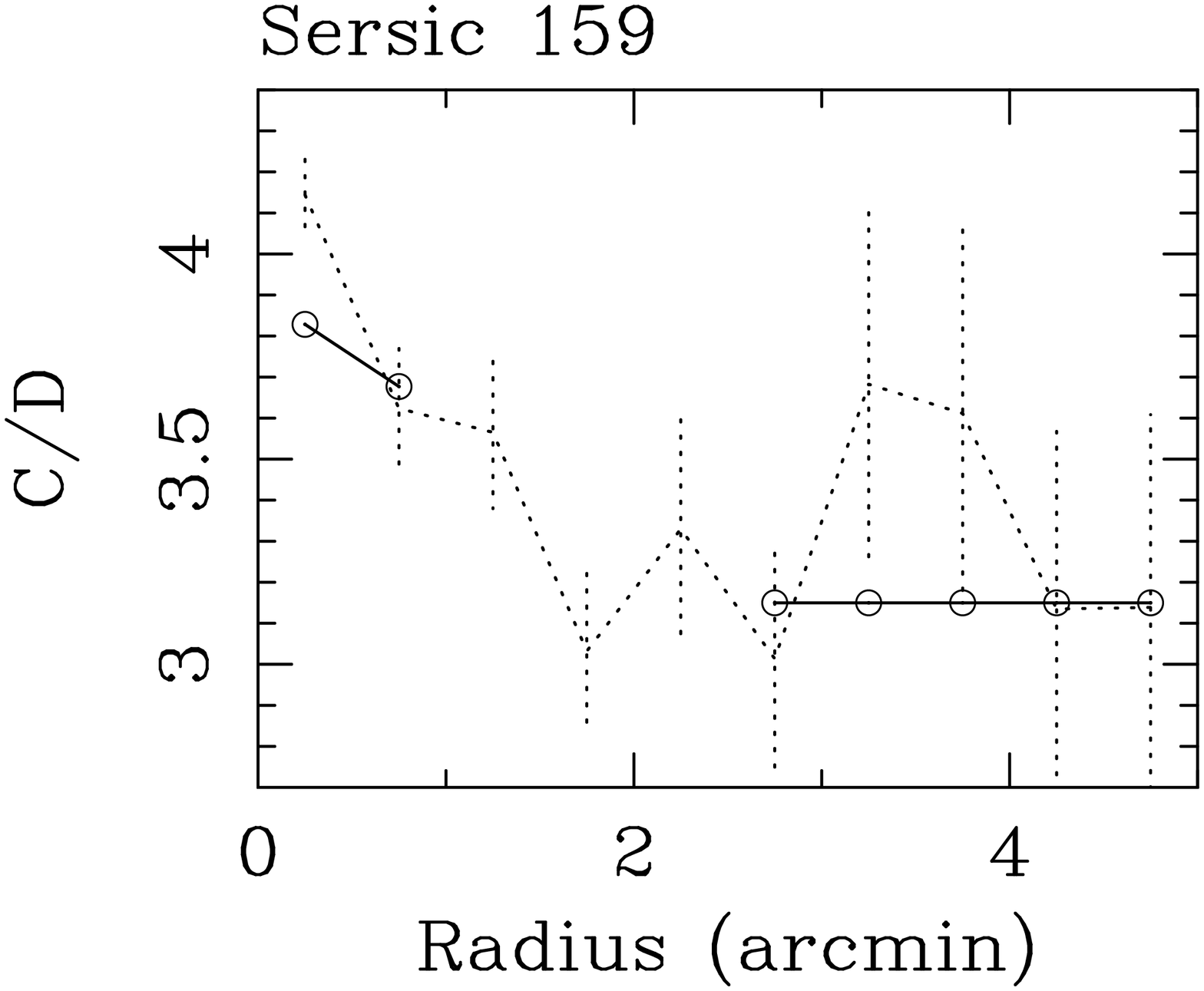,width=0.47\textwidth,angle=270}
\hspace{-0.5cm}\psfig{figure=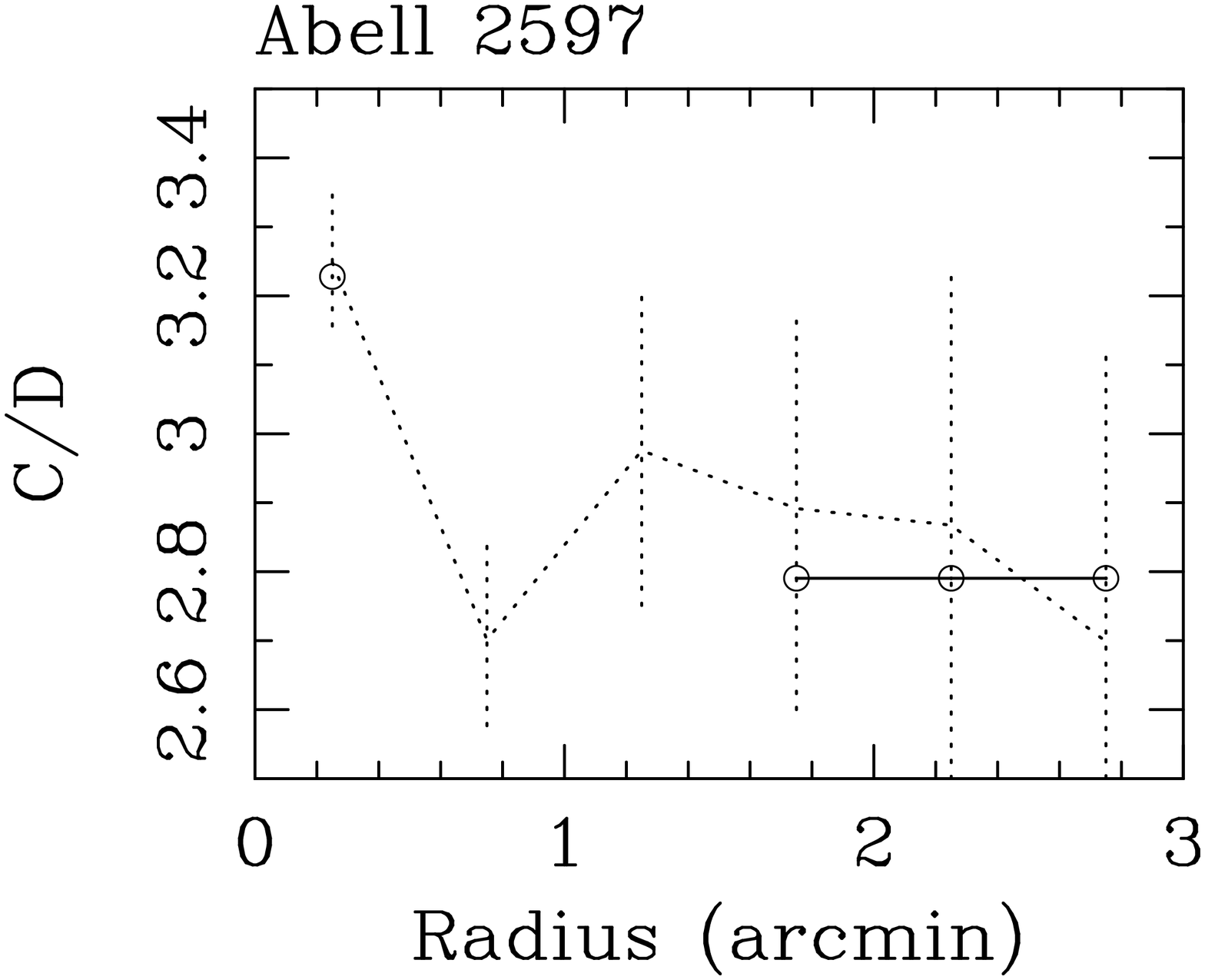,width=0.47\textwidth,angle=270}
}

\caption{ - continued}
\end{figure*}

The C/D profiles shown in Fig. 2 clearly demonstrate the presence of
distributed cool gas in the centres of the cooling flow clusters. The
regions over which the cool gas is distributed are well matched to the cooling
radii inferred from the deprojection analysis (Section 3). Given this
agreement, we have attempted a more sophisticated comparison
of the C/D profiles with the predictions from the deprojection code. 

The deprojection analysis employed in Section 3 divides the X-ray emission from
the cluster into a series of concentric spherical shells. The luminosity
in shell $j$ may be written as the sum of four components.

\begin{equation}
L_j = \Delta{\dot M}_jH_j + \Delta{\dot M}_j \Delta \Phi_j + 
\left[\sum_{i=1}^{j-1} {\Delta{\dot M}_i (\Delta \Phi_j + \Delta H_j
})\right],
\end{equation}

\noindent where $\Delta{\dot M}_j$ is the mass deposited in shell $j$, $H_j$ is the
enthalpy of the gas in shell $j$, and $\Delta \Phi_j$ is the gravitational 
energy released in crossing the shell. $\sum_{i=1}^{j-1}
{\Delta{\dot M}_i }$ is the mass flow rate through shell $j$, and $\Delta
H_j$ the change of enthalpy of the gas as it moves through
the shell.

Thus the first term in equation 1 accounts for the enthalpy of the gas 
deposited in shell $j$. The second term is the gravitational work done on 
the gas deposited in shell $j$. The third and fourth terms account for 
the gravitational work done on gas flowing through shell $j$ to interior radii, and the enthalpy
released by that gas as it passes through the shell. Since we
have constrained the mass-weighted temperatures in the cooling flows
to remain approximately constant with radius (at the values listed in Table
3) the luminosity in the fourth term is negligible.

The luminosity contributed by the second and third terms will have a 
spectrum that can be approximated by an isothermal plasma at the appropriate 
temperature for the cluster ({\it cf.} Fig. 4). The emission accounted for in the 
first term, however,
should have the spectrum appropriate for gas cooling from the ambient
cluster temperature ({\it cf.} Fig. 5). By assigning the appropriate C/D values
from the theoretical curves, corrected for redshift effects, 
to each of the terms in equation 1, the deprojection
code can be used to predict intrinsic and projected C/D colour
profiles for the clusters. These predictions can then be compared to the
observed values.

In Fig. 6 we show the comparison of the observed and predicted C/D
profiles for the cooling flows in the sample. 
Due to the finite ages of the clusters, we limit the comparison
to those radii where the cooling time of the
cluster gas is $\leq 5 \times 10^9$ yr. In the outer regions of the
clusters the emission is assumed to be isothermal, at the 
temperature given in Table 3, and absorbed by the appropriate
Galactic column density. The predicted curves are then re-normalized 
to match the observed C/D values in the outer
regions of the clusters. (Note that the normalization factors, 
which are given by the ratios of the observed to
predicted C/D ratios in the outer regions of the clusters, 
are typically close to unity.) 
We do not attempt any comparison for either the Centaurus
or Virgo clusters since both systems are relatively cool
and are known to contain strong metallicity gradients (Allen \& Fabian 1994;
Fukazawa \etal 1994). The low redshifts of these 
systems also means that some drop in their mass-weighted temperatures, 
as the gravitational potentials of the
clusters merge with those of the central galaxies, are likely in the
innermost 1--2 bins. Indeed, such a temperature drop is required to
explain the large C/D ratio observed in the central bin of the Centaurus data. 
We therefore defer detailed study of the data for the Virgo and Centaurus
clusters to future work.  (The relatively poor quality of the
data for Abell 3112 and Abell 4059 also prohibit any
detailed comparison with the deprojection code.) 
We note that hotter clusters do not generally 
exhibit metallicity gradients (\eg Ohashi \etal 1996) and the effects of
metallicity gradients on the X-ray colour profiles of such systems will,
in any case, be small ({\it cf.} Figs. 4,5).

The agreement between the observed and predicted C/D ratios is remarkably
good. For the best data sets (Abell 426, Abell 1795 and Abell
2199; see also Fig. 13) the agreement is especially good. 
In all other cases the
predicted C/D values either agree with, or slightly under-predict, the
observed values. Under-prediction in the central bin  may again indicate 
a small drop in the ambient cluster temperature.

\subsection{The B/D ratios and intrinsic absorption}

\begin{table*}
\vskip 0.2truein
\begin{center}
\caption{Absorption results from the B/D ratio}
\vskip 0.2truein
\begin{tabular}{ c c c c c c }
 Cluster   & ~ & $N_{\rm H}$ & $\Delta N_{\rm H}'$ & $\Delta N_{\rm H}(1)$ & $\Delta N_{\rm H} (2)$  \\
\hline                                                                                               
&&&&& \\                                                                                               
Abell 85   & ~ & 2.96  & $2.34 \pm 0.60$  & $5.4^{+1.7}_{-1.6}$   & $10.5 \pm 2.7$  \\  
Abell 426$^*$  & ~ & 13.7  & $0.50 \pm 0.55$  & $1.0^{+1.2}_{-1.0}$   & $2.3 \pm 2.5$  \\
Abell 478  & ~ & 20.0  & $7.4 \pm 0.63$   & $20.1^{+2.7}_{-2.4}$  & $33.3 \pm 2.8$  \\
Abell 496  & ~ & 4.27  & $2.57 \pm 0.62$  & $6.0^{+1.9}_{-1.6}$   & $11.6 \pm 2.8$   \\
Hydra A    & ~ & 4.84  & $3.07 \pm 0.41$  & $7.5^{+1.4}_{-1.2}$   & $13.8 \pm 1.8$   \\
Abell 1795 & ~ & 1.11  & $2.20 \pm 0.20$  & $5.0^{+0.6}_{-0.5}$   & $9.9 \pm 0.9$     \\
Abell 2029 & ~ & 3.05  & $0.83 \pm 0.42$  & $1.7^{+1.0}_{-0.9}$   & $3.7 \pm 1.9$    \\
MKW3s      & ~ & 2.86  & $7.29 \pm 0.88$  & $11.7^{+1.9}_{-1.7}$  & $32.8 \pm 4.0$    \\
Abell 2199 & ~ & 0.87  & $2.67 \pm 0.25$  & $6.3^{+0.8}_{-0.7}$   & $12.0 \pm 1.1$   \\
Cygnus A   & ~ & 32.8  & $5.5 \pm 1.0$    & $13.4^{+3.3}_{-2.9}$  & $24.8 \pm 4.5$  \\
Sersic 159 & ~ & 1.80  & $1.87 \pm 0.48$  & $4.2^{+1.2}_{-1.2}$   & $8.4 \pm 2.2$     \\
Abell 2597 & ~ & 2.46  & $2.36 \pm 0.58$  & $5.5^{+1.6}_{-1.6}$   & $10.6 \pm 2.6$  \\
&&&&& \\                                                                   
\end{tabular}                                                              
\end{center}                                                               
                                                                           
\parbox {7in}                                                              
{The intrinsic column densities and            
1$\sigma$ errors (in units of $10^{20}$ \apc) for the central 30 arcsec    
regions of the clusters, inferred from the B/D analysis (Section 5.3).  
Column 2 lists the Galactic column densities. 
Column 3 ($\Delta N_{\rm H}'$) summarizes the excess column densities
determined with the simple uniform absorbing screen model. 
Column 4 [$\Delta N_{\rm H} (1)$] lists the corrected intrinsic column  
densities, determined from equation 4, with the partial covering model
(the preferred model). 
A covering fraction, $f = 0.5$, and 
a photon energy, $E = 0.4$  keV,
have been assumed for all clusters except Abell 478 and Cygnus A (which have 
much higher Galactic column densities) for which $E = 0.5$  keV is adopted.
For MKW3s the observed $\Delta N_{\rm H}'$ is inconsistent with $f=0.5$ and 
$f=0.75$ is used instead.
Column 5 [$\Delta N_{\rm H}(2)$] lists the corrected intrinsic column
densities determined with equation 5, for the multilayer
absorption model. An intermediate correction factor of 4.5 has been assumed. 
$^*$Between radii of 1-2 arcmin in Abell 426, the corrected intrinsic 
column density is $\Delta N_{\rm H} (1) \sim 5-10 \times 
10^{20}$ \apc.
}
\end{table*}

As with the C/D results, the B/D colour profiles shown in Fig. 3 are
essentially flat at large radii ($r > r_{\rm cool}$). The 
observed and predicted values at large radii are generally in good 
agreement (although with a few exceptions detailed below) indicating that the
emission from these regions can be
well-described by a simple isothermal model, with the temperatures and
Galactic column densities listed in Table 3. 
At small radii, many of the B/D profiles increase. 
However, the gradients are not as steep as for the C/D
results, and for a number of the clusters the B/D profiles are essentially 
flat or even decrease slightly in the innermost bins (\eg Abell 478). 

The most notable exception to the agreement between the observed and
predicted colour ratios at large radii is Abell 478, 
for which the B/D (and C/D) ratio implies 
a Galactic column density of $\sim 3 \times 10^{21}$ \apc, 
an excess of $\sim 10^{21}$ \apc~over the value listed in 
Table 3. In contrast, the data for Abell 426 require a
column density of $\sim 1.1 \times 10^{21}$ \apc~,
slightly less than the Galactic value.
For the Coma cluster, the B/D and C/D ratios imply an
emission-weighted temperature at large radii of $\sim 6$ keV, 
somewhat less than the 8 keV listed in Table 3. 
For Abell 2199 the data suggest an
emission weighted temperature of $\sim 3.5$ keV and a column
density of $\sim 5 \times 10^{19}$ \apc. 
Similarly, the data for the Virgo cluster imply a 
temperature of $\sim 2$ keV and a column density of $\sim 1.3 \times
10^{20}$ \apc, both slightly less than the nominal values.

Following our success in applying the deprojection code to the prediction
of the C/D profiles (Section 5.2),  we  have used the same
technique to predict the B/D profiles. The predicted data are shown as
the circled points in Fig. 3. The results demonstrate a clear 
tendency for the predicted values to overestimate the observations in the
central regions. 

The theoretical curves presented in Figs. 4 and 5 show that the
most plausible interpretation for the central deficits in the 
B/D observations, with respect to the predicted data, is the presence of 
intrinsic absorbing material in the clusters. The excess column densities 
required to bring the observed and
predicted B/D values for the central bins into agreement are summarized in
Table 6.  These values have been calculated using XSPEC models for the 
projected spectrum of the central 30 arcsec (radius) in each cluster, as 
implied by the deprojection analysis.
The excess absorption has then been modelled as a uniform screen in front of
the cluster. Note that since the observed C/D ratios for the central bins
sometimes (slightly) exceed the predicted values, the column
densities listed in Table 6 may (slightly) underestimate the true values. 

Where comparisons between the observed and predicted B/D ratios are
possible for more than just the central bin, the data suggest that the
column density increases with decreasing radius within the cooling flow. 
This is consistent with a picture wherein the intrinsic absorber is
accumulated from material deposited by the cooling flow. The only
exception is Abell 426 (the Perseus Cluster) for which the apparent
excess column density at radii of 1--2 arcmin is $5-10$ times
larger than for the central bin. It is likely that the results for
the central regions of Abell 426 are affected by the complex morphology 
of the cluster (B\"ohringer \etal 1993) and emission from the active 
nucleus of the central galaxy. Note also that although the 
presence of temperature and metallicity gradients 
complicate the interpretation of the B/D data for the Virgo and Centaurus 
clusters, the comparison of their B/D ratios, which are only 
observed to rise within the central arcminute or so, with their C/D ratios, 
which in contrast rise steadily within radii of 7 and 3 arcmin respectively, 
again indicates the presence of distributed absorbing material.

Our results confirm that excess absorption is common in cooling flows.
All of the cooling flow clusters in our sample, for which an X-ray colour
profile/deprojection comparison is possible, show evidence for intrinsic
absorption.  Note also that the cooling flow with the 
lowest intrinsic column density, Abell 2029, is also the only
cluster with a Central Cluster Galaxy (CCG) that does not exhibit optical
emission lines (\eg Hu, Cowie \& Wang 1985; Johnstone, Fabian \& Nulsen
1987). This suggests a link between the presence of
X-ray absorbing gas in cooling flows and the production of optical
emission lines in CCGs. Such a result is compatible with models in which
the emission lines are powered by massive stars that collapse out of the
cooled gas accumulated by the cooling flows (Allen 1995). Processes such
as turbulent mixing layers and shocks in the X-ray absorbing medium may
also contribute significantly to the observed emission-line flux
(Crawford \& Fabian 1992). 

Finally, we note that although the analysis of the B/D profiles 
apparently suggests intrinsic column densities 
of only a few $10^{20}$ \apc~through the centres of cooling flows, 
when modelling the absorber as a uniform screen in front of the
clusters, the true intrinsic column
densities are likely to be $2-5$ times higher, once systematic effects
associated due to the spatial distribution of the absorbing gas are accounted for
(see Section 6).

\subsection{The absorption signature in the `A' band}

The B/D ratio provides the most reliable probe of the intrinsic column
density in the clusters. As noted in Section 1, at energies below 0.40 keV 
(the lower limit of the B band) the point spread function of the PSPC 
broadens significantly. In the $A_1$ ($0.1-0.2$ keV) band a significant
fraction of the flux is scattered (somewhat asymmetrically) to radii $>
1$ arcmin. In the $A_2$ band, the PSF is better behaved, but the
FWHM is still $\sim 1.7$ times broader than in
bands B--D. [Note that the B/D ratio also has the advantage of being 
applicable to clusters with
a wide range of Galactic column densities, whereas the A band data
are lost for clusters behind Galactic column densities
$\approxgt 10^{21}$ \apc.]

We have carried out a 
search for the signature of intrinsic absorption in the $A_2$ data, by 
forming profiles in the ratio $A_2$/D$'$,  where D$'$ is the D band
profile convolved with a Gaussian filter of 30 arcsec FWHM. 
For the two best data sets for which this test is applicable, Abell 1795 and 
Abell 2199, we determine excess column densities in the central bin 
of $\sim 1-2 \times 10^{19}$ \apc. These column densities are significantly 
less than those inferred from the B/D analysis. However such results are to 
be expected if the absorber is spatially distributed throughout
the cooling flows, as discussed in Section 6, below. Note also that the
presence of an absorption signature in the $A_2$ data implies that
absorbers other than Oxygen (which  has its K edge at 0.53 keV in low
temperature gas) must be
present in the clusters.

\subsection{The ages of the cooling flows}

\begin{figure*}
\hbox{
\hspace{0.7cm}\psfig{figure=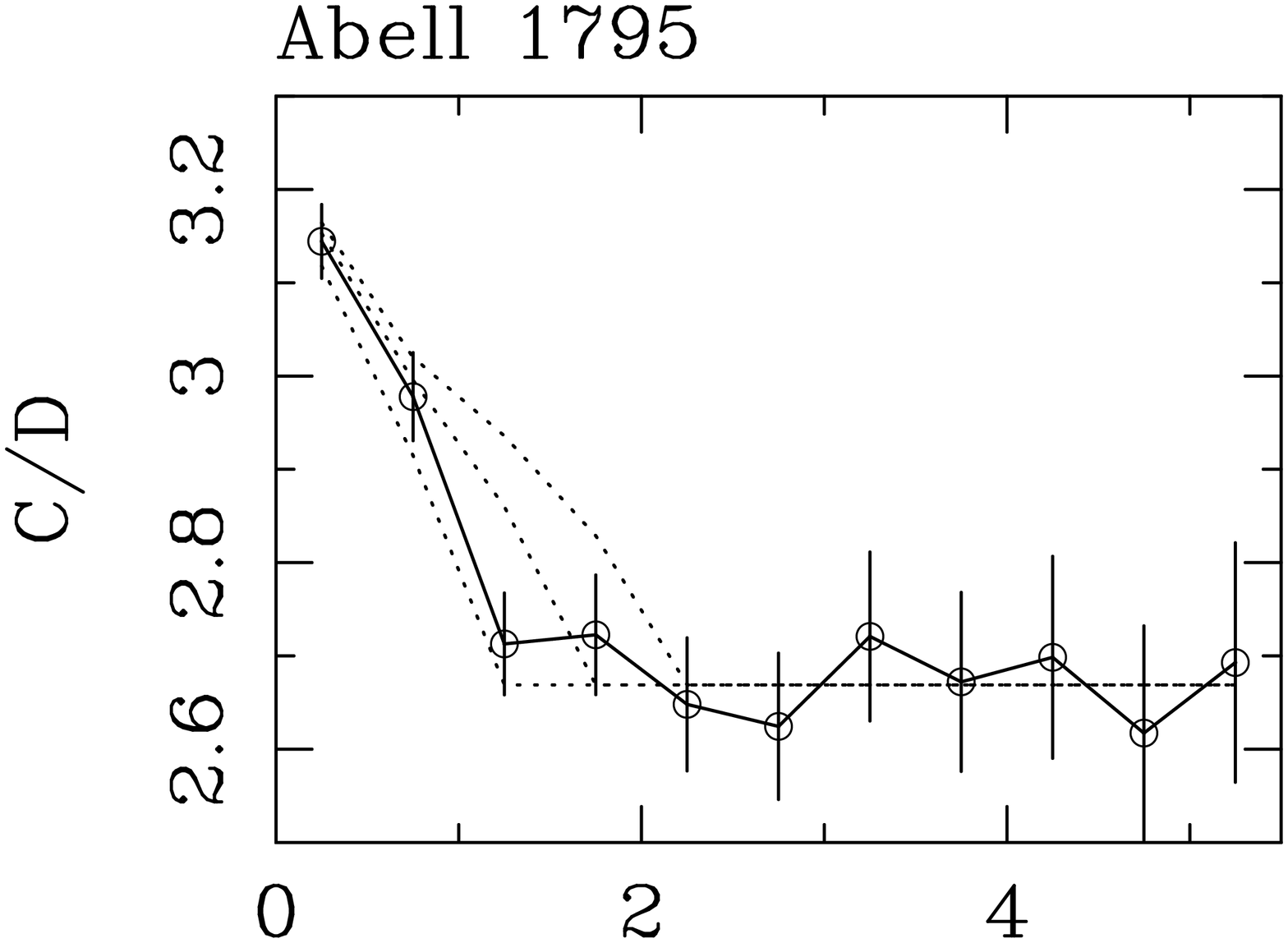,width=0.5\textwidth,angle=270}
\hspace{-0.5cm}\psfig{figure=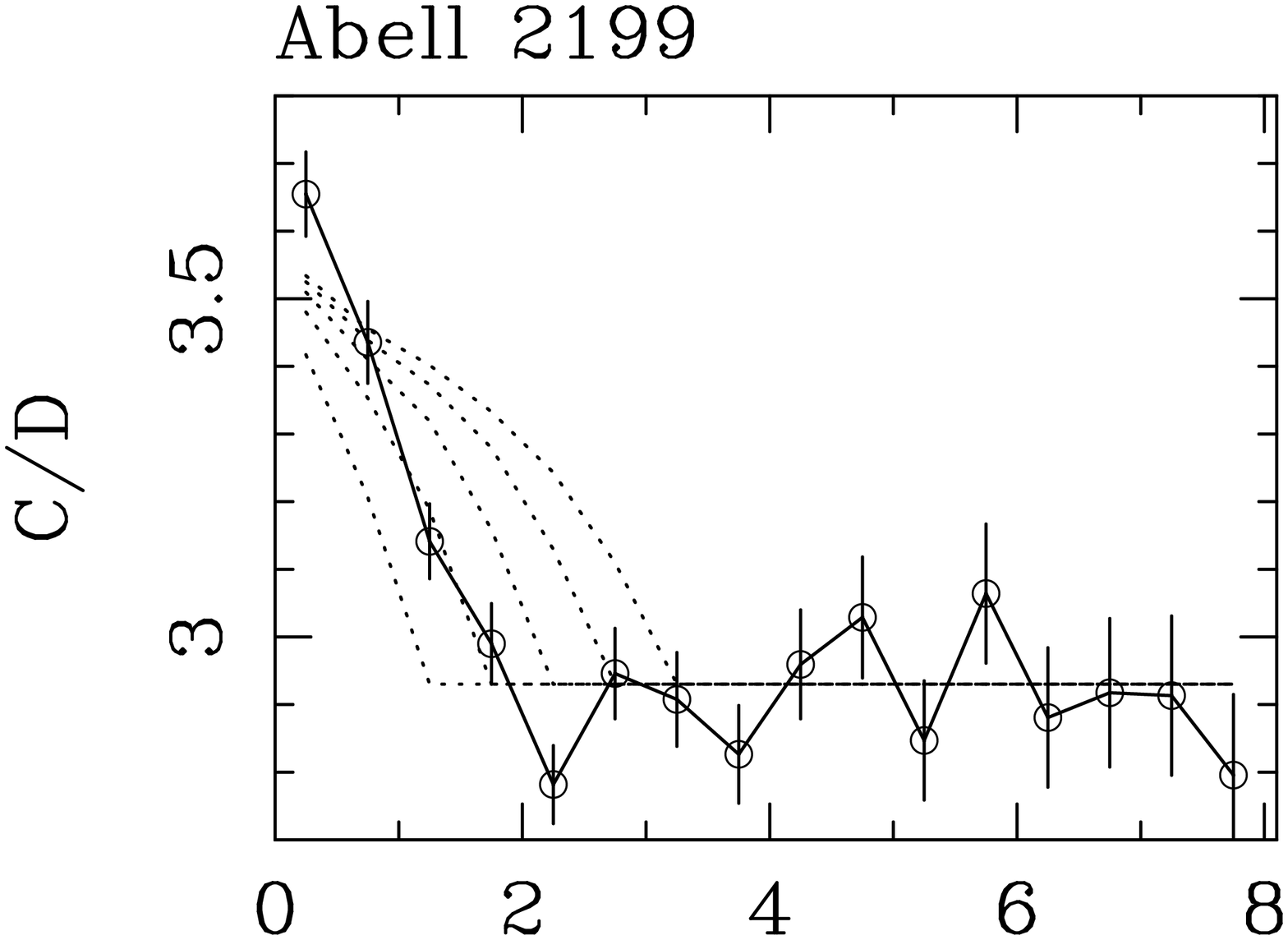,width=0.5\textwidth,angle=270}
}
\vskip -1.7cm
\hbox{
\hspace{0.7cm}\psfig{figure=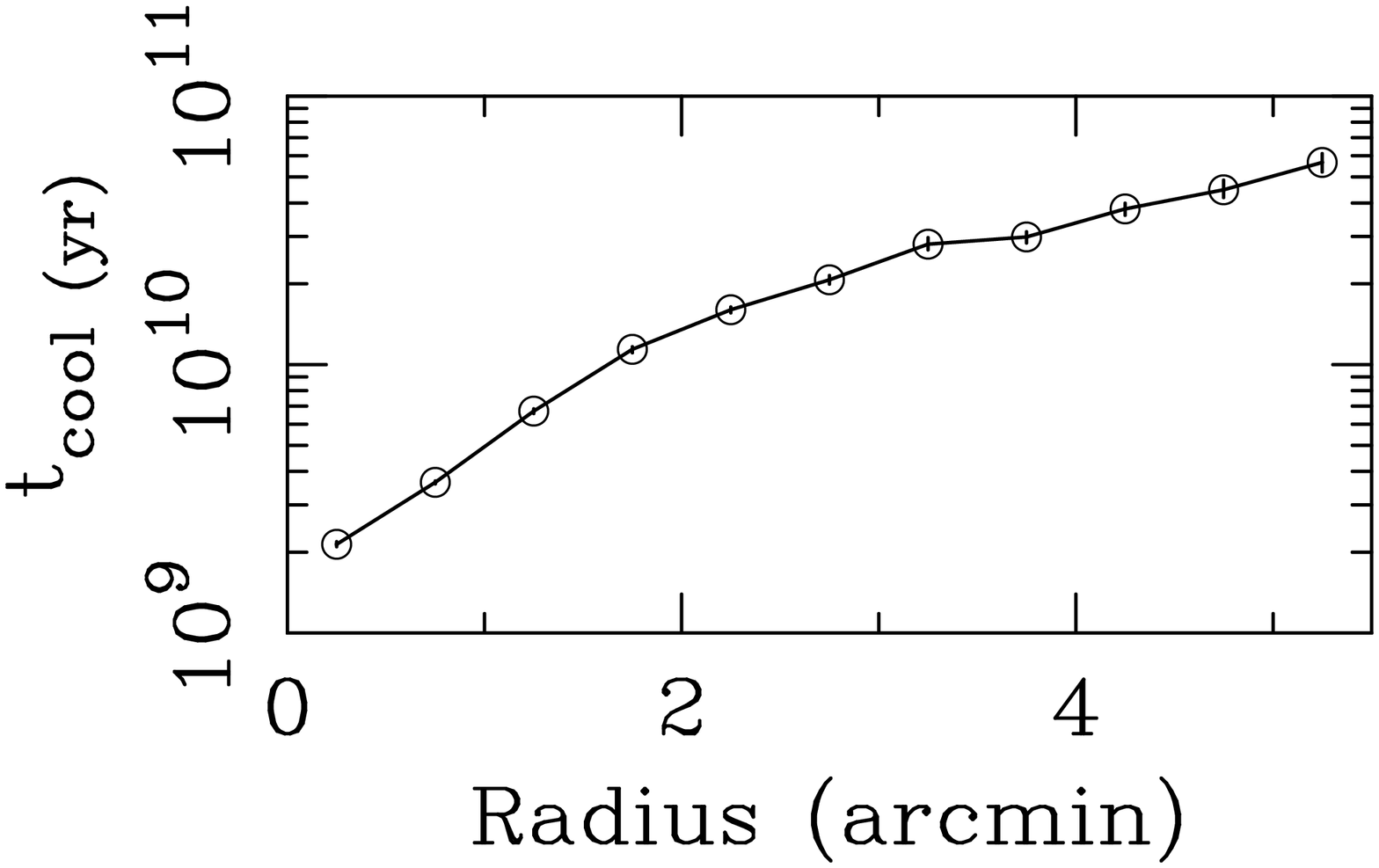,width=0.5\textwidth,angle=270}
\hspace{-0.5cm}\psfig{figure=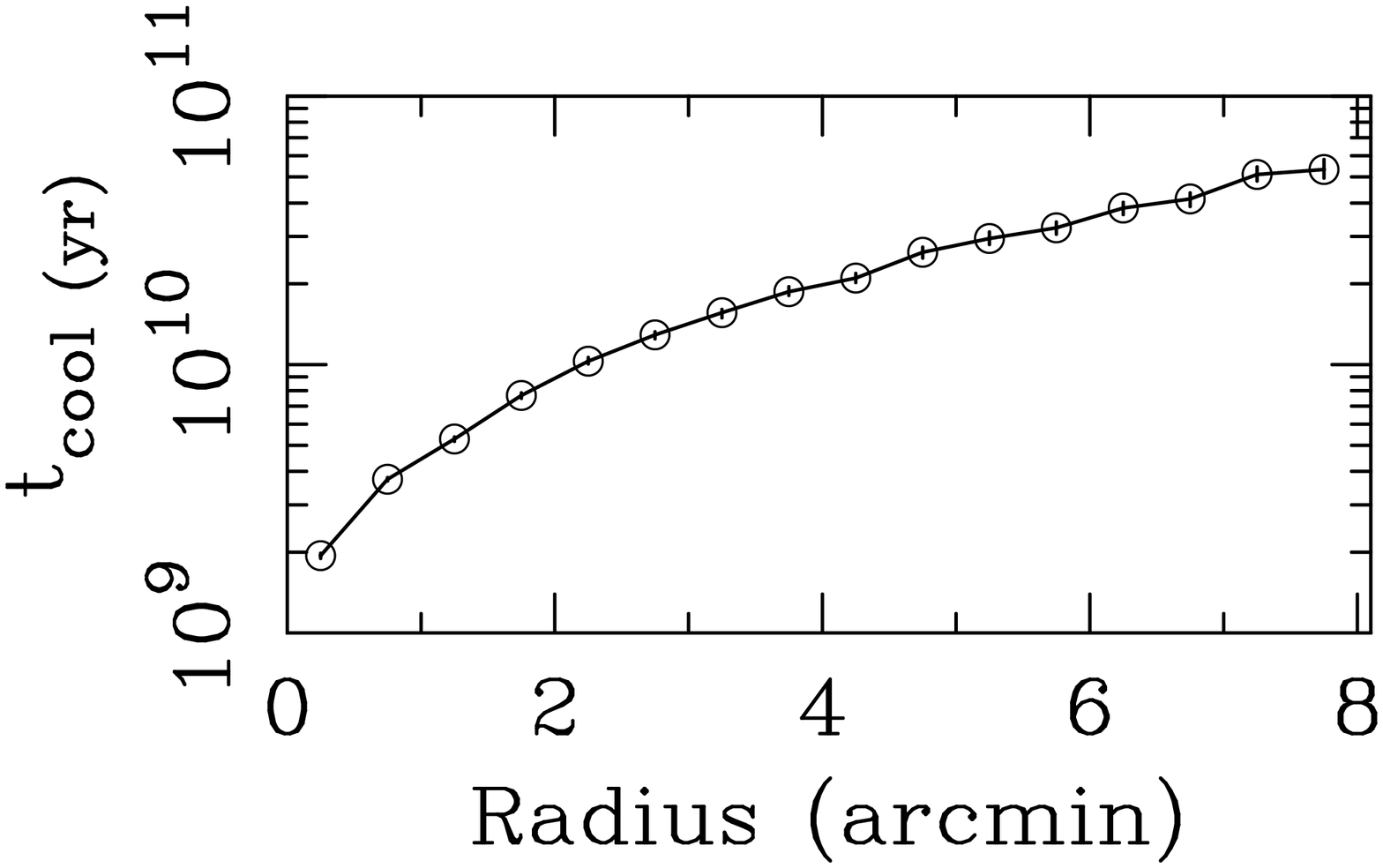,width=0.5\textwidth,angle=270}
}
\vskip -1cm
\caption{(Upper panels) Comparison of the 
observed C/D colour profiles with the predictions (dotted curves) from the 
deprojection code, as a function of the age of the cooling flows. For
Abell 1795, the  predictions are shown for cooling cut-off times of $6.7, 11.4$ and 
$16.0 \times 10^9$ yr, corresponding to ages 
of $3.6-6.7 \times 10^9,
6.7-11.4 \times 10^9$, and $11.4 - 16.0 \times 10^9$ yr, respectively.
For Abell 2199 the cooling cut-off points are $5.3, 7.7, 10.3, 12.9$ and
$15.6 \times 10^9$ yr, implying ages of $3.7-5.3, 5.3-7.7, 7.7-10.3,
10.3-12.9$ and $12.9-15.6 \times 10^9$ yr. 
(Lower panels) The mean cooling time as a function of radius in the
clusters. } 
\end{figure*}

Throughout this Section we have limited our comparison of the 
observed and predicted X-ray colour profiles 
to radii where the mean cooling time of the cluster gas is 
$\leq 5 \times 10^{9}$ yr. However, for the best data sets, 
the data from larger radii allow 
constraints to be placed on the age of the cooling flows 
(where the age is presumably related to the time since the central regions 
of the clusters were last disrupted by a major subcluster merger event). 
For these data sets, the deprojection code has been used to predict C/D 
profiles throughout the clusters, as a function of cooling-flow age. 
At radii where the mean cooling time of the gas is equal to or exceeds 
a specified value, the cluster is assumed to be isothermal with no mass
deposition occurring. At interior radii, where the mean cooling time is less 
than this value, the predictions are calculated in the manner
described in Section 5.2. 

Fig. 7 shows the observed C/D profiles and the predicted results, as
a function of cooling flow age, for the two best data sets in the 
sample, Abell 1795 and Abell 2199. Also plotted are the mean
cooling times as a function of radius in the clusters.
For Abell 1795, the data are
best described by an age of between 3.6 and $6.7 \times 10^9$ yr ($6.7
\times 10^9$ yr being the mean cooling time at the radius where the C/D
ratio rises, and $3.6 \times 10^9$ yr the mean cooling 
time for the next interior bin.)
For Abell 2199, the data suggest an age of $5.3 - 7.7 \times 10^9$ yr. 
The results also imply mass deposition rates in the clusters 
somewhat less than the nominal values quoted in Table 3 (which are
calculated for an assumed age of $1.3 \times 10^{10}$ yr.) 
For Abell 1795 we determine a revised mass
deposition rate of $\sim 250$ \Msunpyr. For Abell 2199 the revised mass
deposition rate is $\sim 80$ \Msunpyr. These results are approximately a 
factor 2 less than the values  of $\sim 450$ and 160 \Msunpyr determined 
under the standard deprojection assumptions, but are in good agreement with
the rates inferred from the analysis of Einstein
SSS spectra for these clusters by White \etal (1991).

In general, the data for the other clusters do not 
place firm constraints on the age of the cooling flows.
However, for Abell 496 the data
suggest an age of between 3.3 and $6.8 \times 10^9$ yr, and a mass
deposition rate, ${\dot M} \sim
50$ \Msunpyr (again in good agreement with the SSS spectral results). 
For Sersic 159, the age and mass deposition rate are 
$6.4-10.5 \times 10^9$ yr and
${\dot M} \sim 220$ \Msunpyr, respectively. For Abell 2597, the values
are  $2.5-5.1 \times 10^9$
yr and ${\dot M} \sim 200$ \Msunpyr. In contrast the data for both Hydra
A and Abell 2029 are consistent with the 
cooling flows having been undisturbed for close to a Hubble time.  

In detail, the relationship between the age of the cooling flow and the 
mean cooling time in the region where the C/D ratio changes 
will be more complex than discussed above, and will depend on 
(undetermined) factors relating 
to the distribution of density inhomogeneities in the cluster gas. However, 
the uncertainties associated with such effects should not much exceed 
those quoted above. 

These results illustrate how standard image deprojection analyses, 
which assume that cooling flows have existed
for a Hubble time, can overestimate the integrated mass 
deposition rates in clusters by up to a factor of 2. 
Future studies, with the improved spectral and spatial resolution of AXAF, 
will allow such properties 
to be explored in greater detail. The aging of cooling flows 
may provide a powerful probe of the evolutionary history of clusters.

\section{The interpretation of fitted column densities}

\begin{figure*}
\hbox{
\hspace{0cm}\psfig{figure=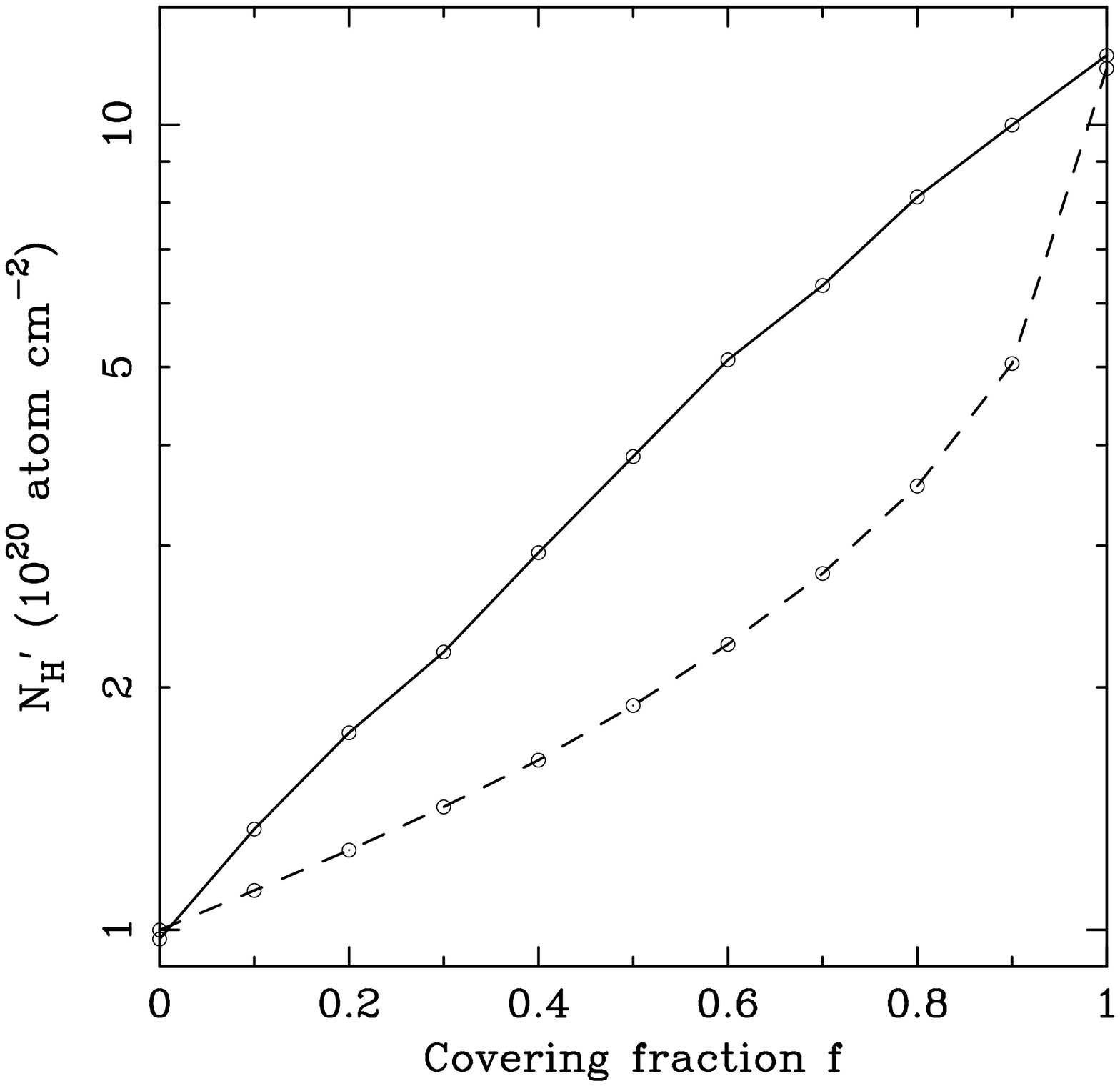,width=0.6\textwidth,angle=270}
\hspace{-2.5cm}\psfig{figure=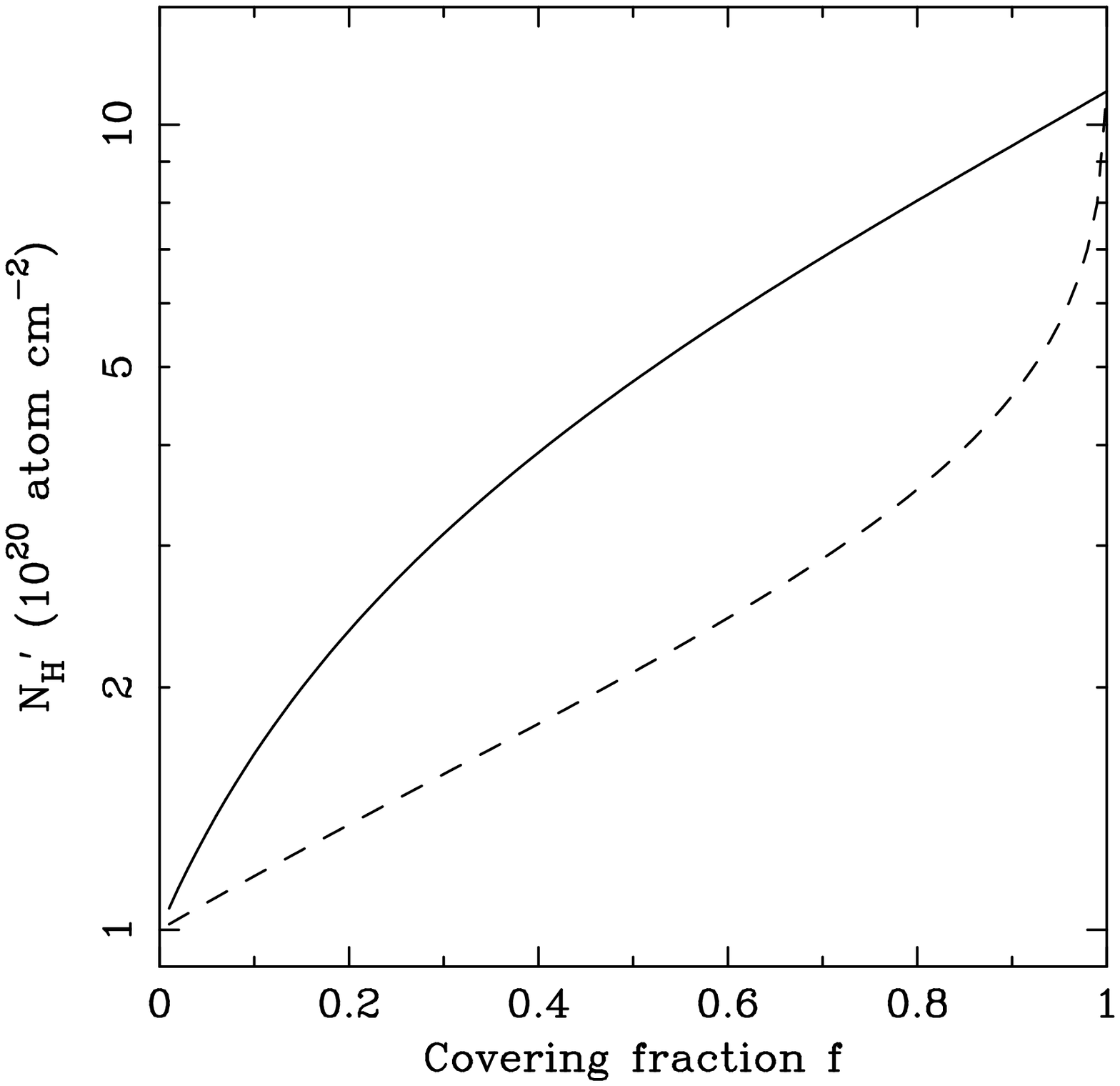,width=0.6\textwidth,angle=270}
}

\vspace{0cm}

\hbox{
\hspace{0cm}\psfig{figure=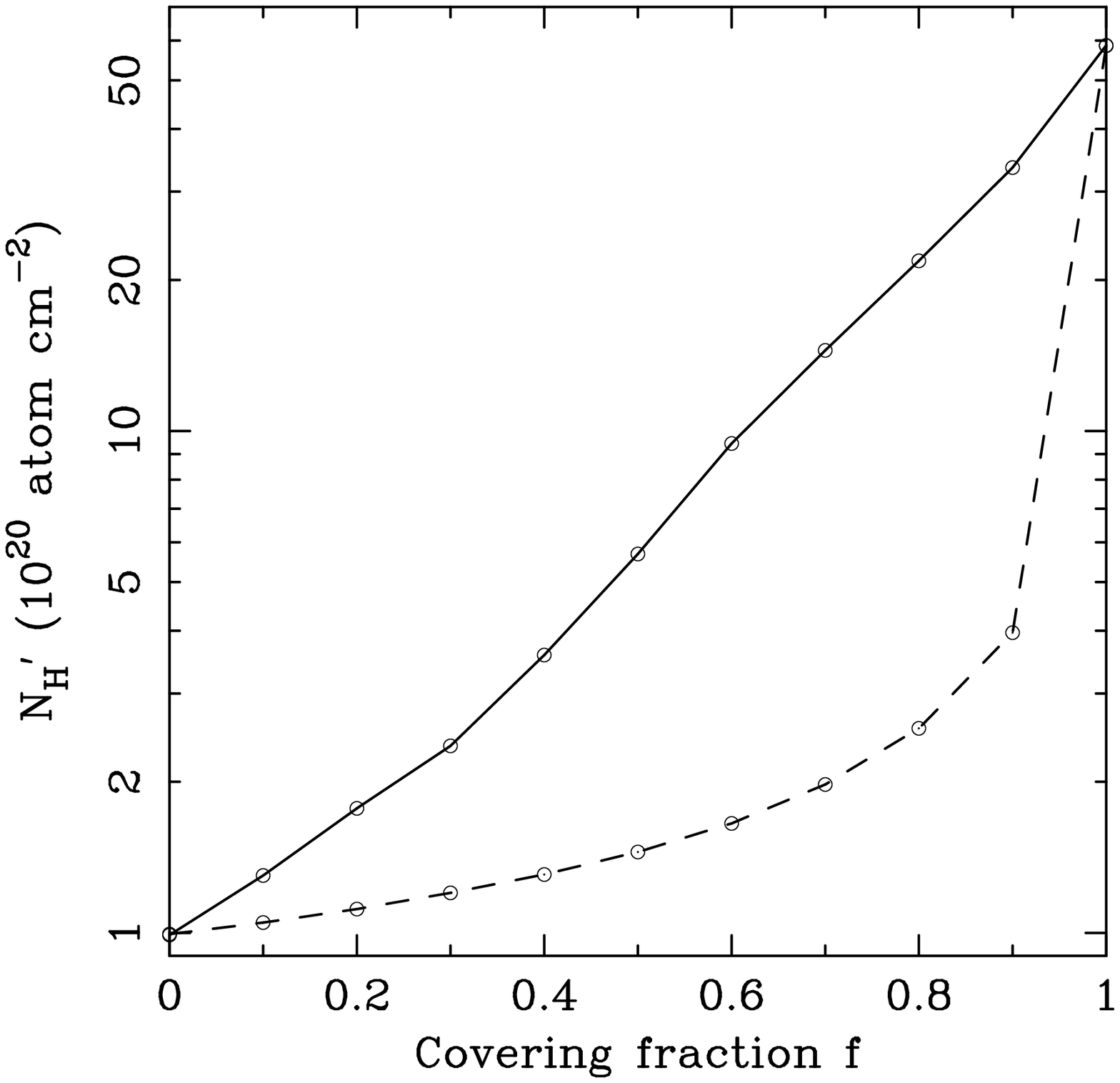,width=0.6\textwidth,angle=270}
\hspace{-2.5cm}\psfig{figure=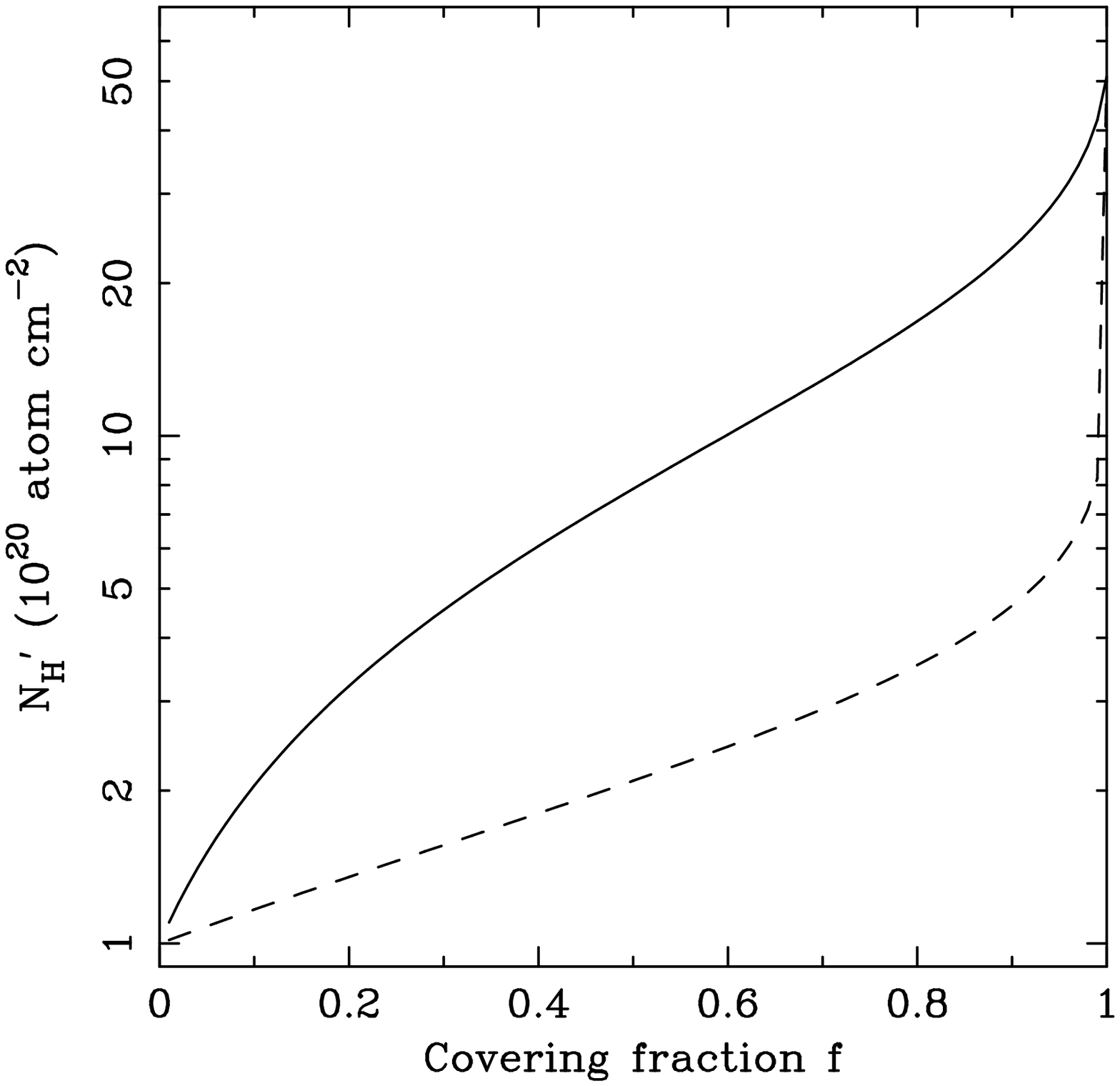,width=0.6\textwidth,angle=270}
}

\caption{(a) The apparent column density, $N_{\rm H}'$,  as a function of the covering
fraction, $f$, of the intrinsic absorber, determined from 
fits to the simulated spectra with simple uniform-foreground-screen
absorption models (Section 6.1). 
A Galactic column density, $N_{\rm H} = 10^{20}$ \apc, and an intrinsic
column density, $\Delta N_{\rm H} = 10^{21}$ \apc, are assumed. 
We adopt a fixed temperature of 5 keV, a metallicity 0.5Z$_\odot$ and a redshift
$z=0.01$. Results are shown for fits to both the full $0.1-2.0$ keV band (lower
dashed curve) and the $0.4-2.0$ keV band (upper solid curve).  (b) The similar
results obtained from the analytical expression in 
equation 3 for $\Delta N_{\rm H} = 10^{21}$ \apc~and photon
energies of 0.4~keV (upper solid line) and 0.2~keV (lower dashed line).
(c) As for (a) but with $\Delta N_{\rm H} = 5 \times 10^{21}$ \apc. 
(d) As for (b) with $\Delta N_{\rm H} = 5 \times 10^{21}$ \apc.
}
\end{figure*}

\begin{figure*}
\hbox{
\hspace{-0.7cm}\psfig{figure=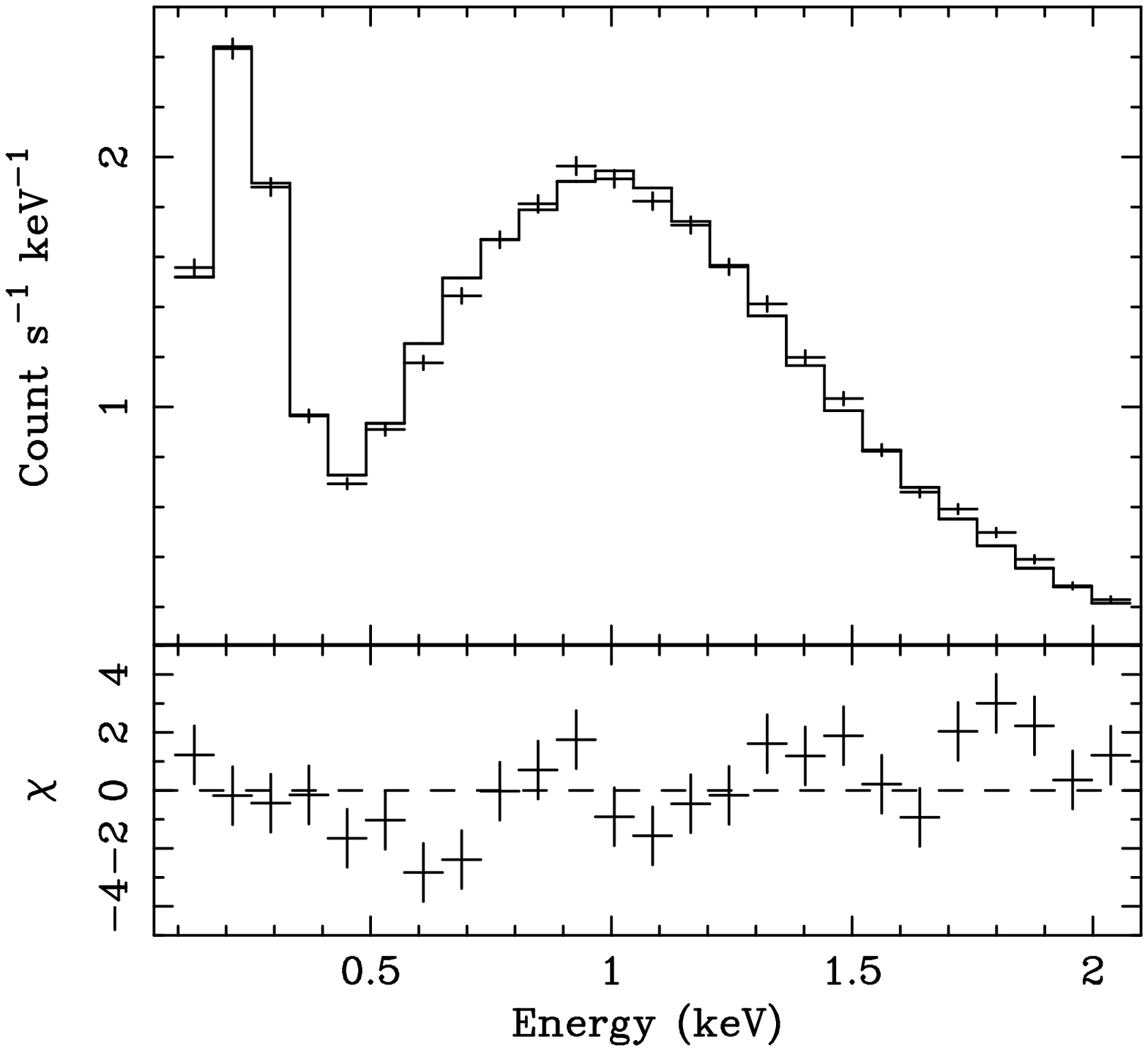,width=0.7\textwidth,angle=270}
\hspace{-3.5cm}\psfig{figure=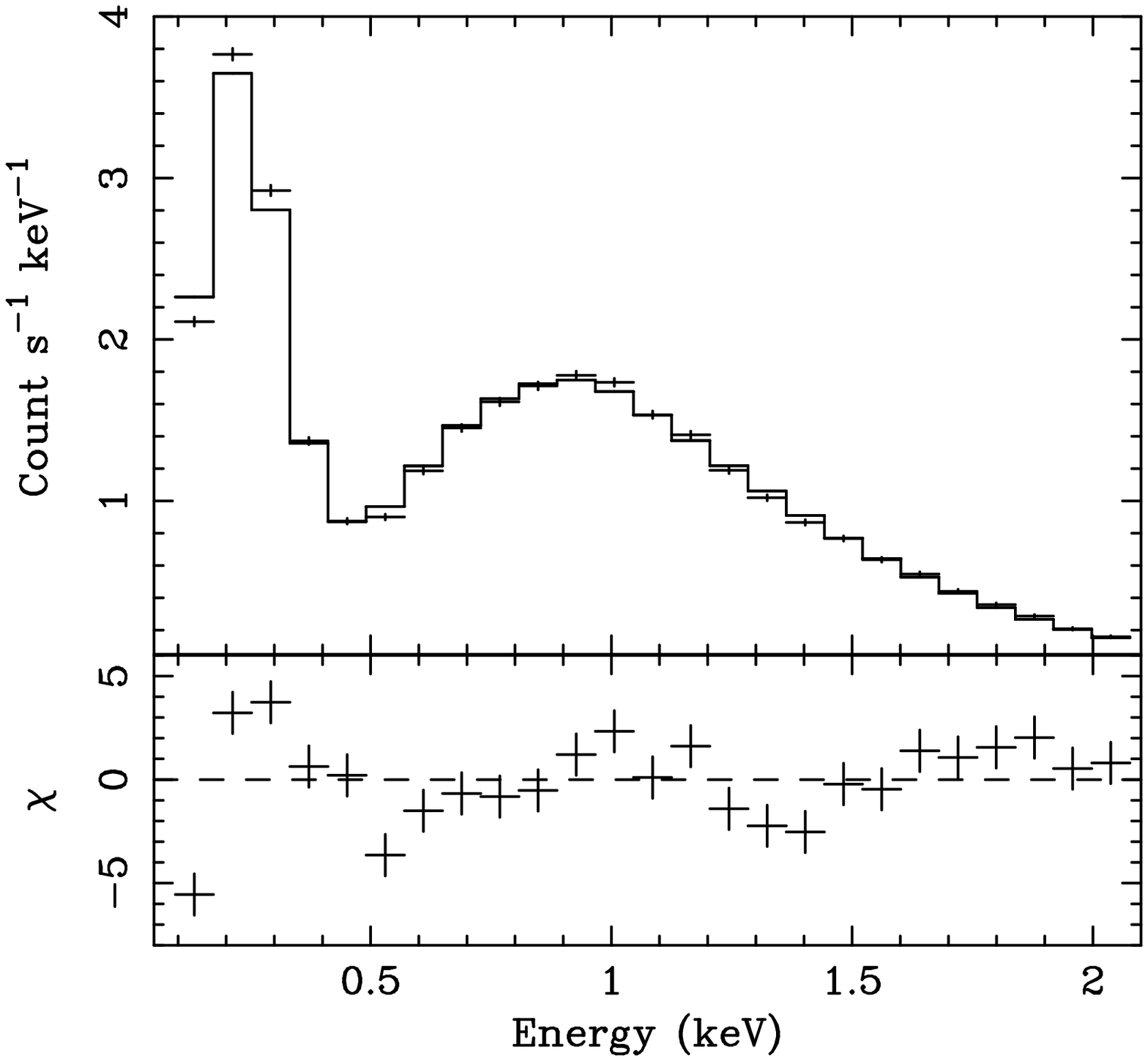,width=0.7\textwidth,angle=270}
}
\caption{(a) Simulated spectrum of a 5 keV cluster at a redshift 
$z=0.01$, absorbed by a Galactic column density of $10^{20}$ \apc~
(covering fraction unity) and an intrinsic column density of $10^{21}$ \apc~ 
(covering fraction $f = 0.5$). The data are shown for the
full $0.1-2.0$ keV range, with the best-fitting 
single temperature, uniform absorbing screen model overlaid. 
The best-fit column density is $1.25^{+0.10}_{-0.11} \times 10^{20}$ \apc, 
implying an excess column density of $2.5^{+1.0}_{-1.1} \times 10^{19}$ \apc. 
The lower panel shows the residuals to the fit. 
(b) The spectrum for the central 2 arcmin of
Abell 1795, at a redshift $z=0.06334$, with the best-fitting 
single temperature, uniform absorbing screen model overlaid. The best
fit column density is $1.27^{+0.04}_{-0.04} \times 10^{20}$ \apc, implying an
excess column density of $1.6^{+0.4}_{-0.4} \times 10^{19}$ \apc. 
The measured temperature of $3.04^{+0.23}_{-0.19}$ keV is lower than the 
5 keV assumed in the simulated spectrum,  due
to the strong cooling flow. The Abell 1795 spectrum has the same 
integrated count rate in the $0.1-2.0$ keV band as the simulated spectrum.
Error bars are $1 \sigma$ confidence limits.} 
\end{figure*}

\begin{figure}
\hbox{
\hspace{-0.7cm}\psfig{figure=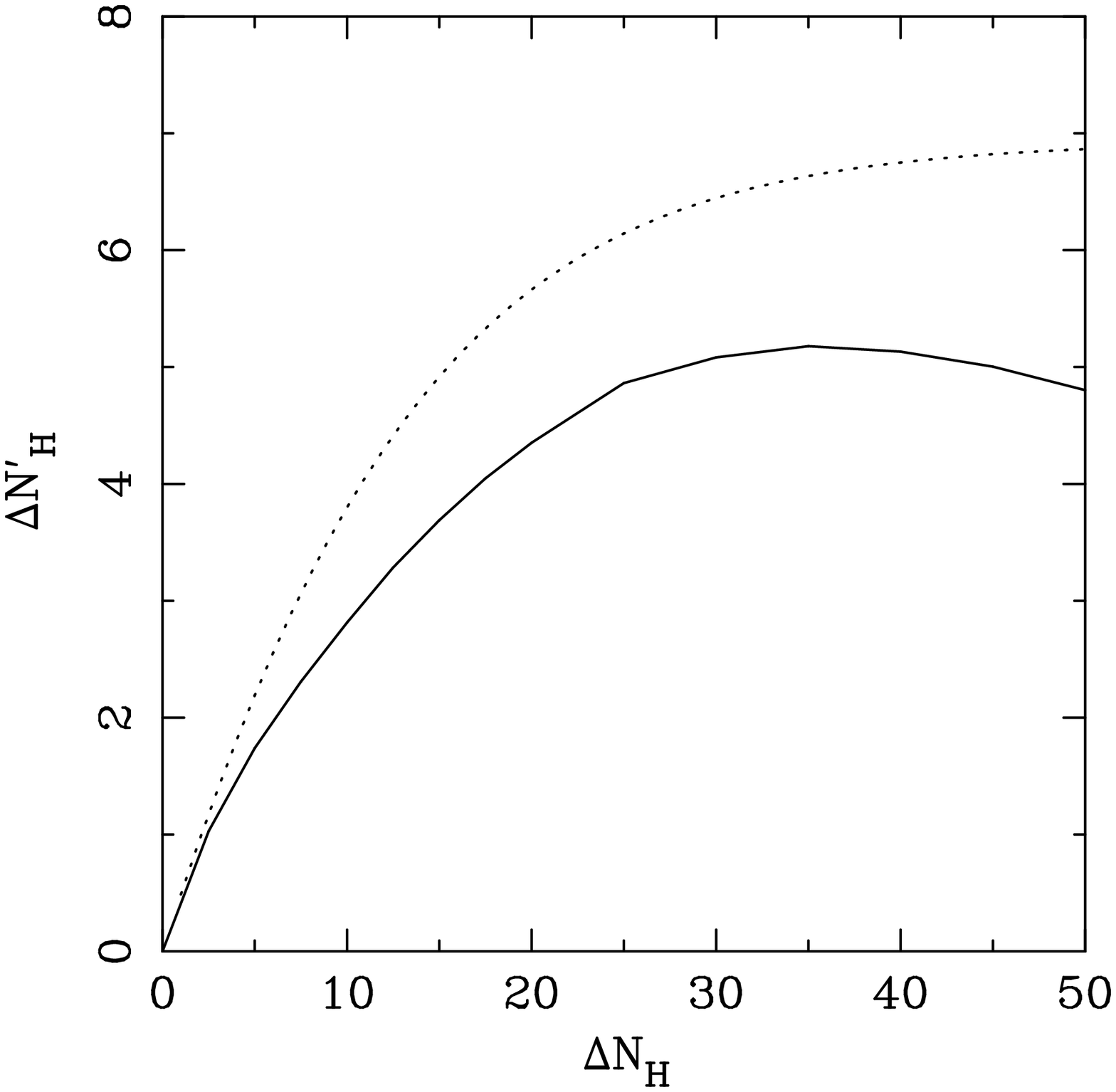,width=0.7\textwidth,angle=270}
}
\caption{The relationship between the apparent excess column 
density, $\Delta N_{\rm H}'$, and the true intrinsic column density, 
$\Delta N_{\rm H}$, determined from the analytical expression in equation
4 (dotted curve) and spectral simulations (solid curve). For the spectral
simulations a Galactic column density of $10^{20}$ \apc~has been assumed. 
Both axes are in units of $10^{20}$ \apc.
}
\end{figure}

In this Section we further examine the results on intrinsic
absorption in the cooling flows. We discuss how measurements of
intrinsic column densities, determined from fits with uniform foreground 
screen absorption models, will naturally 
underestimate the true intrinsic column densities when 
the absorbing gas is spatially distributed throughout the cooling flows. 
In particular, we explore two limiting distributions for the absorbing gas; 
partial covering  (which will apply if the absorbing gas is clumped on large scales)
and a homogeneous distribution of absorbing material
throughout the X-ray emitting region, which we refer to as the `multilayer'
model. We compare the absorption measurements 
from the ROSAT X-ray colour profile analysis to previous results from the 
Einstein SSS (White \etal 1991) and ASCA. Finally in this Section, we
comment on the mass of absorbing gas in the clusters and on the role of
this material as a sink for the cooling gas.

\subsection{The effects of partial covering} 

The use of simple uniform-screen absorption models in the analysis of cluster 
spectra  can lead to significant underestimates of intrinsic column densities.  
A simple illustration of this, in the case where the absorbing material
covers only a fraction, $f$, of the intrinsic emission, is given in 
Figs. 8(a) and (c). 
The figures show the column densities inferred from fits with simple
uniform (foreground) screen 
absorption models to simulated cluster spectra incorporating intrinsic
absorption with partial covering. For the simulations we assume a cluster  temperature of 5 keV, a
metallicity of 0.5 solar, and a redshift, $z = 0.01$.  The cluster is assumed 
to lie behind a Galactic column density, $N_{\rm H} = 10^{20}$ \apc, and to be 
intrinsically absorbed by an excess column density, $\Delta N_{\rm H}$, which 
covers a fraction, $f$, of the  emitted flux. Intrinsic column densities of 
$\Delta N_{\rm H} = 10^{21}$
and $\Delta N_{\rm H} = 5 \times 10^{21}$ \apc, and covering fractions of
$0.1-1.0$, were examined. [These 
column densities span the range of values inferred from Einstein Observatory SSS 
and ASCA observations of cooling flows; White \etal (1991), Fabian \etal (1994), Allen \etal (1996a), 
Allen \etal (1996b),  Fabian \etal (1996).]

The simple partial covering distribution provides a crude approximation to the
likely situation in  an intrinsically 
absorbed cooling flow. In the simplest case of a geometrically thin
absorbing screen 
of uniform column density, the
observed emission can be modelled by  two components, accounting for the
emission from in front of and behind the  absorber, respectively. If
the absorber spans the entire region of interest, the
components should have equal intrinsic fluxes, and be absorbed by column
densities $N_{\rm H}$ and $N_{\rm H} + \Delta N_{\rm H}$.  In this case, the
covering fraction of the  intrinsic absorber is 0.5. 
If the intrinsic absorber is present over 
only part of the region of
interest, as would be the case if the
extent of the absorber were less than the region being analysed, 
the covering fraction will be less than 0.5. (With a real cooling
flow, variations in both column density and 
covering fraction are likely.)

The fits to the simulated spectra were carried out in both 
the $0.1-2.0$ keV and $0.4-2.0$ keV energy bands. The 
temperature and metallicity were fixed at 5 keV and 
0.5 Z$_\odot$, respectively. For covering fractions less than unity, 
the column densities determined from fits with uniform-screen ($f=1$) absorption 
models significantly underestimate the `true' absorption. For an intrinsic column
density, $\Delta N_{\rm H} = 5 \times 10^{21}$ \apc, and a covering
fraction, $f= 0.5$, a fit to the $0.4-2.0$ keV spectrum implies a
column density of $5.7 \times 10^{20}$
\apc~, an excess of only $4.7 \times 10^{20}$ \apc~over the Galactic value.
Thus the column density inferred from the spectral fit to the
$0.4-2.0$ keV band data underestimates 
the true intrinsic column density by a factor of $\sim 10$. Even more
dramatically, when the full $0.1-2.0$ keV energy range is used, 
the spectral fit implies an excess column density of only 
$4.5  \times 10^{19}$ \apc, underestimating the true intrinsic column
density by a factor of $\sim 100$ ({\it cf.} Section 5.4).
In a $\chi^2$ sense, however, the uniform absorption model still provides a 
statistically adequate description of the simulated spectra 
(for signal-to-noise ratios in the simulated spectra typical of PSPC cluster 
observations). Similar results are obtained for the simulations with 
$\Delta N_{\rm H} = 10^{21}$ \apc, although the underestimation of 
intrinsic column densities from the spectral fits is less severe. A fit 
to the $0.4-2.0$ keV spectrum, with $f=0.5$, gives a measured 
column density of $3.9 \times 10^{20}$ \apc, implying an excess column
density a factor of 
$\sim 3$ less than the true value. 

Such results are not
sensitive to the precise emission spectrum assumed. If a cooling flow,  
rather than isothermal, model is used both for generating and fitting
the simulated spectra, results very similar
results to those plotted in Fig. 8 are obtained. 
Significant discrepancies only arise when different emission 
models are used for generating and fitting the simulated data. 

In Fig. 9 we show the results from a consistency check in which 
compare the results
from a simple, uniform-screen fit to a simulated cluster 
spectrum with $N_{\rm H} =
10^{20}$ \apc, $\Delta N_{\rm H} = 10^{21}$ \apc and $f=0.5$, with the
results from a fit
to the observed spectrum for the central 2 arcmin (radius) region of Abell 
1795. (The signal-to-noise ratio in the simulated spectrum has been adjusted 
to approximately match that in the Abell 1795 data.)
In both cases, the fit to the $0.1-2.0$ keV spectrum implies an excess column
density of $\sim 2 \times 10^{19}$  (whereas the B/D results for 
Abell 1795 imply a much larger intrinsic column density of 
$\sim 5 \times 10^{20}$ \apc).

Some intuitive understanding of these results 
may be gained from the
following argument. We can write the observed flux as 

\begin{equation}
A(E)e^{-\sigma(E)N_{\rm H}}[(1-f) + fe^{-\sigma(E)\Delta N_{\rm H}}]
\approx A(E) e^{-\sigma(E)N_{\rm H}'}, 
\end{equation}

\noindent where $A(E)$ is the
emission at energy $E$.  Then the apparent column density
determined from a fit assuming a covering fraction of unity is 

\begin{equation}
N_{\rm H}' = N_{\rm H} - \log [1-f+fe^{-\sigma(E) \Delta N_{\rm
H}}]/\sigma(E), 
\end{equation}

\noindent implying an apparent excess absorption of 

\begin{equation}
\Delta N_{\rm H}' = - \log [1-f+fe^{-\sigma(E) \Delta N_{\rm
H}}]/\sigma(E). 
\end{equation}

The absorption cross-section may be crudely approximated by $\sigma(E) \approx 
10^{-21}(E/0.4)^{-8/3}$ cm$^{-2}$. In Figs. 8(b) and (d) we plot 
$N_{\rm H}'$ as a
function of $f$ for photon energies $E = 0.20$ and 0.40 keV, and
intrinsic column densities of $10^{21}$ and $5 \times 10^{21}$ \apc. 
(Note that the lowest-energy photons included in any fits will dominate the
constraints on the absorption value.) The analytic curves provide a good 
approximation to the observed behaviour. The apparent
excess column density, $\Delta N_{\rm H}'$, remains low so long as significant
signal from low energies (below the energy where $\sigma N_{\rm H} \sim1$) is
included , i.e. $E<0.4 N_{\rm H, 21}^{3/8}\keV,$ where $N_{\rm H}=10^{21}
N_{\rm H, 21}\psqcm$. Where $\sigma \Delta N_{\rm H}$ is small, $\Delta N'_{\rm H}
\rightarrow f \Delta N_{\rm H}$.

Our results imply that simple fits to PSPC spectra
for clusters with low Galactic column densities 
($N_{\rm H} \sim 10^{20}$ \apc), which assume covering fractions for
the absorber of unity, will always
return low values for the excess column density even if, in reality, 
 $f\sim 0.5$ and $\Delta
N_{\rm H} \approxgt 10^{21}$ \apc. 
If the true intrinsic column densities
in clusters are $\approxgt 10^{21}$ \apc, 
we should only obtain reasonable estimates for
$\Delta N_{\rm H}$ with simple spectral fits if we either 
restrict the bandwidth artificially, or if the cluster has a large Galactic 
column density ($N_{\rm H} \sim 10^{21}$ \apc). Such a scenario occurred naturally in 
the PSPC studies of Abell 478 by Allen \etal (1993) and of 2A 0335+096 by Irwin
\& Sarazin (1995), where significant
excess column densities of $\sim 10^{21}$ \apc~were inferred.

Such arguments also imply a weak positive correlation between the apparent excess 
column density, $\Delta N'_{\rm H}$, determined from fits with uniform
foreground screen absorption models, and the
Galactic column density, $N_{\rm H}$. When $N_{\rm H}$ is high then the
photon energy dominating the absorption fit (generally the lowest energy
with a significant detection) is also high and $\sigma \Delta N_{\rm
H}$ tends to be small, so that the inferred column density $\Delta
N'_{\rm H} \approx f \Delta N_{\rm H}$.
Provided that the covering fraction, $f$, is also high, the model
will tend to the `true' answer. This is consistent with the results
presented in Table 6, which exhibit a weak correlation between $\Delta
N'_{\rm H}$ and $N_{\rm H}$, particularly once the radial variation of
$\Delta N'_{\rm H}$ in Abell 426 is taken into account. 

We have used the relationship between $\Delta N'_{\rm H}$ and $\Delta
N_{\rm H}$ in equation 4, and the results from further spectral
simulations, to estimate the correction factors that should be applied to
the $\Delta N'_{\rm H}$ values listed in Table 6, in
determining the true intrinsic column densities in the clusters.  
For these purposes
we assume a covering fraction $f=0.5$ for the intrinsic absorber (which
should approximate the true value in the central regions of the clusters).
The results are shown in Fig. 10. We see that for an apparent excess
column density, $\Delta N'_{\rm H} = 2 \times 10^{20}$ \apc, the
analytical expression and spectral simulations imply true intrinsic column
densities of $4.5 \times 10^{20}$  and $6.1 \times 10^{20}$ \apc,
respectively.  Typically, the corrected intrinsic column densities for the 
clusters in our
sample are a factor of $2-3$ larger than apparent values 
determined directly from the B/D analysis. [The corrected intrinsic column 
densities determined from equation 4 are 
listed in Column 4 of Table 6 as $\Delta N_{\rm H}(1)$.]

Finally, it is also important to note that the use of slightly more complex spectral 
models incorporating both isothermal and cooling-flow components, with the 
excess absorber modelled as a uniform screen in front of the cooling flow, 
will still lead to significant underestimates of the true
levels of intrinsic absorption in clusters. Typically, only a few 
per cent of the X-ray emission along the line of sight to the central 30 arcsec
bin will be due to material in front of the cooling flows. 
If the absorbing material is distributed throughout the cooling flow, as
required by the observations reported here, any model which places the
absorber entirely in front of the cooling flow will underestimate the true
intrinsic column densities in much the same manner as in the simulations
discussed above.

\subsection{The multilayer absorption model} 

We have also explored a second plausible distribution for the 
X-ray absorbing material in which the absorbing gas is assumed to 
reside in many small clouds, each of which has an individual 
column density much less than the total intrinsic
column density, $\Delta N_{\rm H}$. The clouds are assumed to be evenly 
distributed throughout the X-ray emitting plasma. 
For an intrinsic emission
spectrum $A(E)$, the observed spectrum, $A'(E)$, emerging from 
the emitting/absorbing region may be written as

\begin{equation}
A'(E) = A(E) \left ( \frac {1 - e^{-\sigma(E) \Delta N_{\rm H}}}
{\sigma(E) \Delta N_{\rm H}} \right ).
\end{equation}

A series of spectra incorporating absorption as described by
equation 5 (which we hereafter refer to  as the `multilayer' absorption
model)  were simulated. Again, the intrinsic emission spectrum
was taken to be a 5 keV thermal plasma, with a  metallicity, $Z =
0.5Z_\odot$, and a redshift, $z = 0.01$. Total intrinsic column densities
of $\Delta N_{\rm H} = 10^{21}$ and $5 \times 10^{21}$ \apc were examined.

The simulated spectra were then fitted with simple uniform foreground
screen absorption models. As with the partial covering 
model discussed in Section 6.1, the apparent excess column densities
determined with the uniform absorbing
screen model significantly underestimate 
the true intrinsic column density. For a Galactic
column density, $N_{\rm H} = 10^{20}$ \apc, and intrinsic multilayer 
column densities of $\Delta N_{\rm H} = 10^{21}$ and $5 \times 10^{21}$
\apc, the apparent excess column densities determined with the uniform 
absorbing screen model 
are $\Delta N_{\rm H}' = 2.65$ and $10.3\times 10^{20}$ \apc,
respectively. Thus, with a multilayer distribution for the intrinsic absorbing
gas, the simple uniform foreground screen model underestimates 
the true intrinsic column density by
factors of $\sim 4$ and 5 respectively.  Note also that these
factors are essentially independent of the Galactic column density, since
multilayer absorption, with total column densities of 
$\Delta N_{\rm H} = 10^{21}$ and $5 \times 10^{21}$ \apc, can be 
well-approximated by uniform absorbing screens with column densities of $\sim
0.26$ and $1.0 \times 10^{21}$ \apc, respectively.  In Column 5 of Table 6 
[$\Delta N_{\rm H}(2)$], we list
the intrinsic column densities for the sample of 
cooling flows in this paper,
corrected (with an intermediate correction factor of 4.5) 
for an assumed multilayer distribution of the 
absorbing gas.

\subsection{The comparison of results from ROSAT, SSS and ASCA} 

The results presented here confirm that
intrinsic X-ray absorption is a common property of cooling flows. 
All of the clusters with previous positive detections of intrinsic
absorption, from the Einstein SSS study of White \etal (1991), 
also exhibit absorption signatures in their ROSAT X-ray colour profiles.  
The column densities inferred from the two
studies are also similar, ranging from a few $10^{20}$ to a few 
$10^{21}$ \apc. However, although the results reported
here (Table 6 columns 4, 5) include corrections for the effects of 
the likely spatial distribution of absorbing gas on the measured column densities, 
the White \etal (1991) results were all determined under the more simple assumption
that the absorber acts as a uniform screen in front of
the clusters, with a covering
fraction of unity. When the same simple model for the distribution
of the absorbing gas is used with the ROSAT data, (Table 6, Column
3) we obtain results that are typically factors of $2 - 6$ less than the SSS 
results. In this Section we examine whether this discrepancy 
should simply be attributed to calibration uncertainties, 
or if it instead provides further information about the
nature of the absorbing gas.

The SSS observations were made in the $0.6- 4.5$ keV energy band, with a
spectral resolution of $\sim 160$ eV, a higher energy band and a 
higher spectral resolution than that of the PSPC. We have 
examined the variation in the results from fits to 
simulated cluster spectra, incorporating intrinsic
absorption with both partial covering and multilayer 
distributions, when observed with the two instruments. 
(With the SSS simulations, the effects of ice build-up on
the instrument response were accounted for.) Total multilayer and partial 
covering column densities of 
$10^{21} - 10^{22}$ \apc~were examined, with a default Galactic column
density of $10^{20}$ \apc~assumed. 
The simulated spectra were then fitted with simple uniform foreground screen 
models, assuming a covering fraction for the excess absorption of unity. 

With a multilayer distribution for the intrinsic absorbing gas in the 
simulated spectra, the fits with the uniform-screen model give very
similar results for the excess column densities with 
the PSPC and SSS instruments
(in both cases the fit results underestimate the true 
intrinsic column densities by factors of 
$\sim 4-5$; {\it cf.} Section 6.2). 
However, when the intrinsic absorber in the simulated spectra is distributed 
according to the partial covering model, clear differences between the PSPC and 
SSS results become apparent.

\begin{figure}
\hbox{
\hspace{-0.7cm}\psfig{figure=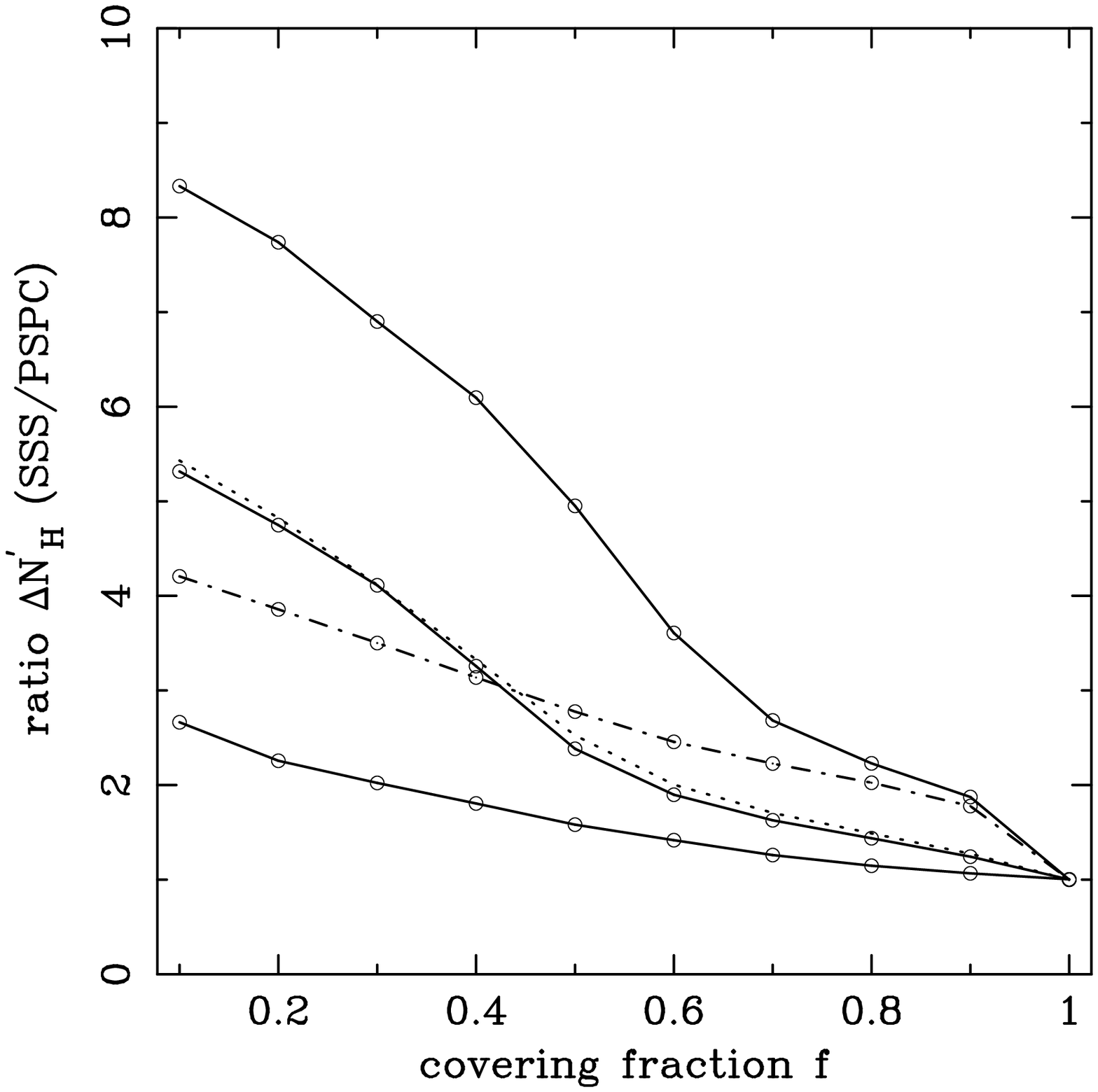,width=0.7\textwidth,angle=270}
}
\caption{The ratio of the apparent excess column densities, $\Delta
N_{\rm H}'$, inferred from fits with simple uniform foreground 
screen absorption models to simulated SSS and 
PSPC spectra incorporating intrinsic absorption with  
a covering fraction, $f$, of between 0.1 and 1.0.
The solid curves show the results (from top to
bottom) for $\Delta N_{\rm H} = 10, 5$ and $1 \times 10^{21}$ \apc, with a
fixed Galactic column density, $N_{\rm H}$, of $10^{20}$ \apc. 
The dot-dashed curve 
shows the results for $\Delta N_{\rm H} = 10^{22}$ \apc~and $N_{\rm H} =
3 \times 10^{20}$ \apc. The dotted curve shows the ratio of the results
for the simulated ASCA SIS and PSPC spectra, for $\Delta N_{\rm H} = 5 \times 
10^{21}$ \apc~and $N_{\rm H} = 10^{20}$ \apc. 
Note the agreement between the SIS/PSPC and SSS/PSPC results.}

\end{figure}

The differences between the PSPC and SSS results, from the 
fits to the simulated spectra, incorporating partial covering, are
summarized in Fig. 11. For a Galactic column density of 
$\sim 10^{20}$ \apc, and intrinsic absorption with a 
column density of a few $10^{21}$ \apc~and a covering fraction $f =
0.2-0.6$, discrepancies of factors of between 
2 and 6 will naturally arise between measurements made with the
SSS and PSPC instruments (in the sense the SSS results exceed the PSPC
results).  Such an effect is sufficient to account for the bulk of the 
discrepancy between the White \etal (1991) SSS results and the ROSAT 
results reported here. 

We have also carried out a further comparison, between the PSPC results 
and the results from simulated spectra generated with an instrument response 
appropriate for the Solid-state Imaging Spectrometers (SISs) on ASCA. 
The dotted curve in Fig. 11 shows the ratio of the excess column density
determined from the ASCA SIS simulations to that with ROSAT PSPC, for a Galactic column
density $N_{\rm H} = 10^{20}$ \apc~and an intrinsic column density of $5
\times 10^{21}$ \apc. The results are very similar to those obtained from the
comparison of the SSS and PSPC data, for the same absorption parameters.

Since the partial covering model can account for many of the
observed properties of the intrinsic absorber, we have made one 
further check on the validity of this model by applying it to observed 
ROSAT PSPC, Einstein SSS and ASCA SIS spectra for Abell 2199 (which has 
the best combined data set of any of the clusters in the present sample). 
Spectra from the central 2 arcmin (radius) region of Abell 2199 were 
extracted from the PSPC and ASCA SIS data sets (2 arcmin radius 
being the smallest region that can be consistently analysed with the two
instruments, given the relevant PSFs). 
The SSS offers no spatial resolution but did have a 3
arcmin (radius) circular aperture, reasonably well-matched to the
regions studied with the other instruments. 
With the ASCA analysis, the data from both SIS detectors were used. With
the SSS analysis, the data from all three observations of the source
(White \etal 1991) were included. 

The spectra were fitted with a model incorporating emission from an
isothermal plasma and a cooling flow of 80 \Msunpyr ({\it cf.} Section
5.5), at a redshift of $z =0.0309$, absorbed by a (fixed) 
Galactic column density of $8.7 \times 10^{19}$ \apc ~(except for the PSPC data
where the Galactic column density was fixed at $5 \times 10^{19}$ \apc; 
see Section 5.3) . A second absorption 
component of variable column density and covering fraction was also
assumed to act on the emission. The cooling flow was forced to cool from 
the ambient cluster temperature, which was a free parameter in the fits 
(with the exception of the ROSAT analysis where, due to the limitations
of the data, the upper temperature of the flow was fixed at 3.5
keV, in agreement with the results of Section 5). 
The metallicity of the X-ray emitting gas and the emission measure of the 
isothermal component were also free parameters in the fits. With the 
ROSAT data, the spectral analysis was limited to the $0.4-2.0$ keV energy 
range. For the ASCA SIS and Einstein SSS data, spectral ranges of $0.6-10.0$ 
keV and $0.6-4.0$ keV were  used, respectively.  All appropriate ice 
corrections were applied to the SSS data.

\begin{figure*}
\hbox{
\hspace{1.5cm}\psfig{figure=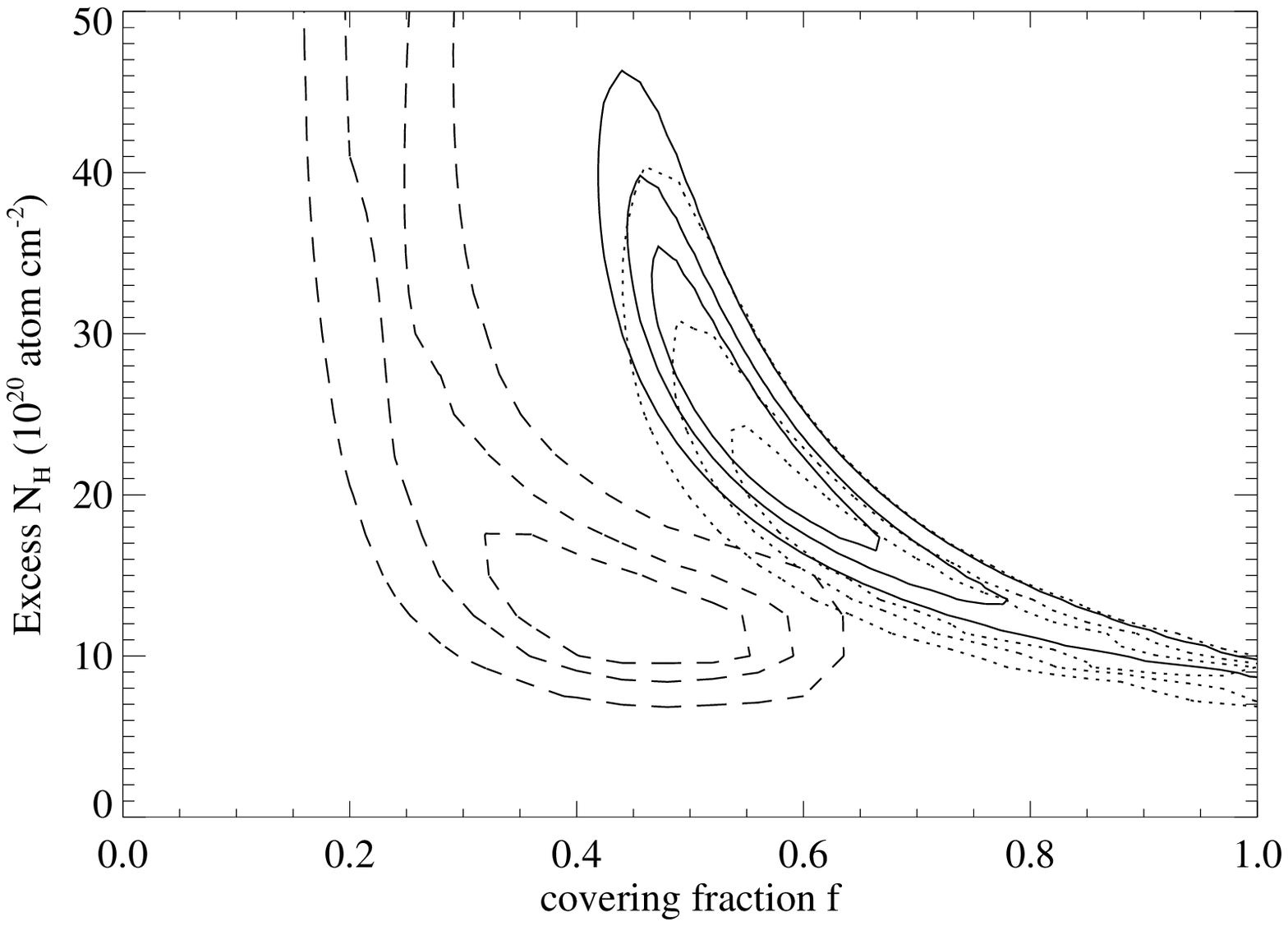,width=0.75\textwidth,angle=0}
}
\caption{The joint confidence contours on the column density and
covering fraction of the intrinsic absorber in Abell 2199 
for the ASCA SIS (solid curves), ROSAT PSPC (dashed curves) and
Einstein SSS (dotted  curves) data. The contours mark the 
$\Delta \chi^2 = 2.30, 4.61$ and 9.21 limits, corresponding 
to 68, 90 and 99 per cent confidence for two interesting parameters.
}

\end{figure*}

\begin{figure*}
\hbox{
\hspace{1cm}\psfig{figure=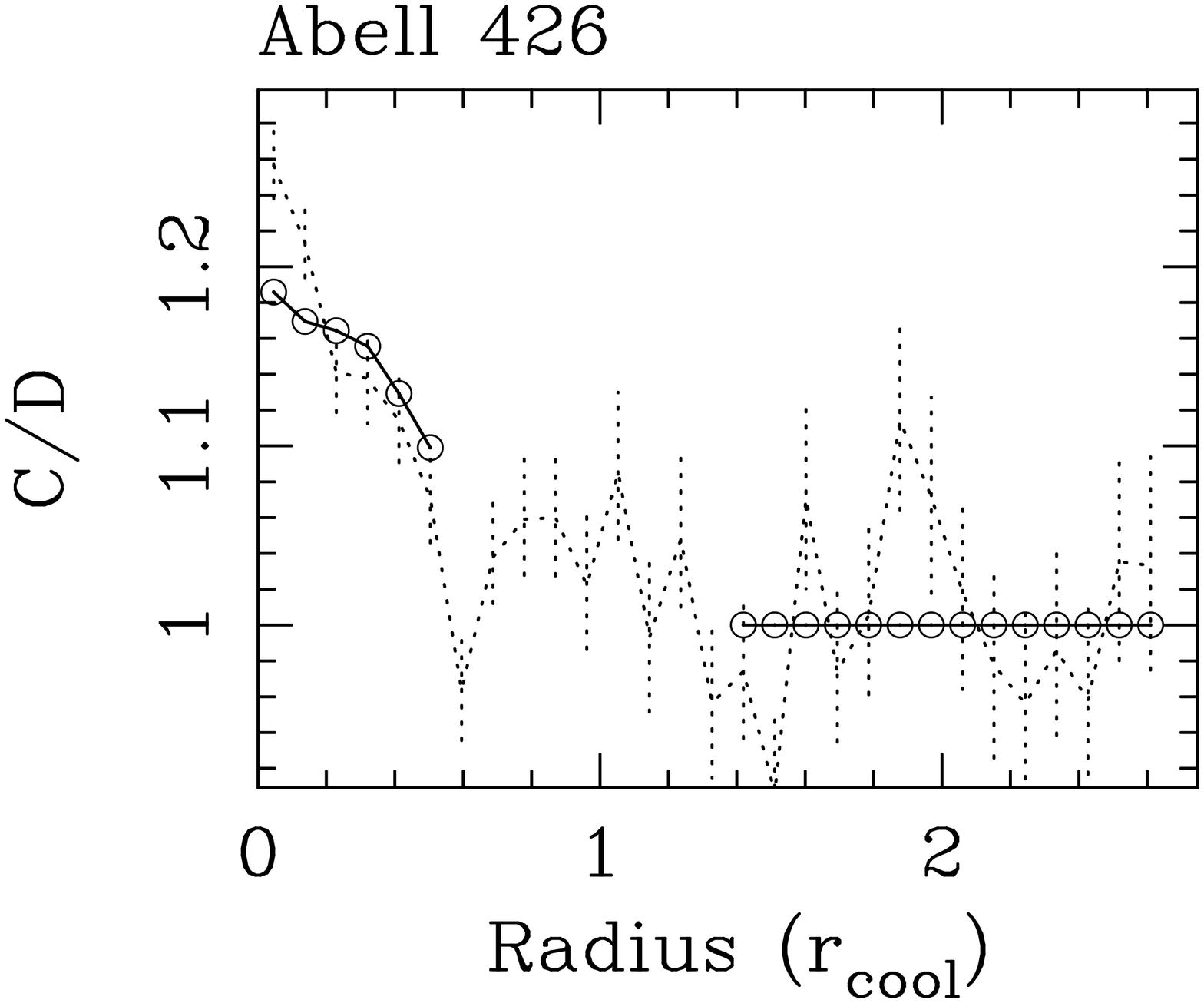,width=0.47\textwidth,angle=270}
\hspace{-0.5cm}\psfig{figure=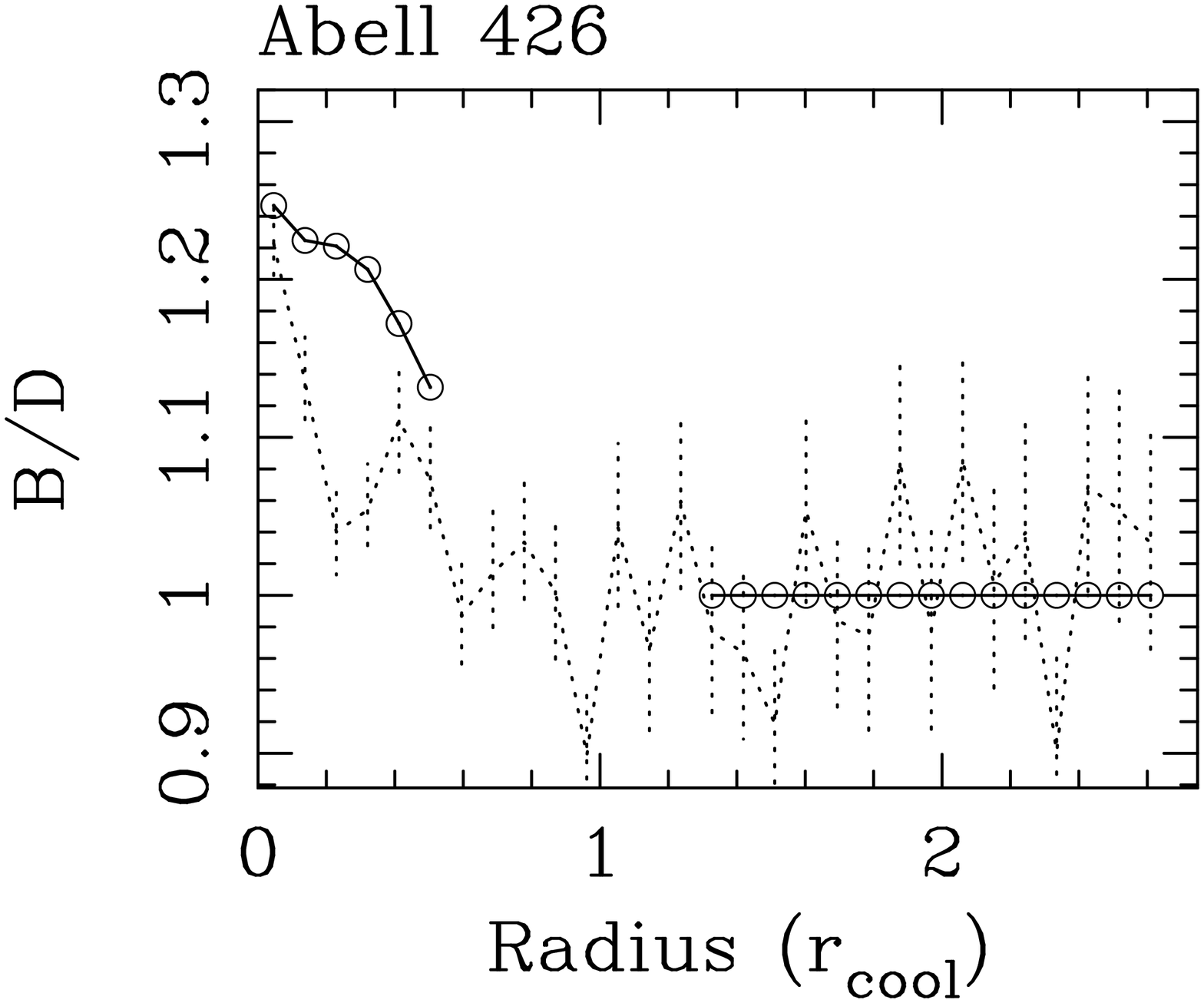,width=0.47\textwidth,angle=270}
}

\vspace{-0.2cm}

\hbox{
\hspace{1cm}\psfig{figure=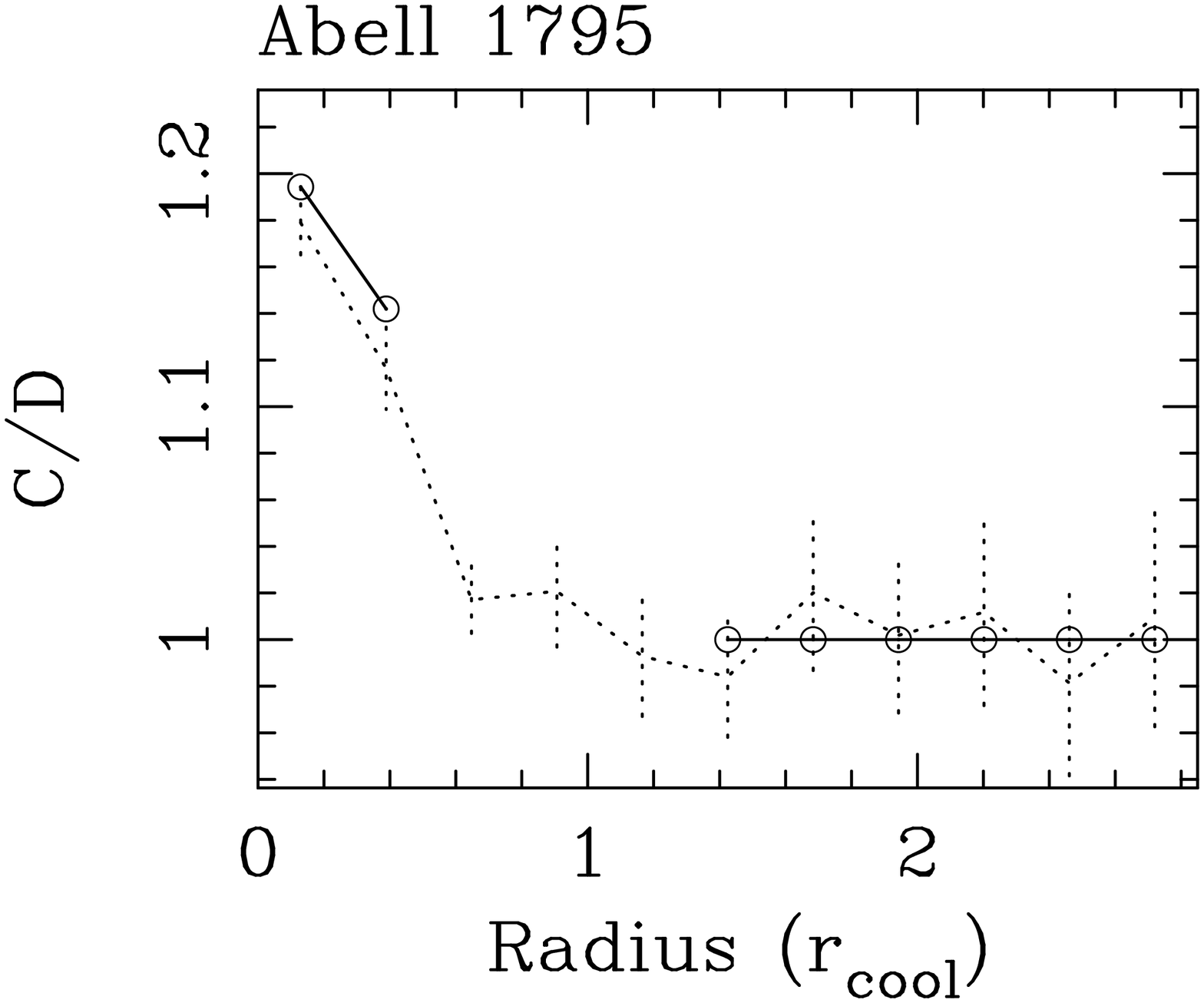,width=0.47\textwidth,angle=270}
\hspace{-0.5cm}\psfig{figure=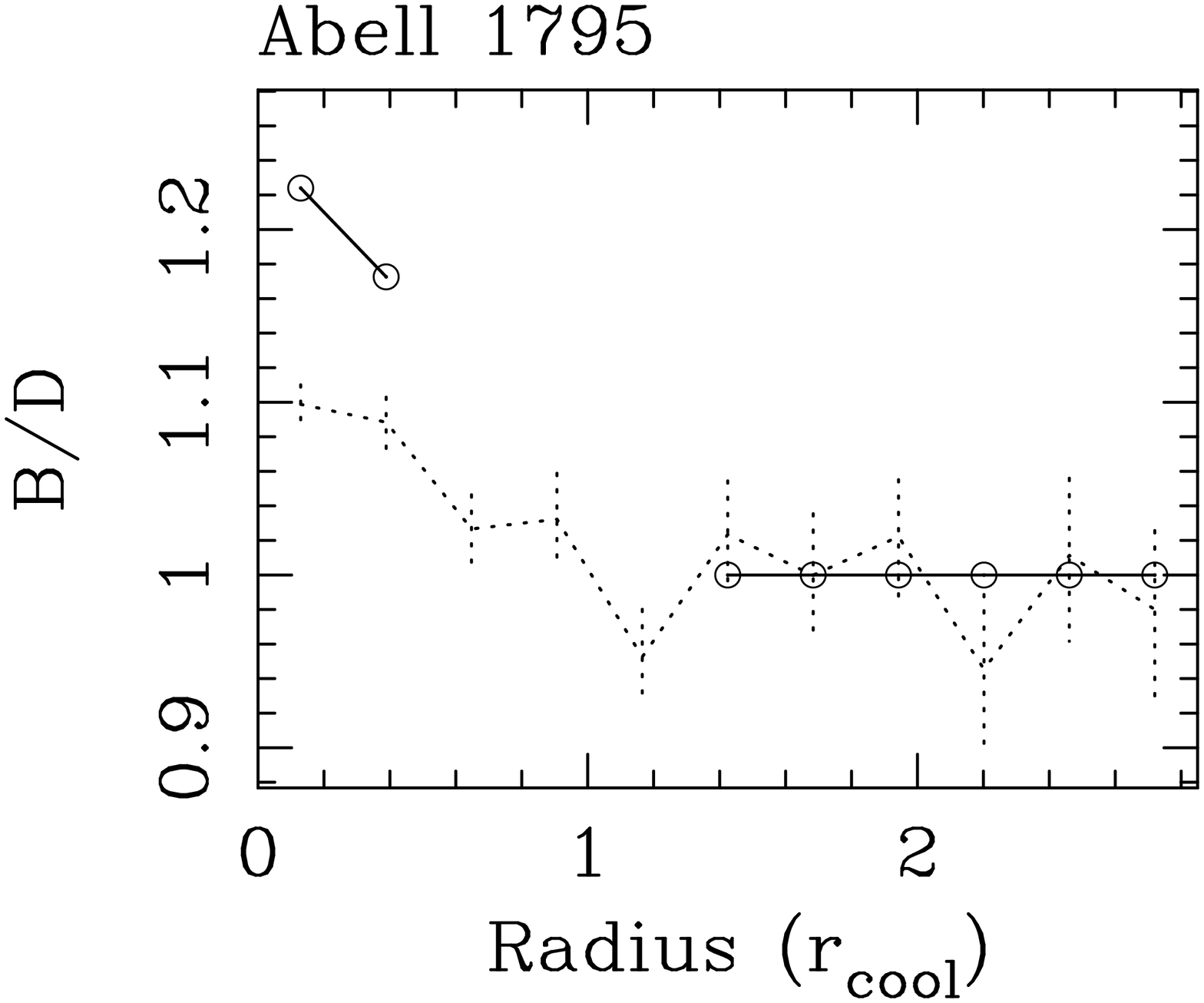,width=0.47\textwidth,angle=270}
}
\vspace{-0.2cm}

\hbox{
\hspace{1cm}\psfig{figure=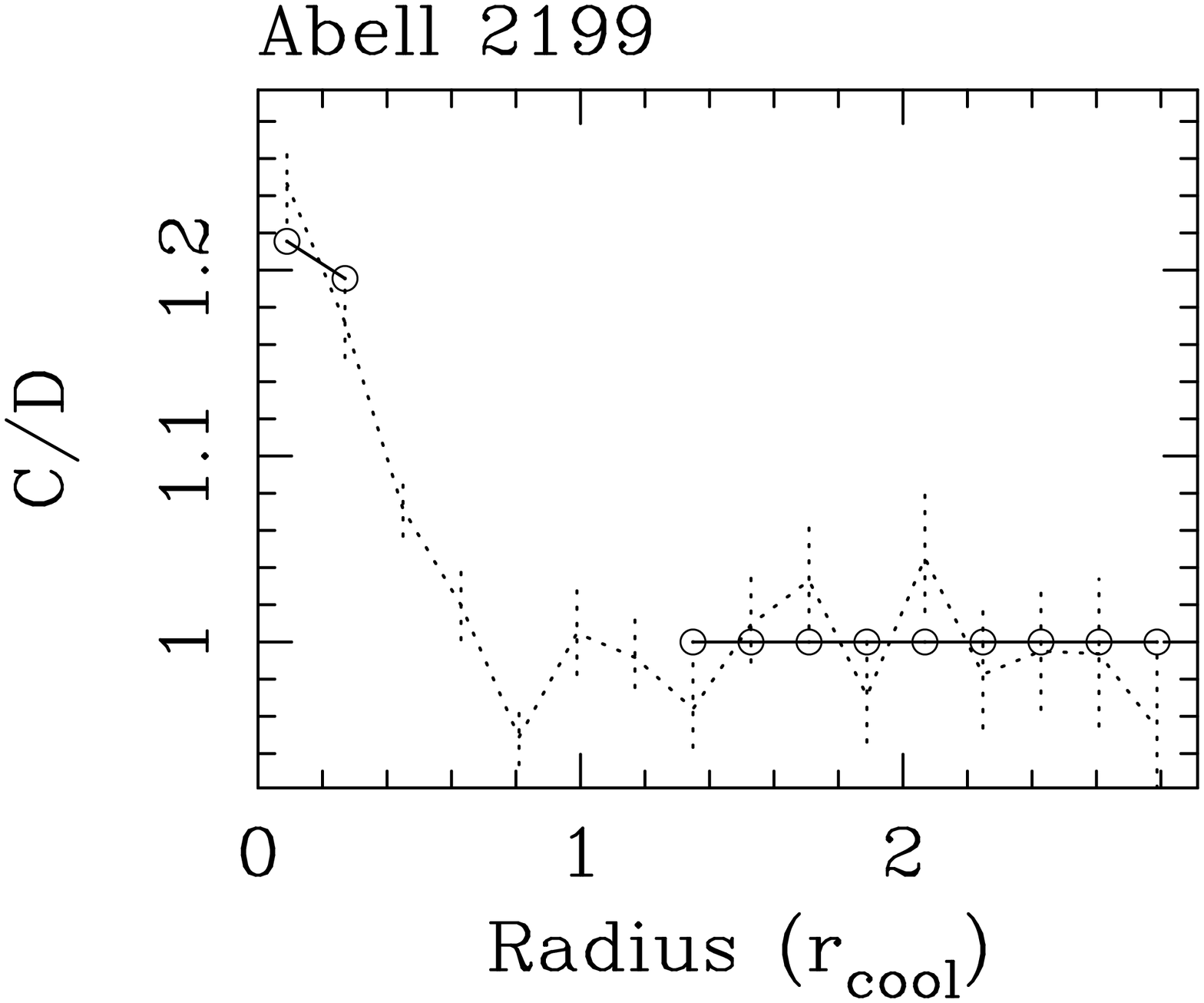,width=0.47\textwidth,angle=270}
\hspace{-0.5cm}\psfig{figure=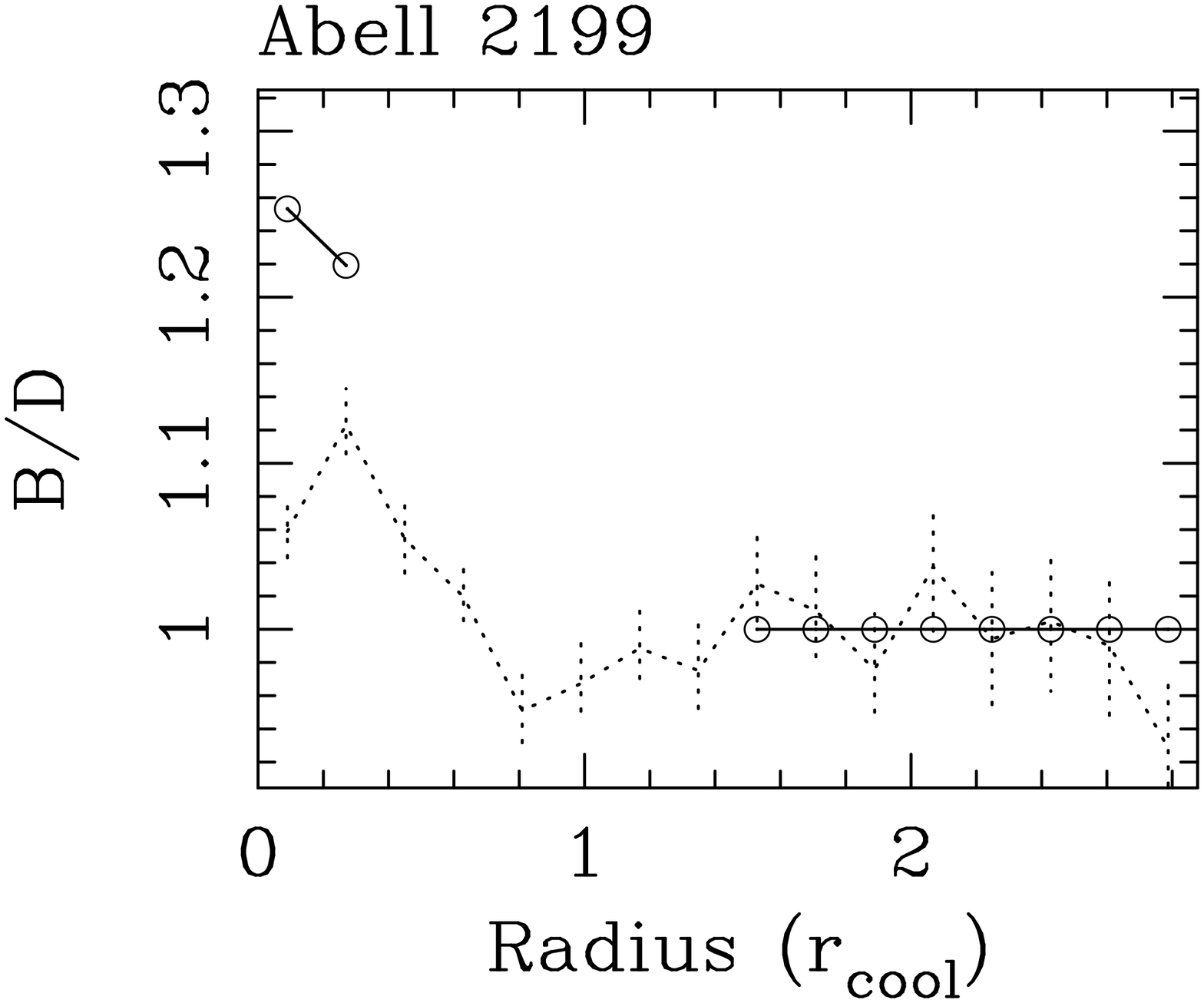,width=0.47\textwidth,angle=270}
}

\caption{A summary of the results on the distributions of cool gas and
intrinsic absorbing material in Abell 426,
Abell 1795 and Abell 2199. The y axes show the C/D and B/D ratios
normalized to the values at large radii (Table 5). The x axes are in units
of the cooling radii, $r_{\rm cool}$ (see Table 3). The C/D profiles 
clearly demonstrate the presence of distributed cool gas within the cooling
radii, and the B/D data the need for excess absorption in the central
regions. Other details as in Figs. 3 and 5.
}
\vskip 2cm
\end{figure*}

The joint confidence contours on the column density and
covering fraction of the excess absorber in Abell 2199 
are shown in Fig. 12. The contours indicate the 
regions of 68, 90 and 99 per cent confidence on two interesting parameters.
The SIS and SSS data, in particular, exhibit good agreement.
For the SIS data, the best-fit 
excess column density is $2.5^{+1.1}_{-1.0} \times 10^{21}$ \apc~and the 
covering fraction, $f = 0.53^{+0.17}_{-0.07}$. 
For the PSPC data, the best-fit values are $\Delta N_{\rm H} =
1.2^{+0.7}_{-0.3} \times 10^{21}$ \apc~and $f=0.43^{+0.14}_{-0.13}$, 
where errors are the 90 per cent ($\Delta \chi^2 = 2.71$) 
confidence limits on a single interesting parameter.

The simple partial covering model thus provides a reasonable
description of the excess absorption in Abell 2199 and leads to 
reasonable agreement in the parameter values inferred from the
different data sets. However, although the errors on the individual 
parameter values overlap, the $\Delta \chi^2$ contours shown in Fig. 12 
indicate an inconsistency between the PSPC data and the other two data sets. 
This implies either a deficiency in the spectral model 
or a calibration problem with the instruments. 
The simple partial-covering model used undoubtedly
underestimates the complexity of the absorption in a real cluster,  where 
the column density acting on the
emission from different regions is likely to span a range of values. 
The abundances of separate elements in the absorber could also
deviate from the assumed solar values, perhaps as a function of radius.
Further complexities would be introduced if the 
absorbing material were dusty, which recent evidence suggests is likely 
to be the case (Fabian, Johnstone \& Daines, Allen \etal 1995, Voit \& Donahue 1995,
Allen \etal 1996). If, however, the origin of the discrepancy is a calibration 
problem, the agreement between the SIS and SSS results would suggest that 
the problem is likely to lie with the PSPC data.~~~~~~~~~~~~~~~~~~~~~

\subsection{The mass of absorbing gas} 

\begin{table*}
\vskip 0.2truein
\begin{center}
\caption{The masses of absorbing gas and accumulation timescales}
\vskip 0.2truein
\begin{tabular}{ c c c c c c }
 Cluster   & ~ & $M_{\rm abs}$        & ${\dot M}$            & $t_{\rm acc}$          &  $M_{\rm hot}$ \\
           & ~ &  ($10^9$ \Msun)      &  (\Msunpyr)           & ($10^{8}$ yr)          &  ($10^{10}$ \Msun)  \\  
\hline                                                                                               
&&&&& \\                                                                                               
Abell 85   & ~ & $15.0^{+4.7}_{-4.4}$ & $36.3^{+1.4}_{-1.9}$  & $4.1^{+1.3}_{-1.1}$    & $20.9 \pm 0.4$ \\  
Abell 426  & ~ & $0.4^{+0.5}_{-0.4}$  & $16.0^{+0.9}_{-1.1}$  & $0.25^{+0.31}_{-0.25}$ & $3.0 \pm 0.1$  \\  
Abell 478  & ~ & $143^{+19}_{-17}$    & $160^{+4}_{-5}$       & $8.9^{+1.2}_{-1.1}$    & $88.6 \pm 1.0$ \\  
Abell 496  & ~ & $6.8^{+2.2}_{-1.8}$  & $17.2^{+0.9}_{-0.7}$  & $4.0^{+1.3}_{-1.1}$    & $6.8 \pm 0.1$  \\  
Hydra A    & ~ & $21.1^{+3.9}_{-3.4}$ & $69.4^{+1.7}_{-1.2}$  & $3.0^{+0.6}_{-0.5}$    & $25.9 \pm 0.3$ \\  
Abell 1795 & ~ & $20.0^{+2.4}_{-2.0}$ & $83.4^{+2.0}_{-2.1}$  & $2.4^{+0.3}_{-0.2}$    & $35.7 \pm 0.3$ \\  
Abell 2029 & ~ & $9.5^{+5.6}_{-5.1}$  & $108^{+3.4}_{-4.3}$   & $0.88^{+0.52}_{-0.47}$ & $57.7 \pm 0.9$ \\  
MKW3s      & ~ & $34.7^{+5.6}_{-5.1}$ & $19.7^{+1.4}_{-1.4}$  & $17.6^{+2.9}_{-2.6}$   & $9.0 \pm 0.3$  \\  
Abell 2199 & ~ & $6.7^{+0.8}_{-0.7}$  & $13.6^{+0.4}_{-0.3}$  & $4.9^{+0.6}_{-0.5}$    & $5.6 \pm 0.1$  \\  
Cygnus A   & ~ & $44^{+11}_{-10}$     & $118^{+5}_{-5}$       & $3.7^{+0.9}_{-0.8}$    & $39.3 \pm 0.6$  \\  
Sersic 159 & ~ & $13.3^{+3.8}_{-3.8}$ & $61.6^{+2.3}_{-1.8}$  & $2.2^{+0.6}_{-0.6}$    & $24.6 \pm 0.3$  \\  
Abell 2597 & ~ & $35^{+10}_{-10}$     &  $126^{+4}_{-5}$      & $2.8^{+0.8}_{-0.8}$    & $72.5 \pm 1.1$  \\  
&&&&& \\                                                                   
\end{tabular}                                                              
\end{center}                                                               
                                                                           
\parbox {7in}                                                              
{The masses of absorbing gas, $M_{\rm abs}$ (calculated assuming solar
metallicity in the absorbing material), mass deposition 
rates, ${\dot M}$, accumulation timescales, $t_{\rm acc}$, 
and masses of hot, X-ray emitting gas, $M_{\rm hot}$,  in the central 30 arcsec 
regions of the clusters. The $M_{\rm abs}$ results are derived from the
[$\Delta N_{\rm H}(1)$]  column density
measurements made with the partial covering model (Table 6, Column 4).
A covering fraction $f=0.5$ has been assumed for all clusters 
except MKW3s, for which $f=0.75$ was used.
The ${\dot M}$ and $M_{\rm hot}$ values apply to the central 30 arcsec (radius) volume.
}
\end{table*}

The results on the intrinsic column densities, summarized in Table 6,
allow a simple estimation of the mass of absorbing gas in the central 30 
arcsec (radius) regions of the clusters to be made. 
Adopting the partial covering
model for the distribution of the absorbing gas,
with $f = 0.5$, and assuming solar metallicity in this material, 
the mass of gas can be
written as $M_{\rm abs} = 1.62 \times 10^6 r^2 \Delta N_{\rm
H,20}$ \Msun, where $r$ is the
radius of the 30 arcsec region in kpc and $\Delta N_{\rm H,20}$ is the
intrinsic column density in units of $10^{20}$ \apc. 
We can also estimate the time, $t_{\rm acc}$, required for the 
cooling flows to accumulate these masses of gas, 
given the observed mass
deposition rates in these regions ($t_{\rm acc} =
M_{\rm abs}/{\dot M}$). [Note that the mass of absorbing gas is 
actually a projected mass through the central 30 arcsec region, whereas
the mass deposition rates apply to the central 30 arcsec volume. Thus,
where significant mass deposition occurs outside the
central 30 arcsec region,  the accumulation timescales, $t_{\rm acc}$,
quoted will slightly overestimate the
true values. Assuming ${\dot M \propto r}$, the projected mass deposition 
rate through the central 30 arcsec region will be $\sim
30$ per cent higher than the rate within that volume, and 
the accumulation timescale will be $\sim 25$ per cent shorter than the
quoted result.]
The masses of absorbing gas, mass deposition rates, accumulation
times, and the masses of hot X-ray emitting gas within the central
30 arcsec (radius) volumes, are summarized in Table 7. 

From Table 7 we see that the accumulation timescales are typically only a
few $10^8$ years, much less than the ages of the cooling flows ({\it
cf.} Section 5.5.) This implies that most of the mass deposited by the
cooling flows in the central regions of the clusters must reside in some
form other than X-ray absorbing gas (perhaps low mass stars or brown dwarfs;
Fabian 1994). Note, however, that if the metallicity of the absorbing
material is substantially sub-solar, the true masses of absorbing gas
are likely to be $\sim 2-3$ times larger than the values listed in Table 7. 
Note also that the mass of absorbing material (assuming solar metallicity) 
is typically
$\approxlt 10$ per cent of the mass of the X-ray emitting gas in the same
regions (although for Abell 478 and
MKW3s the values are somewhat higher.)


\section{Conclusions} 

In this paper we have described and discussed the results from a detailed
X-ray colour profile analysis of a sample of nearby, luminous cooling
flow clusters observed with the ROSAT PSPC. Our analysis of the C/D
($0.80-1.39$ keV/$1.40-2.00$ keV) ratios clearly demonstrates the presence
of distributed cool gas in the central regions ($r \approxlt r_{\rm
cool}$) of the cooling flow clusters. The spatial distributions and
emissivity of the cooling
gas are in excellent agreement with the predictions from standard image
deprojection analyses. In the outer regions of the clusters the data
suggest approximate isothermality. Our results place new
constraints on the ages of the cooling flows and provide
the most accurate determinations of the mass deposition rates in these
systems of any imaging study to date. 

The analysis of the B/D ($0.41-0.79$ keV/$1.40-2.00$ keV) ratios provides
firm evidence for spatially distributed intrinsic absorbing material in
the cooling flows. The absorbing gas generally 
exhibits an increasing column density
with decreasing radius, and is confined to radii $\approxlt r_{\rm
cool}$. Our results are consistent with the large
intrinsic column densities ($\Delta N_{\rm H} \sim$ $10^{21}$ \apc)
inferred from previous studies with the 
Einstein Observatory SSS and ASCA (White \etal 1991; 
Fabian \etal 1994) but require that the absorber 
only partially covers the emitting region. 
This suggests significant clumping of the absorber 
on large scales. 

We have discussed the complexities involved in the measurement 
of intrinsic column densities in clusters and have shown that 
the use of simple spectral models, which treat the absorber as a
uniform foreground screen (in front of the cooling flow or cluster) 
will naturally lead to significant underestimates 
(by factors of $\sim 2-5$) of the true amounts of absorbing material.

Assuming solar metallicity in the absorbing material, the masses of absorbing 
gas in the central 30 arcsec regions of the clusters 
typically correspond to $\approxlt 10$ per cent of the mass of the 
hot, X-ray emitting gas. The timescales required for the 
cooling flows to accumulate such
masses are significantly shorter than ($\sim 10$ per
cent of) the likely ages of the cooling flows, and imply that the bulk of
the matter deposited by the cooling flows (in the central 30 arcsec
radius regions) resides in some form other than
X-ray absorbing gas, presumably stars or sub-stellar objects. 

The new results presented in this paper provide strong support for the
standard inhomogeneous model of cooling flows in clusters of
galaxies. Future studies of the range of temperature/density phases 
in cooling flows (as a function of radius), more precise determinations of 
the `ages' of cooling flows, and more detailed mapping of the intrinsic column
densities in clusters, will become possible following the launch of AXAF.

\section*{Acknowledgments}

We thank Roderick Johnstone and Dave White for software assistance, and 
the Royal Society for support.



\begin{thebibliography}{}
\bibitem{} Allen S.W., 1995, MNRAS, 276, 947
\bibitem{} Allen S.W., Fabian A.C., 1994, MNRAS, 269, 409
\bibitem{} Allen S.W., Fabian A.C., Kneib J.P., 1996a, MNRAS, 279, 615
\bibitem{} Allen S.W., Fabian A.C., Johnstone R.M., Nulsen P.E.J., Edge A.C., 1992, MNRAS, 254, 51
\bibitem{} Allen S.W., Fabian A.C., Johnstone R.M., White D.A., Daines S.J., Edge A.C., Stewart G.C., 1993, MNRAS, 262, 901
\bibitem{} Allen S.W., Fabian A.C., Edge A.C., B\"ohringer H., White D.A., 1995, MNRAS, 275, 741
\bibitem{} Allen S.W., Fabian A.C., Edge, A.C., Bautz M.W., Furuzawa A., Tawara Y., 1996, MNRAS, in press
\bibitem{} Anders E., Grevesse N., 1989, Geochemica et Cosmochimica Acta 53, 197
\bibitem{} B\"ohringer H., Voges W., Fabian A.C., Edge A.C., Neumann D.M., 1993, MNRAS, 264, L25
\bibitem{} Crawford C.S., Fabian A.C., 1992, MNRAS, 259, 265	
\bibitem{} Cowie L.L.,  Binney J., 1977, ApJ, 215, 723
\bibitem{} Edge A.C., Stewart G.C., Fabian A.C., Arnaud K.A., 1990, MNRAS, 245,559
\bibitem{} Edge A.C., Stewart G.C., Fabian A.C., 1992, MNRAS, 258, 177
\bibitem{} Fabian A.C., 1994, A\&AR, 32, 277
\bibitem{} Fabian A.C., Nulsen P.E.J., 1977, MNRAS, 180, 479 
\bibitem{} Fabian A.C., Nulsen P.E.J., Canizares C.R., 1991, A\&AR, 2, 191 
\bibitem{} Fabian A.C., Johnstone R.M., Daines S.J., 1994, MNRAS, 271, 737
\bibitem{} Fabian A.C., Hu E.M., Cowie L.L., Grindlay J., 1981, ApJ, 248, 47
\bibitem{} Fabian A.C., Arnaud K.A., Bautz M.W., Tawara Y., 1994, ApJ, 436, L63
\bibitem{} Fabian A.C. \etal 1996, in preparation
\bibitem{} Fukazawa Y. \etal 1994, PASJ, 46, L55
\bibitem{} Hasinger G., Turner T.J., George I.M., Boese G., 1992, OGIP Calibration Memo CAL/ROS/92-001, NASA
\bibitem{} Hu E.M., Cowie L.L., Wang Z., 1985, ApJS, 59, 447
\bibitem{} Irwin J.A., Sarazin C.L., 1995, ApJ, 355, 497 
\bibitem{} Johnstone R.M., Fabian A.C., Nulsen P.E.J., 1987, MNRAS, 224, 75
\bibitem{} Johnstone R.M., Fabian A.C., Edge A.C., Thomas P.A., 1992, MNRAS, 255, 431
\bibitem{} Kaastra J.S., Mewe R., 1993, Legacy, 3, HEASARC, NASA
\bibitem{} Mathews W.G., Bregman J.N., 1978, ApJ, 244, 308
\bibitem{} Morrison R., McCammon D.M., 1983, ApJ, 270, 119
\bibitem{} Ohashi T. \etal 1996, in preparation
\bibitem{} Shafer R.A., Haberl F., Arnaud K.A., Tennant A.F., 1991, XSPEC User's Guide,. ESA, Noordwijk
\bibitem{} Stark A.A., Gammie C.F., Wilson R.W., Bally J., Linke R.A., Heiles C. \& Hurwitz M., 1992. ApJS, 79, 77
\bibitem{} Voit G.M., Donahue M., 1995, ApJ, 452, 164 
\bibitem{} Thomas P.A., Fabian A.C., Nulsen P.E.J, 1987, MNRAS, 228, 973
\bibitem{} White D.A., Fabian A.C., Johnstone R.M., Mushotzky R.F., Arnaud K.A., 1991, MNRAS, 252, 72
\end{thebibliography}
\end{document}